%%%%%%%%%%%%%%%%%%  tex macros for preprints, cm version %%%%%%%%%%%%%%
%         (P. Ginsparg <ginsparg@lanl.gov>, last updated 7/94)
%         hypertex extensions (still provisional), 7/26/94
%	  Some modifications by C.R.Mafra, 2012

%comment out this line to restore non-hyper functionality
%\input hyperbasics

\input amssym.tex % for blackboard bold

\def\unredoffs{}
\tolerance=1000\hfuzz=2pt
\catcode`\@=11 % This allows us to modify PLAIN macros.
\ifx\hyperdef\UNd@FiNeD\def\hyperdef#1#2#3#4{#4}\def\hyperref#1#2#3#4{#4}\def\href#1#2{#2}\fi
\magnification=1200\unredoffs\baselineskip=16pt plus 2pt minus 1pt
\def\Date#1{\vfill\leftline{#1}\tenpoint\supereject%
\footline={\hss\tenrm\hyperdef\hypernoname{page}\folio\folio\hss}}%
% (restores pagenumbers)

%%%%%% Hour:Minute %%%%%%%%%%%%%%%%%
{\count255=\time\divide\count255 by 60 \xdef\hourmin{\number\count255}
 \multiply\count255 by-60\advance\count255 by\time
 \xdef\hourmin{\hourmin:\ifnum\count255<10 0\fi\the\count255}
}
\def\date{\number\day.\number\month.\number\year\ at \hourmin}

%%%%%%%%%%%% Draft mode %%%%%%%%%%%%%
% puts date/time on each page in big mode, writes labels in margins

% use \nolabels to get rid of eqn, ref, and fig labels in draft mode
\def\nolabels{\def\wrlabeL##1{}\def\eqlabeL##1{}\def\reflabeL##1{}}
\def\writelabels{\def\wrlabeL##1{\leavevmode\vadjust{\rlap{\smash%
{\line{{\escapechar=` \hfill\rlap{\sevenrm\hskip.03in\string##1}}}}}}}%
\def\eqlabeL##1{{\escapechar-1\rlap{\sevenrm\hskip.05in\string##1}}}%
\def\reflabeL##1{\noexpand\llap{\noexpand\sevenrm\string\string\string##1}}}
\nolabels

% tagged sec numbers
\global\newcount\secno \global\secno=0
\global\newcount\meqno \global\meqno=1
\def\s@csym{}

%%%%%%%%% Section %%%%%%%%%%%%%
\def\newsec#1\par{\global\advance\secno by1%
{\toks0{#1}\message{(\the\secno. \the\toks0)}}%
\global\subsecno=0\eqnres@t\let\s@csym\secsym\xdef\secn@m{\the\secno}\noindent
{\bf\hyperdef\hypernoname{section}{\the\secno}{\the\secno.} #1}%
\writetoca{{\string\hyperref{}{section}{\the\secno}{\bf \the\secno\quad}} {\bf #1}}\par%
\nobreak\medskip\nobreak\noindent\ignorespaces}
\def\eqnres@t{\xdef\secsym{\the\secno.}\global\meqno=1\bigbreak\bigskip}
\def\sequentialequations{\def\eqnres@t{\bigbreak}}\xdef\secsym{}

%%%%%%%% Subsection %%%%%%%%%%%
\global\newcount\subsecno \global\subsecno=0
\def\subsec#1\par{\global\advance\subsecno by1%
{\toks0{#1}\message{(\s@csym\the\subsecno. \the\toks0)}}%
\global\subsubsecno=0%
\ifnum\lastpenalty>9000\else\bigbreak\fi
\noindent{\it\hyperdef\hypernoname{subsection}{\secn@m.\the\subsecno}%
{\secn@m.\the\subsecno.} #1}\writetoca{\string\hskip1.45cm
{\string\hyperref{}{subsection}{\secn@m.\the\subsecno}{\secn@m.\the\subsecno.}}
{#1}}\par\nobreak\medskip\nobreak\noindent\ignorespaces}

%%%%%%% Appendix %%%%%%%%%%%%%%
\def\appendix#1#2{\global\meqno=1\global\subsecno=0\xdef\secsym{\hbox{#1.}}%
\bigbreak\bigskip\noindent{\bf Appendix \hyperdef\hypernoname{appendix}{#1}%
{#1.} #2}{\toks0{(#1. #2)}\message{\the\toks0}}%
\xdef\s@csym{#1.}\xdef\secn@m{#1}%
\writetoca{{\string\hyperref{}{appendix}{#1}{\bf {#1}\quad}} {\bf #2}}%
\par\nobreak\medskip\nobreak}

% \eqn\label{a+b=c}   gives displayed equation, numbered consecutively within sections.
% \eqnn, \eqna        define labels in advance, use \eqna\label before an eqalign and
%                     later \label a, \label b etc inside eqalign to get (2.3a), (2.3b) etc
%
\def\checkm@de#1#2{\ifmmode{\def\f@rst##1{##1}\hyperdef\hypernoname{equation}%
{#1}{#2}}\else\hyperref{}{equation}{#1}{#2}\fi}
\def\eqnn#1{\DefWarn#1\xdef #1{(\noexpand\relax\noexpand\checkm@de%
{\s@csym\the\meqno}{\secsym\the\meqno})}%
\wrlabeL#1\writedef{#1\leftbracket#1}\global\advance\meqno by1}
\def\f@rst#1{\c@t#1a\em@ark}\def\c@t#1#2\em@ark{#1}
\def\eqna#1{\DefWarn#1\wrlabeL{#1$\{\}$}%
\xdef #1##1{(\noexpand\relax\noexpand\checkm@de%
{\s@csym\the\meqno\noexpand\f@rst{##1}1}{\hbox{$\secsym\the\meqno##1$}})}
\writedef{#1\numbersign1\leftbracket#1{\numbersign1}}\global\advance\meqno by1}
\def\eqn#1#2{\DefWarn#1%
\xdef #1{(\noexpand\hyperref{}{equation}{\s@csym\the\meqno}%
{\secsym\the\meqno})}$$#2\eqno(\hyperdef\hypernoname{equation}%
{\s@csym\the\meqno}{\secsym\the\meqno})\eqlabeL#1$$%
\writedef{#1\leftbracket#1}\global\advance\meqno by1}
\def\xeqn{\expandafter\xe@n}\def\xe@n(#1){#1}
\def\xeqna#1{\expandafter\xe@n#1}
\def\eqns#1{(\e@ns #1{\hbox{}})}
\def\e@ns#1{\ifx\UNd@FiNeD#1\message{eqnlabel \string#1 is undefined.}%
\xdef#1{(?.?)}\fi{\let\hyperref=\relax\xdef\next{#1}}%
\ifx\next\em@rk\def\next{}\else%
\ifx\next#1\xeqn#1\else\def\n@xt{#1}\ifx\n@xt\next#1\else\xeqna#1\fi
\fi\let\next=\e@ns\fi\next}

\def\DefWarn#1{\ifx\UNd@FiNeD#1\else
\immediate\write16{*** WARNING: the label \string#1 is already defined ***}\fi}
%
% footnotes
\newskip\footskip\footskip14pt plus 1pt minus 1pt %sets footnote baselineskip
\def\footnotefont{\ninepoint}\def\f@t#1{\footnotefont #1\@foot}
\def\f@@t{\baselineskip\footskip\bgroup\footnotefont\aftergroup\@foot\let\next}
\setbox\strutbox=\hbox{\vrule height9.5pt depth4.5pt width0pt}
\global\newcount\ftno \global\ftno=0
\def\foot{\global\advance\ftno by1\def\foot@rg{\hyperref{}{footnote}%
{\the\ftno}{\the\ftno}\xdef\foot@rg{\noexpand\hyperdef\noexpand\hypernoname%
{footnote}{\the\ftno}{\the\ftno}}}\footnote{$^{\foot@rg}$}}
%
%
%     \ref\label{text}
% generates a number, assigns it to \label, generates an entry.
% To list the refs on a separate page,  \listrefs
%
\global\newcount\refno \global\refno=1
\newwrite\rfile
\def\ref{[\hyperref{}{reference}{\the\refno}{\the\refno}]\nref}
\def\nref#1{\DefWarn#1%
\xdef#1{[\noexpand\hyperref{}{reference}{\the\refno}{\the\refno}]}%
\writedef{#1\leftbracket#1}%
\ifnum\refno=1\immediate\openout\rfile=\jobname.refs\fi
\chardef\wfile=\rfile\immediate\write\rfile{\noexpand\item{[\noexpand\hyperdef%
\noexpand\hypernoname{reference}{\the\refno}{\the\refno}]\ }%
\reflabeL{#1\hskip.31in}\pctsign}\global\advance\refno by1\findarg}
%	horrible hack to sidestep tex \write limitation
\def\findarg#1#{\begingroup\obeylines\newlinechar=`\^^M\pass@rg}
{\obeylines\gdef\pass@rg#1{\writ@line\relax #1^^M\hbox{}^^M}%
\gdef\writ@line#1^^M{\expandafter\toks0\expandafter{\striprel@x #1}%
\edef\next{\the\toks0}\ifx\next\em@rk\let\next=\endgroup\else\ifx\next\empty%
\else\immediate\write\wfile{\the\toks0}\fi\let\next=\writ@line\fi\next\relax}}
\def\striprel@x#1{} \def\em@rk{\hbox{}}
\def\lref{\begingroup\obeylines\lr@f}
\def\lr@f#1#2{\DefWarn#1\gdef#1{\let#1=\UNd@FiNeD\ref#1{#2}}\endgroup\unskip}
\def\semi{;\hfil\break}
\def\addref#1{\immediate\write\rfile{\noexpand\item{}#1}} %now unnecessary
\def\listrefs{\vfill\supereject\immediate\closeout\rfile\writestoppt
\baselineskip=\footskip\centerline{{\bf References}}\bigskip{\parindent=20pt%
\frenchspacing\escapechar=` \input \jobname.refs\vfill\eject}\nonfrenchspacing}
\def\startrefs#1{\immediate\openout\rfile=\jobname.refs\refno=#1}
\def\xref{\expandafter\xr@f}\def\xr@f[#1]{#1}
\def\refs#1{\count255=1[\r@fs #1{\hbox{}}]}
\def\r@fs#1{\ifx\UNd@FiNeD#1\message{reflabel \string#1 is undefined.}%
\nref#1{need to supply reference \string#1.}\fi%
\vphantom{\hphantom{#1}}{\let\hyperref=\relax\xdef\next{#1}}%
\ifx\next\em@rk\def\next{}%
\else\ifx\next#1\ifodd\count255\relax\xref#1\count255=0\fi%
\else#1\count255=1\fi\let\next=\r@fs\fi\next}
%

%
% this is ugly, but moore insists
\newwrite\ffile\global\newcount\figno \global\figno=1
\def\fig{fig.~\hyperref{}{figure}{\the\figno}{\the\figno}\nfig}
\def\nfig#1{\DefWarn#1%
\xdef#1{fig.~\noexpand\hyperref{}{figure}{\the\figno}{\the\figno}}%
\writedef{#1\leftbracket fig.\noexpand~\xfig#1}%
\ifnum\figno=1\immediate\openout\ffile=\jobname.figs\fi\chardef\wfile=\ffile%
{\let\hyperref=\relax
\immediate\write\ffile{\noexpand\medskip\noexpand\item{Fig.\ %
\noexpand\hyperdef\noexpand\hypernoname{figure}{\the\figno}{\the\figno}. }
\reflabeL{#1\hskip.55in}\pctsign}}\global\advance\figno by1\findarg}
\def\xfig{\expandafter\xf@g}\def\xf@g fig.\penalty\@M\ {}
\def\figs#1{figs.~\f@gs #1{\hbox{}}}
\def\f@gs#1{{\let\hyperref=\relax\xdef\next{#1}}\ifx\next\em@rk\def\next{}\else
\ifx\next#1\xfig #1\else#1\fi\let\next=\f@gs\fi\next}
%
%% because TeXlive 2011 is buggy wrt to tikz pictures with plain TeX..
\def\figin{\epsfcheck\figin}\def\figins{\epsfcheck\figins}
\def\epsfcheck{\ifx\epsfbox\UnDeFiNeD
\message{(NO epsf.tex, FIGURES WILL BE IGNORED)}
\gdef\figin##1{\vskip2in}\gdef\figins##1{\hskip.5in}% blank space instead
\else\message{(FIGURES WILL BE INCLUDED)}%
\gdef\figin##1{##1}\gdef\figins##1{##1}\fi}
\def\DefWarn#1{}
\def\figinsert{\goodbreak\topinsert}
\def\ifig#1#2#3{\DefWarn#1\xdef#1{fig.~\the\figno}
\writedef{#1\leftbracket fig.\noexpand~\the\figno}%
\figinsert\figin{\centerline{#3}}
\smallskip
\leftskip=0pt \rightskip=0pt
\baselineskip12pt\noindent
{{\bf Fig.~\the\figno}\ \ninepoint #2}
\medskip
\global\advance\figno by1\par\endinsert}
%%%%%%%%%%%%%%%%%%%%%%%%%%%%%%%%%%%%%%%%%%%%%%%%%%%%%%%%%
\newwrite\lfile
{\escapechar-1\xdef\pctsign{\string\%}\xdef\leftbracket{\string\{}
\xdef\rightbracket{\string\}}\xdef\numbersign{\string\#}}
\def\writedefs{\immediate\openout\lfile=label.defs \def\writedef##1{%
{\let\hyperref=\relax\let\hyperdef=\relax\let\hypernoname=\relax
 \immediate\write\lfile{\string\def\string##1\rightbracket}}}}%
\def\writestop{\def\writestoppt{\immediate\write\lfile{\string\pageno
 \the\pageno\string\startrefs\leftbracket\the\refno\rightbracket
 \string\def\string\secsym\leftbracket\secsym\rightbracket
 \string\secno\the\secno\string\meqno\the\meqno}\immediate\closeout\lfile}}
\def\writestoppt{}\def\writedef#1{}

% Section, subsection and appendix labels %
% Note that there must be a blanck line after \newsec,\subsec and before \seclab,\subseclab!
\def\seclab#1{\DefWarn#1%
\xdef #1{\noexpand\hyperref{}{section}{\the\secno}{\the\secno}}%
\writedef{#1\leftbracket#1}\wrlabeL{#1=#1}\par%
\nobreak\medskip\nobreak\noindent\ignorespaces}
\def\subseclab#1\par{\DefWarn#1%
\xdef #1{\noexpand\hyperref{}{subsection}{\the\secno.\the\subsecno}%
{\the\secno.\the\subsecno}}\writedef{#1\leftbracket#1}\wrlabeL{#1=#1}\par%
\nobreak\medskip\nobreak\noindent\ignorespaces}
\def\applab#1{\DefWarn#1%
\xdef #1{\noexpand\hyperref{}{appendix}{\secn@m}{\secn@m}}%
\writedef{#1\leftbracket#1}\wrlabeL{#1=#1}}
\def\appsublab#1{\DefWarn#1%
\xdef #1{\noexpand\hyperref{}{appendix}{\secn@m.\the\subsecno}{\secn@m.\the\subsecno}}%
\writedef{#1\leftbracket#1}\wrlabeL{#1=#1}}
\newwrite\tfile \def\writetoca#1{}
\def\leaderfill{\leaders\hbox to 1em{\hss.\hss}\hfill}
% use this to write file with table of contents
\def\writetoc{\immediate\openout\tfile=\jobname.toc
   \def\writetoca##1{{\edef\next{\write\tfile{\noindent ##1
   \string\leaderfill{
% comment this line if you don't want hyperlinked page numbers on TOC
   \string\hyperref{}{page}{\noexpand\number\pageno}%
   {\noexpand\number\pageno}} \par}}\next}}
}
% and this lists table of contents on second pass
\newread\ch@ckfile
\def\listtoc{\immediate\closeout\tfile\immediate\openin\ch@ckfile=\jobname.toc
\ifeof\ch@ckfile\message{no file \jobname.toc, no table of contents this pass}%
\else\closein\ch@ckfile\centerline{\bf Contents}\nobreak\medskip%
{\baselineskip=16pt\footnotefont\parskip=0pt\catcode`\@=11\input\jobname.toc
\catcode`\@=12\bigbreak\bigskip}\fi}
\catcode`\@=12 % at signs are no longer letters
\def\tenpoint{\def\rm{\fam0\tenrm}% switch back to 10-point type
\textfont0=\tenrm \scriptfont0=\sevenrm \scriptscriptfont0=\fiverm
\textfont1=\teni  \scriptfont1=\seveni  \scriptscriptfont1=\fivei
\textfont2=\tensy \scriptfont2=\sevensy \scriptscriptfont2=\fivesy
\textfont\itfam=\tenit \def\it{\fam\itfam\tenit}\def\footnotefont{\ninepoint}%
\textfont\bffam=\tenbf \def\bf{\fam\bffam\tenbf}\def\sl{\fam\slfam\tensl}\rm}
\font\ninerm=cmr9 \font\sixrm=cmr6 \font\ninei=cmmi9 \font\sixi=cmmi6
\font\ninesy=cmsy9 \font\sixsy=cmsy6 \font\ninebf=cmbx9
\font\nineit=cmti9 \font\ninesl=cmsl9 \skewchar\ninei='177
\skewchar\sixi='177 \skewchar\ninesy='60 \skewchar\sixsy='60
\def\ninepoint{\def\rm{\fam0\ninerm}% switch to footnote font
\textfont0=\ninerm \scriptfont0=\sixrm \scriptscriptfont0=\fiverm
\textfont1=\ninei \scriptfont1=\sixi \scriptscriptfont1=\fivei
\textfont2=\ninesy \scriptfont2=\sixsy \scriptscriptfont2=\fivesy
\textfont\itfam=\ninei \def\it{\fam\itfam\nineit}\def\sl{\fam\slfam\ninesl}%
\textfont\bffam=\ninebf \def\bf{\fam\bffam\ninebf}\rm}
%
%---------------------------------------------------------------------
\hyphenation{anom-aly anom-alies coun-ter-term coun-ter-terms}

%%%%%%%%%%%%%%% Subsubsection %%%%%%%%%%%%%%%%%%%%%%%%%%%%%%%%%%%%
\global\newcount\subsubsecno \global\subsubsecno=0
\def\subsubsec#1\par{\global\advance\subsubsecno by1%
{\toks0{#1}\message{(\the\secno\the\subsecno\the\subsubsecno. \the\toks0)}}%
\ifnum\lastpenalty>9000\else\bigbreak\fi
\noindent{\it\hyperdef\hypernoname{subsubsection}{\the\secno.\the\subsecno\the\subsubsecno}%
{\the\secno.\the\subsecno.\the\subsubsecno.} #1}
%%% Add Subsubsections to Index
%\writetoca{\string\quad{\string\hyperref{}{subsubsection}{\the\secno\the\subsecno\the
%\subsubsecno}{\baselineskip=9pt\it\the\secno.\the\subsecno.\the\subsubsecno.}}
% {\baselineskip=9pt\it\ #1}}
\par\nobreak\medskip\nobreak\noindent\ignorespaces}

% Caption for inline tikzpictures
\def\DefWarn#1{}
\def\tikzcaption#1#2{\DefWarn#1\xdef#1{Fig.~\the\figno}
\writedef{#1\leftbracket Fig.\noexpand~\the\figno}%
{
\smallskip
\leftskip=20pt \rightskip=20pt \baselineskip12pt\noindent
{{\bf Fig.~\the\figno}\ \ninepoint #2}
\bigskip
\global\advance\figno by1 \par}}

% convert numbers [1-9] to upper case letters [A-I]
\def\ntoalpha#1{%
\ifcase#1%
@%
\or A\or B\or C\or D\or E\or F\or G\or H\or I
\fi
}

% Appendix label
\global\newcount\appno \global\appno=1
\def\applab#1{\xdef #1{\ntoalpha\appno}\writedef{#1\leftbracket#1}\wrlabeL{#1=#1}
\global\advance\appno by1}

% Clean up the title page definitions
\def\preprint#1 #2\par{\rightline{\vbox{\baselineskip12pt\hbox{#1}\hbox{#2}}}\vskip2cm}
% title with more than one line (note the blanck line in between)
%\title some line
%
%\tile another line
\def\title#1\par{\centerline{\bf #1}\nopagenumbers\pageno=0}
\def\author#1\par{\bigskip\bigskip\centerline{#1}}

\newcount\addressno

\def\email#1#2{%\unskip$^#1$
\footnote{\null}{\kern-\parindent \llap{$^#1$\hskip1pt}email: #2}}

% centermode for address lines
\def\startcenter{%
  \par
  \begingroup
  \leftskip=0pt plus 1fil
  \rightskip=\leftskip
  \parindent=0pt
  \parfillskip=0pt
}
\def\stopcenter{\endgroup}

\def\address{\bigskip%
  \ifnum\the\addressno=0\else\stopcenter\endgroup\fi
  \advance\addressno by 1%
  \begingroup
  \startcenter
  \it
  \obeylines
  \addressAux
}
\def\addressAux#1{#1}

% need to stop center mode and obeylines from address
\def\abstract{\stopcenter\endgroup\bigskip\bigskip\noindent}

% some sample definitions
\def\Dsl{\,\raise.15ex\hbox{/}\mkern-13.5mu D} %this one can be subscripted
\def\dsl{\raise.15ex\hbox{/}\kern-.57em\partial}
 
\def\boxeqn#1{\vcenter{\vbox{\hrule\hbox{\vrule\kern3pt\vbox{\kern3pt
	\hbox{${\displaystyle #1}$}\kern3pt}\kern3pt\vrule}\hrule}}}

 %pound sterling

\def\ap{{\alpha^{\prime}}}

\def\a{\alpha}
\def\b{{\beta}}
\def\g{{\gamma}}

\def\l{\lambda}

\def\s{{\sigma}}
\def\t{{\theta}}

\def\half{{1\over 2}}
\def\p{{\partial}}

\def\({\left(}
\def\){\right)}

\def\cA{{\cal A}}
\def\cF{{\cal F}}

\def\cW{{\cal W}}

\def\cZ{{\cal Z}}

% blackboard bold

% primed summation symbol

% length of words, |P|
\def\len#1{{%
\def\Dlen{\left|\mkern-1mu #1\mkern -0.5mu\right|}%
\def\Sslen{\left|\mkern-1.3mu #1\mkern -1.3mu\right|}%
\def\SSlen{\left|\mkern-2.8mu #1\mkern-1.3mu\right|}%
\mathchoice{\Dlen}{\Dlen}{\Sslen}{\SSlen}}}

 %redefine plain TeX \Im..
% small inlined fractions, from the TeXbook
\def\sfrac#1/#2{\kern.1em\raise.5ex\hbox{\the\scriptfont0 #1}%
\kern-.1em/\kern-.15em\lower.25ex\hbox{\the\scriptfont0 #2}}

%shuffle product
\font\tenshuffle=shuffle10 \font\sevenshuffle=shuffle7 \font\fiveshuffle=shuffle7 at 5pt
\def\shuffle{{%
\def\Dshuffle{\mathbin{\hbox{\tenshuffle\char'001}}}%
\def\Sshuffle{\mathbin{\hbox{\sevenshuffle\char'001}}}%
\def\SSshuffle{\mathbin{\hbox{\fiveshuffle\char'001}}}%
\mathchoice{\Dshuffle}{\Dshuffle}{\Sshuffle}{\SSshuffle}}}

%\owedge

% From Knuth's \pfbox macro
\def\qed{\hbox{\hskip 3pt
%\lower2pt
\vbox{\hrule\hbox to 7pt{\vrule height 7pt\hfill\vrule}
\hrule}}\hskip3pt}

% do not display overfull marks
\overfullrule=0pt\relax

\frenchspacing

% define labels in advance
\newread\instream \openin\instream= label.defs
\ifeof\instream \message{No labels in advance yet. Wait till next pass.}
\else \closein\instream \input label.defs
\fi
\writedefs

%%% References with hyperlinks to arxiv.org; both styles accepted
% Change arXiv to \arXiv ie
% [arXiv:hep-th/1234567].     --> [\arXiv:hep-th/1234567].
% [arXiv:1234.5678 [hep-th]]. --> [\arXiv:1234.5678 [hep-th]].
% Need to strip trailing [hep-th] (if present) to define valid URL
\def\arXiv:#1].{\hepthStrip#1 \nil}
\def\hepthStrip#1 #2\nil{\href{http://arxiv.org/abs/#1}{arXiv:#1 #2\unskip}].}

\input amssym.tex
\input color

\def\cA{{\cal A}}
\def\cW{{\cal W}}

\def\dd{{\rm d}}
\def\PT{{\rm PT}}
\def\teight#1,#2,#3,#4{t_8(\cf_{#1},\cf_{#2},\cf_{#3},\cf_{#4})}
\def\sign{{\rm sgn}}

\font\frakfont=eufm10 at 10pt
\def\ce{\mathord{\hbox{\frakfont e}}}
\def\cf{\mathord{\hbox{\frakfont f}}}

\preprint {}

\title New BCJ representations for one-loop amplitudes in gauge theories and gravity

\author Song He$^{a,b,c}$\email{a}{songhe@itp.ac.cn, yongzhang@itp.ac.cn},
Oliver Schlotterer$^{b,d}$\email{d}{olivers@aei.mpg.de} and
Yong Zhang$^{a,c,e}$%\email{d}{yongzhang@itp.ac.cn}

\address
$^{a}$CAS Key Laboratory of Theoretical Physics, Institute of Theoretical Physics
Chinese Academy of Sciences, Beijing 100190, China

\address
$^{b}$Kavli Institute for Theoretical Physics
University of California, Santa Barbara, CA 93106, USA

\address
$^{c}$School of Physical Sciences, University of Chinese Academy of Sciences,
 No.19A Yuquan Road, Beijing 100049, China

\address
$^{d}$Max--Planck--Institut f\"ur Gravitationsphysik
Albert--Einstein--Institut, 14476 Potsdam, Germany

\address
$^{e}$Department of Physics, Beijing Normal University, Beijing 100875, China

%\author Song He$^{a,c,d}$\email{a}{songhe@itp.ac.cn, yongzhang@itp.ac.cn},
%Oliver Schlotterer$^{b,c}$\email{b}{olivers@aei.mpg.de} and
%Yong Zhang$^{a,d,e}$%\email{d}{yongzhang@itp.ac.cn}
%
%\address
%$^{a}$CAS Key Laboratory of Theoretical Physics, Institute of Theoretical Physics
%Chinese Academy of Sciences, Beijing 100190, China
%
%\address
%$^{b}$Max--Planck--Institut f\"ur Gravitationsphysik
%Albert--Einstein--Institut, 14476 Potsdam, Germany
%
%\address
%$^{c}$Kavli Institute for Theoretical Physics
%University of California, Santa Barbara, CA 93106, USA
%
%\address
%$^{d}$School of Physical Sciences, University of Chinese Academy of Sciences,
% No.19A Yuquan Road, Beijing 100049, China
%
%
%\address
%$^{e}$Department of Physics, Beijing Normal University, Beijing 100875, China
%

\abstract
We explain a procedure to manifest the Bern--Carrasco--Johansson duality between
color and kinematics in $n$-point one-loop amplitudes of a variety of supersymmetric gauge theories.
Explicit amplitude representations are constructed through a systematic reorganization of the
integrands in the Cachazo--He--Yuan formalism. 
%Our construction is independent on the amount of
%supersymmetry or the number of spacetime dimensions.
Our construction holds for any nonzero number of supersymmetries and does not depend
on the number of spacetime dimensions.
%Moreover, it manifests the cancellations from
%supersymmetry multiplets in the loop as well as the resulting power counting of loop momenta.
The cancellations from supersymmetry multiplets in the loop as well as the resulting power
counting of loop momenta is manifested along the lines of the corresponding superstring computations.
The setup is used to derive the one-loop version of the Kawai--Lewellen--Tye formula for the
loop integrands of gravitational amplitudes.

\Date{June 2017}

\newif\iffig
\figfalse
% to speed up compilation comment the following line
\input tikz \figtrue

%**************************************

\lref\GeyerELA{
  Y.~Geyer and R.~Monteiro,
  ``Gluons and gravitons at one loop from ambitwistor strings,''
[arXiv:1711.09923 [hep-th]].
%%CITATION = arXiv:1711.09923%%
}

%\HeIQI
\lref\HeIQI{
  S.~He and Y.~Zhang,
  ``New Formulas for Amplitudes from Higher-Dimensional Operators,''
JHEP {\bf 1702}, 019 (2017).
[arXiv:1608.08448 [hep-th]].
%%CITATION = arXiv:1608.08448%%
}

%\ZhangRZB
\lref\ZhangRZB{
  Y.~Zhang,
  ``CHY formulae in 4d,''
[arXiv:1610.05205 [hep-th]].
%%CITATION = arXiv:1610.05205%%
}

%\HeDOL
\lref\HeDOL{
  S.~He and Y.~Zhang,
  ``Connected formulas for amplitudes in standard model,''
JHEP {\bf 1703}, 093 (2017).
[arXiv:1607.02843 [hep-th]].
%%CITATION = arXiv:1607.02843%%
}

%\BaadsgaardVOA
\lref\BaadsgaardVOA{
  C.~Baadsgaard, N.~E.~J.~Bjerrum-Bohr, J.~L.~Bourjaily and P.~H.~Damgaard,
  ``Integration Rules for Scattering Equations,''
JHEP {\bf 1509}, 129 (2015).
[arXiv:1506.06137 [hep-th]].
%%CITATION = arXiv:1506.06137%%
}

%\BaadsgaardHIA
\lref\BaadsgaardHIA{
  C.~Baadsgaard, N.~E.~J.~Bjerrum-Bohr, J.~L.~Bourjaily, P.~H.~Damgaard and B.~Feng,
  ``Integration Rules for Loop Scattering Equations,''
JHEP {\bf 1511}, 080 (2015).
[arXiv:1508.03627 [hep-th]]. \semi
%%CITATION = arXiv:1508.03627%%
  B.~Feng,
  ``CHY-construction of Planar Loop Integrands of Cubic Scalar Theory,''
JHEP {\bf 1605}, 061 (2016).
[arXiv:1601.05864 [hep-th]].
%%CITATION = arXiv:1601.05864%%
}

\lref\wipL{
C.R.~Mafra, O.~Schlotterer, work in progress
}

\lref\PSS{
  C.R.~Mafra,
  ``PSS: A FORM Program to Evaluate Pure Spinor Superspace Expressions,''
[arXiv:1007.4999 [hep-th]].
%%CITATION = arXiv:1007.4999%%
}

\lref\BerkovitsBK{
  N.~Berkovits and C.~R.~Mafra,
  ``Some Superstring Amplitude Computations with the Non-Minimal Pure Spinor Formalism,''
JHEP {\bf 0611}, 079 (2006).
[hep-th/0607187].
%%CITATION = hep-th/0607187%%
}

\lref\teightMG{
	M.B.~Green and J.~H.~Schwarz,
  	``Supersymmetrical Dual String Theory. 2. Vertices and Trees,''
	Nucl.\ Phys.\ B {\bf 198}, 252 (1982).
	%%CITATION = CALT-68-872%%
}

\lref\MafraKJ{
	C.R.~Mafra, O.~Schlotterer and S.~Stieberger,
  	``Explicit BCJ Numerators from Pure Spinors,''
	JHEP {\bf 1107}, 092 (2011).
	[arXiv:1104.5224 [hep-th]].
	%%CITATION = arXiv:1104.5224%%
}

\lref\MafraVCA{
	C.R.~Mafra and O.~Schlotterer,
  	``Berends-Giele recursions and the BCJ duality in superspace and components,''
	JHEP {\bf 1603}, 097 (2016).
	[arXiv:1510.08846 [hep-th]].
	%%CITATION = DAMTP-2015-69%%
}

\lref\BroedelTTA{
  J.~Broedel, O.~Schlotterer and S.~Stieberger,
  ``Polylogarithms, Multiple Zeta Values and Superstring Amplitudes,''
Fortsch.\ Phys.\  {\bf 61}, 812 (2013).
[arXiv:1304.7267 [hep-th]].
%%CITATION = DAMTP-2013-22%%
}

\lref\BergFUI{
  M.~Berg, I.~Buchberger and O.~Schlotterer,
  ``String-motivated one-loop amplitudes in gauge theories with half-maximal supersymmetry,''
[arXiv:1611.03459 [hep-th]].
%%CITATION = arXiv:1611.03459%%
}

\lref\EOMBBs{
  C.~R.~Mafra and O.~Schlotterer,
  ``Multiparticle SYM equations of motion and pure spinor BRST blocks,''
JHEP {\bf 1407}, 153 (2014).
[arXiv:1404.4986 [hep-th]].
%%CITATION = AEI-2014-011%%
}

%\CaronHuotZT
\lref\CaronHuotZT{
  S.~Caron-Huot,
  %``Loops and trees,''
JHEP {\bf 1105}, 080 (2011).
[arXiv:1007.3224 [hep-ph]].
%%CITATION = arXiv:1007.3224%%
}

\lref\LeeUPY{
  S.~Lee, C.R.~Mafra and O.~Schlotterer,
  ``Non-linear gauge transformations in $D=10$ SYM theory and the BCJ duality,''
JHEP {\bf 1603}, 090 (2016).
[arXiv:1510.08843 [hep-th]].
%%CITATION = DAMTP-2015-68%%
}

\lref\MafraGSA{
  C.~R.~Mafra and O.~Schlotterer,
  ``Cohomology foundations of one-loop amplitudes in pure spinor superspace,''
[arXiv:1408.3605 [hep-th]].
%%CITATION = arXiv:1408.3605%%
}

\lref\HeMZD{
S.~He and O.~Schlotterer,
  ``Loop-level KLT, BCJ and EYM amplitude relations,''
Phys.\ Rev.\ Lett.\  {\bf 118}, no. 16, 161601 (2017).
[arXiv:1612.00417 [hep-th]].
%%CITATION = arXiv:1612.00417%%
}

\lref\AlvarezGaumeIG{
  L.~Alvarez-Gaume and E.~Witten,
  ``Gravitational Anomalies,''
Nucl.\ Phys.\ B {\bf 234}, 269 (1984).
%%CITATION = HUTP-83/A039%%
}

\lref\fivetree{
	C.R.~Mafra,
	``Simplifying the Tree-level Superstring Massless Five-point Amplitude,''
	JHEP {\bf 1001}, 007 (2010).
	[arXiv:0909.5206 [hep-th]].
	%%CITATION = arXiv:0909.5206%%
}
\lref\nptMethod{
	C.R.~Mafra, O.~Schlotterer, S.~Stieberger and D.~Tsimpis,
	``A recursive method for SYM n-point tree amplitudes,''
	Phys.\ Rev.\ D {\bf 83}, 126012 (2011).
	[arXiv:1012.3981 [hep-th]].
	%%CITATION = arXiv:1012.3981%%
}
\lref\nptTree{
C.~R.~Mafra, O.~Schlotterer and S.~Stieberger,
  ``Complete N-Point Superstring Disk Amplitude I. Pure Spinor Computation,''
Nucl.\ Phys.\ B {\bf 873}, 419 (2013).
[arXiv:1106.2645 [hep-th]].
%%CITATION = arXiv:1106.2645%%
}
\lref\wittentwistor{
	E.~Witten,
	``Twistor-Like Transform In Ten-Dimensions,''
	Nucl.\ Phys.\  B {\bf 266}, 245 (1986).
	%%CITATION = NUPHA,B266,245;%%
}
\lref\psf{
 	N.~Berkovits,
	``Super-Poincare covariant quantization of the superstring,''
	JHEP {\bf 0004}, 018 (2000)
	[arXiv:hep-th/0001035].
	%%CITATION = JHEPA,0004,018;%%
}
\lref\MPS{
  N.~Berkovits,
  ``Multiloop amplitudes and vanishing theorems using the pure spinor formalism for the superstring,''
JHEP {\bf 0409}, 047 (2004).
[hep-th/0406055].
%%CITATION = hep-th/0406055%%
}
\lref\oneloopMichael{
  M.~B.~Green, C.~R.~Mafra and O.~Schlotterer,
  ``Multiparticle one-loop amplitudes and S-duality in closed superstring theory,''
[arXiv:1307.3534].
%%CITATION = AEI-2013-219%%
}

\lref\BroedelVLA{
  J.~Broedel, C.~R.~Mafra, N.~Matthes and O.~Schlotterer,
  ``Elliptic multiple zeta values and one-loop superstring amplitudes,''
JHEP {\bf 1507}, 112 (2015).
[arXiv:1412.5535 [hep-th]].
}
\lref\TsuchiyaVA{
  A.~Tsuchiya,
  ``More on One Loop Massless Amplitudes of Superstring Theories,''
Phys.\ Rev.\ D {\bf 39}, 1626 (1989).
%%CITATION = TIT/HEP-135%%
}

\lref\MafraKH{
  C.~R.~Mafra and O.~Schlotterer,
  ``The Structure of n-Point One-Loop Open Superstring Amplitudes,''
JHEP {\bf 1408}, 099 (2014).
[arXiv:1203.6215 [hep-th]].
%%CITATION = AEI-2012-032%%
}

\lref\KawaiXQ{
  H.~Kawai, D.~C.~Lewellen and S.~H.~H.~Tye,
  ``A Relation Between Tree Amplitudes of Closed and Open Strings,''
Nucl.\ Phys.\ B {\bf 269}, 1 (1986).
%%CITATION = CLNS-85/667%%
}

\lref\CachazoGNA{
  F.~Cachazo, S.~He and E.~Y.~Yuan,
  ``Scattering equations and Kawai-Lewellen-Tye orthogonality,''
Phys.\ Rev.\ D {\bf 90}, no. 6, 065001 (2014).
[arXiv:1306.6575 [hep-th]].
%%CITATION = arXiv:1306.6575%%
}

\lref\CachazoHCA{
  F.~Cachazo, S.~He and E.~Y.~Yuan,
  ``Scattering of Massless Particles in Arbitrary Dimensions,''
Phys.\ Rev.\ Lett.\  {\bf 113}, no. 17, 171601 (2014).
[arXiv:1307.2199 [hep-th]].
%%CITATION = arXiv:1307.2199%%
}

\lref\CachazoIEA{
  F.~Cachazo, S.~He and E.~Y.~Yuan,
  ``Scattering of Massless Particles: Scalars, Gluons and Gravitons,''
JHEP {\bf 1407}, 033 (2014).
[arXiv:1309.0885 [hep-th]].
%%CITATION = arXiv:1309.0885%%
}

\lref\GomezLHY{
H.~Gomez,
  ``Quadratic Feynman Loop Integrands From Massless Scattering Equations,''
Phys.\ Rev.\ D {\bf 95}, no. 10, 106006 (2017).
[arXiv:1703.04714 [hep-th]].
%%CITATION = arXiv:1703.04714%%
}

\lref\CachazoNSA{
  F.~Cachazo, S.~He and E.~Y.~Yuan,
  ``Einstein-Yang-Mills Scattering Amplitudes From Scattering Equations,''
JHEP {\bf 1501}, 121 (2015).
[arXiv:1409.8256 [hep-th]].
%%CITATION = arXiv:1409.8256%%
}

\lref\CachazoXEA{
  F.~Cachazo, S.~He and E.~Y.~Yuan,
  ``Scattering Equations and Matrices: From Einstein To Yang-Mills, DBI and NLSM,''
JHEP {\bf 1507}, 149 (2015).
[arXiv:1412.3479 [hep-th]].
%%CITATION = arXiv:1412.3479%%
}

\lref\CachazoNJL{
  F.~Cachazo, P.~Cha and S.~Mizera,
  ``Extensions of Theories from Soft Limits,''
JHEP {\bf 1606}, 170 (2016).
[arXiv:1604.03893 [hep-th]].
%%CITATION = arXiv:1604.03893%%
}

\lref\BerkovitsXBA{
  N.~Berkovits,
  ``Infinite Tension Limit of the Pure Spinor Superstring,''
JHEP {\bf 1403}, 017 (2014).
[arXiv:1311.4156 [hep-th]].
%%CITATION = ICTP-SAIFR-2013-13%%
}

\lref\GomezWZA{
  H.~Gomez and E.~Y.~Yuan,
  ``N-point tree-level scattering amplitude in the new Berkovits` string,''
JHEP {\bf 1404}, 046 (2014).
[arXiv:1312.5485 [hep-th]].
%%CITATION = arXiv:1312.5485%%
}

\lref\AdamoTSA{
  T.~Adamo, E.~Casali and D.~Skinner,
  ``Ambitwistor strings and the scattering equations at one loop,''
JHEP {\bf 1404}, 104 (2014).
[arXiv:1312.3828 [hep-th]].
%%CITATION = DAMTP-2013-74%%
}

\lref\CasaliHFA{
  E.~Casali and P.~Tourkine,
  ``Infrared behaviour of the one-loop scattering equations and supergravity integrands,''
JHEP {\bf 1504}, 013 (2015).
[arXiv:1412.3787 [hep-th]].
%%CITATION = arXiv:1412.3787%%
}

\lref\AdamoHOA{
  T.~Adamo and E.~Casali,
  ``Scattering equations, supergravity integrands, and pure spinors,''
JHEP {\bf 1505}, 120 (2015).
[arXiv:1502.06826 [hep-th]].
%%CITATION = DAMTP-2015-14%%
}

\lref\GeyerBJA{
  Y.~Geyer, L.~Mason, R.~Monteiro and P.~Tourkine,
  ``Loop Integrands for Scattering Amplitudes from the Riemann Sphere,''
Phys.\ Rev.\ Lett.\  {\bf 115}, no. 12, 121603 (2015).
[arXiv:1507.00321 [hep-th]].
%%CITATION = arXiv:1507.00321%%
}

\lref\GeyerJCH{
  Y.~Geyer, L.~Mason, R.~Monteiro and P.~Tourkine,
  ``One-loop amplitudes on the Riemann sphere,''
JHEP {\bf 1603}, 114 (2016).
[arXiv:1511.06315 [hep-th]].
%%CITATION = CERN-PH-TH-2015-267%%
}

\lref\GeyerWJX{
  Y.~Geyer, L.~Mason, R.~Monteiro and P.~Tourkine,
  ``Two-Loop Scattering Amplitudes from the Riemann Sphere,''
Phys.\ Rev.\ D {\bf 94}, no. 12, 125029 (2016).
[arXiv:1607.08887 [hep-th]].
%%CITATION = CERN-TH-2016-172%%
}

\lref\HeYUA{
  S.~He and E.~Y.~Yuan,
  ``One-loop Scattering Equations and Amplitudes from Forward Limit,''
Phys.\ Rev.\ D {\bf 92}, no. 10, 105004 (2015).
[arXiv:1508.06027 [hep-th]].
%%CITATION = arXiv:1508.06027%%
}

\lref\CardonaBPI{
  C.~Cardona and H.~Gomez,
  ``Elliptic scattering equations,''
JHEP {\bf 1606}, 094 (2016).
[arXiv:1605.01446 [hep-th]]. \semi
%%CITATION = NCTS-TH-1602%%
  C.~Cardona and H.~Gomez,
  ``CHY-Graphs on a Torus,''
JHEP {\bf 1610}, 116 (2016).
[arXiv:1607.01871 [hep-th]].
%%CITATION = arXiv:1607.01871%%
}

%\lref\CardonaWCR{
%
%}

\lref\MasonSVA{
  L.~Mason and D.~Skinner,
  ``Ambitwistor strings and the scattering equations,''
JHEP {\bf 1407}, 048 (2014).
[arXiv:1311.2564 [hep-th]].
%%CITATION = arXiv:1311.2564%%
}

\lref\CasaliVTA{
  E.~Casali, Y.~Geyer, L.~Mason, R.~Monteiro and K.~A.~Roehrig,
  ``New Ambitwistor String Theories,''
JHEP {\bf 1511}, 038 (2015).
[arXiv:1506.08771 [hep-th]].
%%CITATION = arXiv:1506.08771%%
}

\lref\MafraLTU{
  C.~R.~Mafra,
  ``Berends-Giele recursion for double-color-ordered amplitudes,''
JHEP {\bf 1607}, 080 (2016).
[arXiv:1603.09731 [hep-th]].
%%CITATION = arXiv:1603.09731%%
}

\lref\CarrascoLDY{
  J.~J.~M.~Carrasco, C.~R.~Mafra and O.~Schlotterer,
  ``Abelian Z-theory: NLSM amplitudes and alpha'-corrections from the open string,''
[arXiv:1608.02569 [hep-th]].
%%CITATION = arXiv:1608.02569%%
}

\lref\MafraMCC{
  C.~R.~Mafra and O.~Schlotterer,
  ``Non-abelian $Z$-theory: Berends-Giele recursion for the $\alpha'$-expansion of disk integrals,''
JHEP {\bf 1701}, 031 (2017).
[arXiv:1609.07078 [hep-th]].
%%CITATION = arXiv:1609.07078%%
}

\lref\CarrascoYGV{
  J.~J.~M.~Carrasco, C.~R.~Mafra and O.~Schlotterer,
  ``Semi-abelian Z-theory: NLSM+$\phi^3$ from the open string,''
[arXiv:1612.06446 [hep-th]].
%%CITATION = arXiv:1612.06446%%
}

\lref\BaadsgaardTWA{
  C.~Baadsgaard, N.~E.~J.~Bjerrum-Bohr, J.~L.~Bourjaily, S.~Caron-Huot, P.~H.~Damgaard and B.~Feng,
  ``New Representations of the Perturbative S-Matrix,''
Phys.\ Rev.\ Lett.\  {\bf 116}, no. 6, 061601 (2016).
[arXiv:1509.02169 [hep-th]].
%%CITATION = arXiv:1509.02169%%
}

\lref\BernQJ{
  Z.~Bern, J.~J.~M.~Carrasco and H.~Johansson,
  ``New Relations for Gauge-Theory Amplitudes,''
Phys.\ Rev.\ D {\bf 78}, 085011 (2008).
[arXiv:0805.3993 [hep-ph]].
%%CITATION = arXiv:0805.3993%%
}

\lref\BernUE{
  Z.~Bern, J.~J.~M.~Carrasco and H.~Johansson,
  ``Perturbative Quantum Gravity as a Double Copy of Gauge Theory,''
Phys.\ Rev.\ Lett.\  {\bf 105}, 061602 (2010).
[arXiv:1004.0476 [hep-th]].
%%CITATION = UCLA-10-TEP-102%%
}

\lref\BernYG{
  Z.~Bern, T.~Dennen, Y.~t.~Huang and M.~Kiermaier,
  ``Gravity as the Square of Gauge Theory,''
Phys.\ Rev.\ D {\bf 82}, 065003 (2010).
[arXiv:1004.0693 [hep-th]].
%%CITATION = UCLA-TEP-10-103%%
}

\lref\BernSV{
  Z.~Bern, L.~J.~Dixon, M.~Perelstein and J.~S.~Rozowsky,
  ``Multileg one loop gravity amplitudes from gauge theory,''
Nucl.\ Phys.\ B {\bf 546}, 423 (1999).
[hep-th/9811140].
%%CITATION = hep-th/9811140%%
}

\lref\BjerrumBohrTA{
  N.~E.~J.~Bjerrum-Bohr, P.~H.~Damgaard, B.~Feng and T.~Sondergaard,
  ``Gravity and Yang-Mills Amplitude Relations,''
Phys.\ Rev.\ D {\bf 82}, 107702 (2010).
[arXiv:1005.4367 [hep-th]] \semi
%%CITATION = arXiv:1005.4367%%
N.~E.~J.~Bjerrum-Bohr, P.~H.~Damgaard, B.~Feng and T.~Sondergaard,
  ``New Identities among Gauge Theory Amplitudes,''
Phys.\ Lett.\ B {\bf 691}, 268 (2010).
[arXiv:1006.3214 [hep-th]] \semi
%%CITATION = arXiv:1006.3214%%
N.~E.~J.~Bjerrum-Bohr, P.~H.~Damgaard, T.~Sondergaard and P.~Vanhove,
  ``The Momentum Kernel of Gauge and Gravity Theories,''
JHEP {\bf 1101}, 001 (2011).
[arXiv:1010.3933 [hep-th]].
%%CITATION = arXiv:1010.3933%%
}

\lref\BjerrumBohrRD{
  N.~E.~J.~Bjerrum-Bohr, P.~H.~Damgaard and P.~Vanhove,
  ``Minimal Basis for Gauge Theory Amplitudes,''
Phys.\ Rev.\ Lett.\  {\bf 103}, 161602 (2009).
[arXiv:0907.1425 [hep-th]] \semi
%%CITATION = arXiv:0907.1425%%
S.~Stieberger,
  ``Open \& Closed vs. Pure Open String Disk Amplitudes,''
[arXiv:0907.2211 [hep-th]].
%%CITATION = MPP-2008-01%%
}

\lref\BernZX{
  Z.~Bern, L.~J.~Dixon, D.~C.~Dunbar and D.~A.~Kosower,
  ``One loop n point gauge theory amplitudes, unitarity and collinear limits,''
Nucl.\ Phys.\ B {\bf 425}, 217 (1994).
[hep-ph/9403226].
%%CITATION = hep-ph/9403226%%
}

\lref\BLK{
F.~Brown, A.~Levin,
``Multiple elliptic polylogarithms.''
}

\lref\CachazoNWA{
  F.~Cachazo and H.~Gomez,
  ``Computation of Contour Integrals on ${\cal M}_{0,n}$,''
JHEP {\bf 1604}, 108 (2016).
[arXiv:1505.03571 [hep-th]] \semi
%%CITATION = arXiv:1505.03571%%
C.~Cardona, B.~Feng, H.~Gomez and R.~Huang,
  ``Cross-ratio Identities and Higher-order Poles of CHY-integrand,''
JHEP {\bf 1609}, 133 (2016).
[arXiv:1606.00670 [hep-th]].
%%CITATION = arXiv:1606.00670%%
}

\lref\MafraGJA{
  C.~R.~Mafra and O.~Schlotterer,
  ``Towards one-loop SYM amplitudes from the pure spinor BRST cohomology,''
Fortsch.\ Phys.\  {\bf 63}, no. 2, 105 (2015).
[arXiv:1410.0668 [hep-th]].
%%CITATION = AEI-2014-053%%
}

\lref\BernYXU{
 Z.~Bern, J.~J.~Carrasco, W.~M.~Chen, H.~Johansson and R.~Roiban,
  ``Gravity Amplitudes as Generalized Double Copies of Gauge-Theory Amplitudes,''
Phys.\ Rev.\ Lett.\  {\bf 118}, no. 18, 181602 (2017).
[arXiv:1701.02519 [hep-th]].
%%CITATION = UCLA-17-TEP-101%%
}

\lref\BergWUX{
M.~Berg, I.~Buchberger and O.~Schlotterer,
  ``From maximal to minimal supersymmetry in string loop amplitudes,''
JHEP {\bf 1704}, 163 (2017).
[arXiv:1603.05262 [hep-th]].
%%CITATION = arXiv:1603.05262%%
}

\lref\MinahanHA{
  J.~A.~Minahan,
  ``One Loop Amplitudes on Orbifolds and the Renormalization of Coupling Constants,''
Nucl.\ Phys.\ B {\bf 298}, 36 (1988).
%%CITATION = PUPT-1063%%
}

\lref\ChenEVA{
  W.~M.~Chen, Y.~t.~Huang and D.~A.~McGady,
  ``Anomalies without an action,''
[arXiv:1402.7062 [hep-th]].
%%CITATION = PUPT-2459%%
}

\lref\KleissNE{
  R.~Kleiss and H.~Kuijf,
  ``Multi - Gluon Cross-sections and Five Jet Production at Hadron Colliders,''
Nucl.\ Phys.\ B {\bf 312}, 616 (1989) \semi
%%CITATION = Print-88-0425 (LEIDEN)%%
V.~Del Duca, L.~J.~Dixon and F.~Maltoni,
  ``New color decompositions for gauge amplitudes at tree and loop level,''
Nucl.\ Phys.\ B {\bf 571}, 51 (2000).
[hep-ph/9910563].
%%CITATION = hep-ph/9910563%%
}

\lref\GreenSW{
  M.~B.~Green, J.~H.~Schwarz and L.~Brink,
  ``N=4 Yang-Mills and N=8 Supergravity as Limits of String Theories,''
Nucl.\ Phys.\ B {\bf 198}, 474 (1982).
%%CITATION = CALT-68-880%%
}

\lref\BianchiNF{
  M.~Bianchi and A.~V.~Santini,
  ``String predictions for near future colliders from one-loop scattering amplitudes around D-brane worlds,''
JHEP {\bf 0612}, 010 (2006).
[hep-th/0607224].
%%%CITATION = hep-th/0607224%%
}

\lref\DHokerVCH{
  E.~D'Hoker and D.~H.~Phong,
  ``Two-loop superstrings VI: Non-renormalization theorems and the 4-point function,''
Nucl.\ Phys.\ B {\bf 715}, 3 (2005).
[hep-th/0501197]. \semi
%%CITATION = hep-th/0501197%%
N.~Berkovits,
  ``Super-Poincare covariant two-loop superstring amplitudes,''
JHEP {\bf 0601}, 005 (2006).
[hep-th/0503197].
%%CITATION = hep-th/0503197%%
}

\lref\redSUSYamps{
  P.~Tourkine and P.~Vanhove,
  ``One-loop four-graviton amplitudes in ${\cal N}=4$ supergravity models,''
Phys.\ Rev.\ D {\bf 87}, no. 4, 045001 (2013).
[arXiv:1208.1255 [hep-th]].
%%CITATION = IHES-P-12-21%%
\semi
A.~Ochirov and P.~Tourkine,
  ``BCJ duality and double copy in the closed string sector,''
JHEP {\bf 1405}, 136 (2014).
[arXiv:1312.1326 [hep-th]].
%%CITATION = IPHT-T13-196%%
\semi
 M.~Bianchi and D.~Consoli,
  ``Simplifying one-loop amplitudes in superstring theory,''
JHEP {\bf 1601}, 043 (2016).
[arXiv:1508.00421 [hep-th]].
%%CITATION = ROM2F-2015-10%%
}

\lref\BoelsTP{
  R.~H.~Boels and R.~S.~Isermann,
  ``New relations for scattering amplitudes in Yang-Mills theory at loop level,''
Phys.\ Rev.\ D {\bf 85}, 021701 (2012).
[arXiv:1109.5888 [hep-th]] \semi
%%CITATION = arXiv:1109.5888%%
R.~H.~Boels and R.~S.~Isermann,
  ``Yang-Mills amplitude relations at loop level from non-adjacent BCFW shifts,''
JHEP {\bf 1203}, 051 (2012).
[arXiv:1110.4462 [hep-th]] \semi
%%CITATION = arXiv:1110.4462%%
Y.~J.~Du and H.~Luo,
  ``On General BCJ Relation at One-loop Level in Yang-Mills Theory,''
JHEP {\bf 1301}, 129 (2013).
[arXiv:1207.4549 [hep-th]] \semi
%%CITATION = arXiv:1207.4549%%
A.~Primo and W.~J.~Torres Bobadilla,
  ``BCJ Identities and $d$-Dimensional Generalized Unitarity,''
JHEP {\bf 1604}, 125 (2016).
[arXiv:1602.03161 [hep-ph]].
%%CITATION = arXiv:1602.03161%%
}

%\GomezCQB
\lref\GomezCQB{
  H.~Gomez, S.~Mizera and G.~Zhang,
  ``CHY Loop Integrands from Holomorphic Forms,''
JHEP {\bf 1703}, 092 (2017).
[arXiv:1612.06854 [hep-th]].
%%CITATION = arXiv:1612.06854%%
}

\lref\TourkineBAK{
  P.~Tourkine and P.~Vanhove,
  ``Higher-loop amplitude monodromy relations in string and gauge theory,''
Phys.\ Rev.\ Lett.\  {\bf 117}, no. 21, 211601 (2016).
[arXiv:1608.01665 [hep-th]] \semi
%%CITATION = arXiv:1608.01665%%
S.~Hohenegger and S.~Stieberger,
  ``Monodromy Relations in Higher-Loop String Amplitudes,''
[arXiv:1702.04963 [hep-th]].
%%CITATION = MPP-2017-001%%
}

\lref\CachazoAOL{
  F.~Cachazo, S.~He and E.~Y.~Yuan,
  ``One-Loop Corrections from Higher Dimensional Tree Amplitudes,''
JHEP {\bf 1608}, 008 (2016).
[arXiv:1512.05001 [hep-th]].
%%CITATION = arXiv:1512.05001%%
}

\lref\CachazoUQ{
  F.~Cachazo,
  ``Fundamental BCJ Relation in N=4 SYM From The Connected Formulation,''
[arXiv:1206.5970 [hep-th]].
%%CITATION = arXiv:1206.5970%%
}

\lref\StiebergerWK{
  S.~Stieberger and T.~R.~Taylor,
  ``NonAbelian Born-Infeld action and type 1. - heterotic duality 2: Nonrenormalization theorems,''
Nucl.\ Phys.\ B {\bf 648}, 3 (2003).
[hep-th/0209064];
%%CITATION = hep-th/0209064%%
}

\lref\canceltriag{
N.~E.~J.~Bjerrum-Bohr and P.~Vanhove,
  ``Explicit Cancellation of Triangles in One-loop Gravity Amplitudes,''
JHEP {\bf 0804}, 065 (2008).
[arXiv:0802.0868 [hep-th]].
%%CITATION = arXiv:0802.0868%%
}
\lref\pssnorm{
  N.~Berkovits and B.C.~Vallilo,
  ``Consistency of superPoincare covariant superstring tree amplitudes,''
JHEP {\bf 0007}, 015 (2000).
[hep-th/0004171].
%%CITATION = hep-th/0004171%%
}

\lref\BernUF{
  Z.~Bern, J.~J.~M.~Carrasco, L.~J.~Dixon, H.~Johansson and R.~Roiban,
  ``Simplifying Multiloop Integrands and Ultraviolet Divergences of Gauge Theory and Gravity Amplitudes,''
Phys.\ Rev.\ D {\bf 85}, 105014 (2012).
[arXiv:1201.5366 [hep-th]].
%%CITATION = arXiv:1201.5366%%
}

\lref\BernUFMORE{
  Z.~Bern, S.~Davies, T.~Dennen and Y.~t.~Huang,
  ``Absence of Three-Loop Four-Point Divergences in N=4 Supergravity,''
Phys.\ Rev.\ Lett.\  {\bf 108}, 201301 (2012).
[arXiv:1202.3423 [hep-th]].
%%CITATION = UCLA-12-TEP-101%%
\semi
  Z.~Bern, S.~Davies and T.~Dennen,
  ``The Ultraviolet Structure of Half-Maximal Supergravity with Matter Multiplets at Two and Three Loops,''
Phys.\ Rev.\ D {\bf 88}, 065007 (2013).
[arXiv:1305.4876 [hep-th]].
%%CITATION = UCLA-13-TEP-105%%
\semi
  Z.~Bern, S.~Davies, T.~Dennen, A.~V.~Smirnov and V.~A.~Smirnov,
  ``Ultraviolet Properties of N=4 Supergravity at Four Loops,''
Phys.\ Rev.\ Lett.\  {\bf 111}, no. 23, 231302 (2013).
[arXiv:1309.2498 [hep-th]].
%%CITATION = arXiv:1309.2498%%
\semi
  Z.~Bern, S.~Davies and T.~Dennen,
  ``Enhanced ultraviolet cancellations in $N=5$ supergravity at four loops,''
Phys.\ Rev.\ D {\bf 90}, no. 10, 105011 (2014).
[arXiv:1409.3089 [hep-th]].
%%CITATION = UCLA-14-TEP-106%%
}

\lref\thetaExp{
  	J.P.~Harnad and S.~Shnider,
	``Constraints And Field Equations For Ten-Dimensional Superyang-Mills
  	Theory,''
  	Commun.\ Math.\ Phys.\  {\bf 106}, 183 (1986).
  	%%CITATION = CMPHA,106,183;%%
\semi
	H.~Ooguri, J.~Rahmfeld, H.~Robins and J.~Tannenhauser,
        ``Holography in superspace,''
        JHEP {\bf 0007}, 045 (2000)
        [arXiv:hep-th/0007104].
        %%CITATION = HEP-TH 0007104;%%
\semi
	P.A.~Grassi and L.~Tamassia,
        ``Vertex operators for closed superstrings,''
        JHEP {\bf 0407}, 071 (2004)
        [arXiv:hep-th/0405072].
        %%CITATION = HEP-TH 0405072;%%
\semi
	G.~Policastro and D.~Tsimpis,
  	``$R^4$, purified,''
	Class.\ Quant.\ Grav.\  {\bf 23}, 4753 (2006).
	[hep-th/0603165].
	%%CITATION = hep-th/0603165%%
}
\lref\BerendsME{
	F.~A.~Berends and W.~T.~Giele,
  	``Recursive Calculations for Processes with n Gluons,''
	Nucl.\ Phys.\ B {\bf 306}, 759 (1988).
	%%CITATION = Print-88-0100 (LEIDEN)%%
}
\lref\MafraNWR{
	C.R.~Mafra and O.~Schlotterer,
  	``One-loop superstring six-point amplitudes and anomalies in pure spinor superspace,''
	JHEP {\bf 1604}, 148 (2016).
	[arXiv:1603.04790 [hep-th]].
	%%CITATION = arXiv:1603.04790%%
}

\lref\BernYYA{
  Z.~Bern, S.~Davies, T.~Dennen, Y.~t.~Huang and J.~Nohle,
  ``Color-Kinematics Duality for Pure Yang-Mills and Gravity at One and Two Loops,''
Phys.\ Rev.\ D {\bf 92}, no. 4, 045041 (2015).
[arXiv:1303.6605 [hep-th]].
%%CITATION = MCTP-13-06%%
}

\lref\MafraMJA{
  C.~R.~Mafra and O.~Schlotterer,
  ``Two-loop five-point amplitudes of super Yang-Mills and supergravity in pure spinor superspace,''
JHEP {\bf 1510}, 124 (2015).
[arXiv:1505.02746 [hep-th]].
%%CITATION = DAMTP-2015-25%%
}

\lref\HeWGF{
  S.~He, R.~Monteiro and O.~Schlotterer,
  ``String-inspired BCJ numerators for one-loop MHV amplitudes,''
JHEP {\bf 1601}, 171 (2016).
[arXiv:1507.06288 [hep-th]].
%%CITATION = arXiv:1507.06288%%
}

\lref\GomezUHA{
  H.~Gomez, C.~R.~Mafra and O.~Schlotterer,
  ``Two-loop superstring five-point amplitude and $S$-duality,''
Phys.\ Rev.\ D {\bf 93}, no. 4, 045030 (2016).
[arXiv:1504.02759 [hep-th]].
%%CITATION = DAMTP-2015-20%%
}
\lref\GomezSLA{
  H.~Gomez and C.~R.~Mafra,
  ``The closed-string 3-loop amplitude and S-duality,''
JHEP {\bf 1310}, 217 (2013).
[arXiv:1308.6567 [hep-th]].
%%CITATION = DAMTP-2013-50%%
}

\lref\TsuchiyaNF{
  A.~G.~Tsuchiya,
  ``On the pole structures of the disconnected part of hyper elliptic g loop M point super string amplitudes,''
[arXiv:1209.6117 [hep-th]].
%%CITATION = arXiv:1209.6117%%
}

\lref\ClavelliFJ{
  L.~Clavelli, P.~H.~Cox and B.~Harms,
  ``Parity Violating One Loop Six Point Function in Type I Superstring Theory,''
Phys.\ Rev.\ D {\bf 35}, 1908 (1987)..
%%CITATION = UA-HEP-861%%
}

\lref\CarrascoMN{
  J.~J.~Carrasco and H.~Johansson,
  ``Five-Point Amplitudes in N=4 Super-Yang-Mills Theory and N=8 Supergravity,''
Phys.\ Rev.\ D {\bf 85}, 025006 (2012).
[arXiv:1106.4711 [hep-th]]. \semi
%%CITATION = arXiv:1106.4711%%
R.~H.~Boels, R.~S.~Isermann, R.~Monteiro and D.~O'Connell,
  ``Colour-Kinematics Duality for One-Loop Rational Amplitudes,''
JHEP {\bf 1304}, 107 (2013).
[arXiv:1301.4165 [hep-th]]. \semi
%%CITATION = arXiv:1301.4165%%
  N.~E.~J.~Bjerrum-Bohr, T.~Dennen, R.~Monteiro and D.~O'Connell,
  ``Integrand Oxidation and One-Loop Colour-Dual Numerators in N=4 Gauge Theory,''
JHEP {\bf 1307}, 092 (2013).
[arXiv:1303.2913 [hep-th]].
%%CITATION = arXiv:1303.2913%%
}

\lref\oeis{
	{\tt https://oeis.org}
}
%%%%%%%%%%%%

\listtoc
\writetoc
%\filbreak

%************************************************************************************************
%************************************************************************************************
\newsec{Introduction}
%************************************************************************************************
%************************************************************************************************

Recent progress on the study of scattering amplitudes has uncovered novel properties and symmetries of individual theories, as well as surprising connections between them. A notable example is the Bern--Carrasco--Johansson (BCJ) duality between color and kinematics in gauge theories, and double-copy relations to corresponding gravity theories \refs{\BernQJ, \BernUE}.

The BCJ duality states that gauge-theory amplitudes can be expressed such that their kinematic dependence closely mirrors their color dependence, in which case the kinematic contributions from trivalent diagrams are known as {\it BCJ numerators}.  Their most remarkable property is that gravity amplitudes can be obtained from gauge-theory ones by simply substituting color factors for another copy of BCJ numerators. This procedure to construct gravity amplitudes was known as the double copy, and  has led to great advances in the study of the ultraviolet behavior of supergravity amplitudes \refs{\BernUE, \BernUF, \BernUFMORE}.

The BCJ duality has been proved at tree level \BernYG, where the double copy is equivalent to the field-theory limit of the famous Kawai-Lewellen-Tye (KLT) relations between open- and closed-string amplitudes \KawaiXQ. At loop level, despite of strong evidence \refs{\BernUE, \BernUF, \BernYYA, \CarrascoMN, \MafraGJA, \MafraMJA, \HeWGF}, the duality remains a conjecture and the principle behind it is poorly understood. More recently, there has been progress in trying to double copy without explicit BCJ numerators \BernYXU, which may shed light on the longstanding problem of finding the five-loop four-point integrand of maximal supergravity.

At one-loop level, a generalized KLT formula has been proposed for all-multiplicity integrands in gauge and gravity theories \HeMZD, which does not rely on BCJ numerators or any particular representation of these integrands. An important goal of the current paper is to prove this proposal for the one-loop KLT formula. And it will be shown that the formula is in fact equivalent to a cubic-diagram expansion involving double copies of BCJ numerators in a new representation of Feynman integrals.

Moreover, we will describe an algorithmic procedure to obtain all-multiplicity BCJ numerators for one-loop amplitudes in supersymmetric gauge and gravity theories. This becomes possible thanks to the interplay of two closely-related approaches to tree and loop amplitudes: the approach based on scattering equations and that based on string amplitudes.
The first approach has been originally proposed by Cachazo, Yuan and one of the present authors (CHY) as a new formulation for tree amplitudes in gauge theory and gravity \refs{\CachazoGNA, \CachazoHCA}. It expresses tree amplitudes as localized integrals over the moduli space of punctured Riemann spheres, and the prescription turned out to extend flexibly to a variety of other theories\foot{More formulae have been found for gauge-theory and gravity amplitudes with insertions of higher-dimensional operators \refs{\HeIQI} as well as QCD and Higgs amplitudes \refs{\HeDOL} etc..}, such as the bi-adjoint scalars \CachazoIEA, Einstein--Yang--Mills (EYM) \CachazoNSA, Born--Infeld, non-linear sigma models (NLSM) and special Galileons \CachazoXEA\ as well as couplings thereof \CachazoNJL.  Elegant worldsheet models that underpin the CHY formulation have been proposed, based on ambitwistor strings \refs{\MasonSVA, \AdamoTSA, \CasaliVTA} including a manifestly supersymmetric pure-spinor version \refs{\BerkovitsXBA, \AdamoHOA}.

Already at tree level, it has become clear that the CHY approach is very closely related to the string-theory approach to field-theory amplitudes. The reduced Pfaffian, which is the central object of the CHY integrand for gauge theory and gravity \CachazoIEA, can be recast in a form that coincides with open-superstring correlators \nptMethod. This can be seen at the level of operator product expansions of vertex operators, where the pure-spinor CHY setup of \BerkovitsXBA\ is equivalent to superstring result \refs{\nptMethod, \nptTree} as shown in \GomezWZA. More recently, the CHY formulation for the NLSM \refs{\CachazoXEA, \CachazoNJL}  has found a natural counterpart in form of low-energy limits of the disk integrals in open-string amplitudes \refs{\CarrascoLDY, \MafraMCC}, including couplings to biadjoint scalars \CarrascoYGV. Both approaches have provided important insights to the BCJ duality and double copy at tree level. The first explicit local expressions for BCJ numerators of gauge theories were derived in \MafraKJ\ from the pure-spinor formulation of superstring theory \psf. As shown in \CachazoIEA, in the CHY formulation, BCJ duality and double copy, as well as the KLT formulae for tree amplitudes become completely natural, which has also led to a variety of new theories related by double copy \CachazoXEA.

The ambitwistor theory was first generalized to higher genus in \AdamoTSA\ (see also \AdamoHOA\ for a pure-spinor version). This has led the extension of scattering equations and the CHY formulation to loop level using nodal Riemann spheres \GeyerBJA, which yield loop amplitudes in a new representation of their Feynman integrals with propagators linear in loop momenta. The equivalence with usual Feynman-integral representations can be seen via partial-fraction manipulations and shifts in the loop momenta \refs{\CasaliHFA, \GeyerBJA}, which can also be naturally understood as forward limits of tree amplitudes~\CachazoAOL. In this way, CHY-like formulae have been written down for one-loop gauge and gravity theories \GeyerJCH, for biadjoint scalars \HeYUA\ and more recently for two-loop amplitudes of super-Yang--Mills (SYM) and supergravity \GeyerWJX\ as well as scalar theories \GomezCQB. See \CardonaBPI\ and \BaadsgaardHIA\ for closely-related constructions for loop-level scattering equations and CHY formulae.

In this paper, we exploit that the close interplay of the two approaches continues at loop level and apply a variety of results from the recent string-theory literature to supersymmetric one-loop gauge-theory and gravity amplitudes. Significant progress on loop amplitudes of the pure-spinor superstring has been driven by the framework of multiparticle superfields \refs{\EOMBBs, \LeeUPY} which gave rise to explicit BCJ numerators at loop level \refs{\MafraGJA, \MafraMJA, \HeWGF}. Previously, these building blocks have been used  to determine one-loop amplitudes for a BRST-invariant subsector of ten-dimensional open superstring \MafraKH\ which yields the complete all-multiplicity results for four-dimensional MHV helicities as well \HeWGF.  Moreover, multiparticle superfields have been used to determine complete one-loop six-point \MafraMJA\ results and partial two-loop five-point \GomezUHA\ and three-loop four-point \GomezSLA\ results for open and closed strings. Likewise, a component version of multiparticle superfields has been used to streamline the kinematic factors in one-loop open- and closed-string amplitudes with reduced supersymmetry \refs{\BergWUX, \BergFUI}.

However, in one-loop six- and four-point amplitudes with maximal and reduced supersymmetry, respectively, the above approach faced difficulties in constructing BCJ numerators \refs{\MafraGJA, \BergFUI}. It will be shown how the new representation of Feynman integrals emerging from the CHY formulation of loop amplitudes surpasses these obstacles and reconciles the BCJ duality with the hexagon anomaly of ten-dimensional SYM.

The main results of the current paper are threefold
 and may be summarized as follows.
\medskip
\item{(A)} Based on the CHY-inspired representations of supersymmetric gauge-theory and gravity amplitudes, we present a general proof of one-loop BCJ and KLT relations proposed in \HeMZD.
\item{(B)} An all-multiplicity procedure to determine BCJ numerators for one-loop amplitudes is derived from the RNS version of ambitwistor-string and superstring correlators on a nodal Riemann sphere. Our method works for external bosons in presence of any nonzero number of supercharges as well as for both parity-even and parity-odd sectors. The powercounting of loop momenta $\ell$ is manifested in a manner that is well-known from superstrings: Correlators with maximal and reduced supersymmetry are identified as degree-$(n{-}4)$ and degree-$(n{-}2)$ polynomials in $\ell$ and the Green function on the nodal sphere, respectively.
\item{(C)} At multiplicities $n\leq 6$, these BCJ numerators are supersymmetrized such as to address any combination of external bosons and fermions. These expressions are obtained from the field-theory limit of the pure-spinor superstring.

%****************
\subsec{Outline}
%****************

The paper is organized as follows. In section 2, we review the color-kinematics duality and KLT relations in the tree-level CHY setup, as well as the one-loop CHY prescription and the resulting representations of Feynman integrals. Section 3 is devoted to our main result (A): The notion of ``partial integrands" for gauge-theory amplitudes is introduced, and their BCJ relations as well as their combinations to yield one-loop KLT relations are derived from the scattering equations.

The proof of one-loop KLT relations relies on new representations of correlators on a nodal Riemann sphere which are obtained within the RNS formalism in section 4: For external bosons, all-multiplicity techniques are introduced to simplify supersymmetric correlators and to derive the BCJ numerators of (B) along with their powercounting in $\ell$. Some of the steps are known from the superstring literature \refs{\TsuchiyaVA, \StiebergerWK, \canceltriag, \BroedelVLA} but nevertheless spelt out in a CHY context for the sake of a self-contained presentation.

In section 5, we proceed to (C) and derive supersymmetric generalizations of the CHY correlator from the pure-spinor superstring. Particular emphasis will be placed on the resolution of earlier difficulties in finding six-point BCJ numerators in ten-dimensional SYM. An analogous discussion of correlators and BCJ numerators with reduced supersymmetry (along with a suitable infrared regularization scheme) is given in section 6.

%************************************************************************************************
%************************************************************************************************
\newsec{Review}
%************************************************************************************************
%************************************************************************************************

In this section, we first review, within the tree-level CHY setup,
the color-kinematics duality and double copy, as well as the BCJ and
KLT amplitude relations. The presentation is kept very explicit to
later on connect with the analogous structures at one loop.
Furthermore, a brief reminder of the one-loop CHY prescription as
well as the new form of Feynman integrals therein will be given.
Throughout this work, our conventions for Mandelstam invariants $s_{12\ldots p}$
and multiparticle momenta $k_{12\ldots p}$ are as follows:
\eqn\mand{
k_{12\ldots p} \equiv k_1 +k_2+\ldots +k_p \ , \ \ \ \
s_{12\ldots p} \equiv \sum_{i<j}^p k_i \cdot k_j \ , \ \ \ \
s_{12\ldots p,\pm \ell} \equiv \sum_{i<j}^p k_i \cdot k_j  \pm \ell \cdot k_{12\ldots p}
}

%****************
\subsec{CHY at tree level and doubly-partial amplitudes}
%****************

Tree-level scattering amplitudes in the CHY formulation are
represented by integrals over the moduli space of punctured Riemann
spheres \refs{\CachazoGNA, \CachazoHCA, \CachazoIEA} parametrized by
$\sigma_{i} \in \Bbb C$
\eqn\treeCHY{
{\cal M}^{\rm tree}_{L\otimes R} = \int {\rm d} \mu_n^{\rm tree} \ {\cal I}^{\rm tree} _L \, {\cal I}^{\rm tree} _R
\ , \ \ \ \ \ \
 {\rm d} \mu_n^{\rm tree}  \equiv  { {\rm d} \sigma_1 \,{\rm d} \sigma_2\ldots {\rm d} \sigma_n \over {\rm vol} \, SL(2,\Bbb C)} \prod_{i=1}^n \, \! \! ' \delta \Big( \sum_{j=1 \atop{j\neq i}}^{n} {k_{i} \cdot k_{j} \over \sigma_{ij}} \Big) \ .
}
This formula applies to theories $L\otimes R$ that exhibit a double-copy
structure such that the integrand factorizes into two pieces ${\cal I}^{\rm tree} _L$
and ${\cal I}^{\rm tree} _R$ which depend on the scattering data
(momenta or polarizations) as well as the punctures
$\sigma_{ij} \equiv \sigma_i - \sigma_j$. The delta functions
in the measure $ {\rm d} \mu_n^{\rm tree} $ impose the scattering equations
\eqn\scatteq{
E_i \equiv \sum_{j=1 \atop{j\neq i}}^{n} {k_{i} \cdot k_{j} \over \sigma_{ij}}  = 0\ ,
}
and thereby localize the integrals to their $(n{-}3)!$ solutions.
Both of the theory-dependent ``half-integrands" ${\cal I}^{\rm tree}
_L$ and ${\cal I}^{\rm tree} _R$ are designed to transform with
weight two under M\"obius transformations $\sigma_i \rightarrow {a
\sigma_i + b \over c\sigma_i+d}$, with $a,b,c,d $ forming an
$SL(2,\Bbb C)$ matrix. As indicated by $({\rm vol} \, SL(2,\Bbb
C))^{-1}$ and $\prod \, \! '$, this symmetry is taken into account
by fixing any three punctures to $(0,1,\infty)$ and by dropping
three redundant scattering equations, see \refs{\CachazoGNA,
\CachazoHCA} for details.

The generic theory $L\otimes R$ in the CHY prescription \treeCHY\
can be adapted to biadjoint scalars with gauge group $U(N) \times
U(\tilde N)$ by choosing ${\cal I}^{\rm tree} _L$ and ${\cal I}^{\rm
tree} _R$ as \CachazoIEA
\eqn\colint{
{\cal I}^{\rm tree} _{U(N)} = \sum_{\rho \in S_{n-1}} {\rm Tr}(t^{a_1} t^{a_{\rho(2)}} t^{a_{\rho(3)}}  \ldots t^{a_{\rho(n)}} ) \PT(1,\rho(2,3,\ldots,n)) \ ,
}
where $t^{a_{j}}$ denotes the $U(N)$ generator associated with the
$j^{\rm th}$ leg. Given that the dependence on the punctures is
captured by the Parke--Taylor factors
\eqn\parketaylor{
\PT(1,2,3,\ldots,n{-}1,n) \equiv {1\over \sigma_{12} \sigma_{23} \ldots \sigma_{n-1,n} \sigma_{n1}} \ ,
}
the most general integral appearing in the tree-level S-matrix of
the $U(N) \times U(\tilde N)$ theory is the doubly-partial amplitude
\eqn\doublypartial{
m^{\rm tree}[\rho(1,2,\ldots,n) \, | \,\tau(1,2,\ldots,n)  ] \equiv  \int {\rm d} \mu_n^{\rm tree} \ \PT(\rho(1,2,\ldots,n)) \PT(\tau(1,2,\ldots,n))  \ .
}
It accompanies the product of traces ${\rm
Tr}(t^{a_{\rho(1)}}t^{a_{\rho(2)}} \ldots t^{a_{\rho(n)}} ) {\rm
Tr}(\tilde t^{b_{\tau(1)}}\tilde t^{b_{\tau(2)}} \ldots \tilde
t^{b_{\tau(n)}} ) $ (with possibly distinct permutations $\rho,\tau
\in S_n$) in the expression \treeCHY\ for ${\cal M}^{\rm tree}_{U(N)
\times U(\tilde N)}$. The doubly-partial amplitude $m^{\rm
tree}[\rho(\ldots) \, | \,\tau(\ldots) ]$ assembles the propagators
$s_{i_1 i_2\ldots i_p}^{-1}$ of all the cubic diagrams compatible
with the cyclic orderings $\rho$ and $\tau$ and can be computed
through the algorithm in \CachazoIEA\ or a Berends--Giele recursion
\MafraLTU\ (see also \BaadsgaardVOA).

%****************
\subsec{Tree-level BCJ numerators from CHY}
%****************

The CHY formula \treeCHY\ describes (possibly supersymmetric)
Yang--Mills theory and gravity if one or both of the half-integrands
${\cal I}^{\rm tree} _L$ and ${\cal I}^{\rm tree} _R$ are identified
with a gauge invariant function ${\cal I}^{\rm tree} _{\rm SYM}
\equiv {\cal K}_n^{\rm tree} $ of the polarizations in the gauge
multiplet. For external bosons, the realization of ${\cal K}_n^{\rm
tree} $ as the (reduced) Pfaffian of an antisymmetric $2n\times 2n$
matrix was presented in \CachazoHCA. Despite the lack of a Pfaffian-like
representation, the supersymmetric completion is known from
the pure-spinor version of the CHY setup \BerkovitsXBA.

As pointed out in the ambitwistor setting in \MasonSVA, and detailed
in \GomezWZA\ in a pure-spinor context, ${\cal K}_n^{\rm tree} $ is
identical to the field-theory limit $\ap\to0$ of the $n$-point correlation function of open-string
vertex operators (which sets the Koba--Nielsen factor to the identity).
This equivalence of CHY integrands and superstring correlators holds on
the support of scattering equations, or integration-by-parts
relations of the string worldsheet. Hence, one can import the
manifestly supersymmetric results on the superstring tree-level
correlators obtained in \refs{\nptMethod, \MafraKJ, \nptTree}, and we will
later use the analogous correspondence at one loop.

The superstring version of ${\cal K}_n^{\rm tree} $ was shown in
\MafraKJ\ to be organized in terms of $(n{-}2)!$ Parke--Taylor factors \parketaylor,
\eqn\treecorr{
{\cal K}_n^{\rm tree} = \sum_{\rho \in S_{n-2}} \PT(1,\rho(2,3,\ldots,n{-}1),n) \, N^{\rm tree} _{1|\rho(2,3,\ldots,n{-}1)|n} \ ,
}
and the same form can be attained in the CHY setting \CachazoIEA\ by
applying scattering equations to the Pfaffian representation of its
bosonic components \CachazoHCA. The kinematic numerators $N^{\rm
tree} _{1|\rho(2,3,\ldots,n{-}1)|n}$ refer to the cubic diagrams of
half-ladder topology with fixed endpoints $1$ and $n$, see
\figmaster. Their explicit realization in pure-spinor superspace
\MafraKJ\ is based on superfields of ten-dimensional SYM
\wittentwistor, and the components involving gluon polarization
vectors $e^m$ and gaugino wave functions $\chi^\alpha$ can be
conveniently extracted using the streamlined $\theta$-expansions of
\MafraVCA, also see section \secPS\ for more details.

 \tikzpicture[scale=0.8, line width=0.30mm]
 \draw(0,0) --(-1,-1) node[left]{$1$};
 \draw(0,0) --(-1,1) node[left]{$\rho(2)$};
 \draw(0,0) -- (4.5,0);
 \draw(1,0) -- (1,1) node[above]{$\rho(3)$};
  \draw(2,0) -- (2,1) node[above]{$\rho(4)$};
 \draw (2.75,0.5)node{$\ldots$};
  \draw(3.5,0) -- (3.5,1) node[above]{$\rho(n{-}2)$};
 \draw(4.5,0) --(5.5,-1) node[right]{$n$};
 \draw(4.5,0) --(5.5,1) node[right]{$\rho(n{-}1)$};
 \draw[<->] (7.5,0) -- (9,0);
 \draw(11.4,0)node{$N^{\rm tree} _{1|\rho(2,3,\ldots,n{-}1)|n}$};
 \endtikzpicture
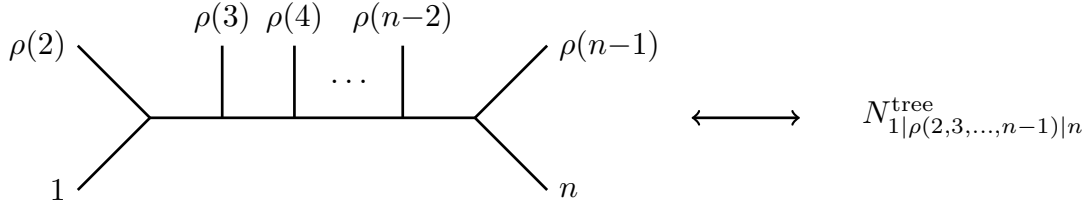
  \tikzcaption\figmaster{Half-ladder diagrams with legs 1 and $n$ attached to opposite endpoints and BCJ master numerators $N^{\rm tree} _{1|\rho(2,3,\ldots,n{-}1)|n}$ determine any other cubic diagram via kinematic Jacobi relations.}

 \tikzpicture[scale=1.1, line width=0.30mm]
 \scope[yshift=-0.5cm, xshift=0.0cm]
 \draw  (2,0.5) -- (1.5,1) node[left]{$a_2$};
 \draw   (2,0.5) -- (1.5,0) node[left]{$a_1$};
 \draw   (2,0.5) -- (2.5,0.5) ;
 \draw   (2.5,0.5) -- (3,1) node[right]{$a_3$};
 \draw   (2.5,0.5) -- (3,0) node[right]{$a_4$};
 \draw (2.2, -0.5) node{$C_i$};
 \draw (3.9,0.5) node{$+$};
 \endscope
 \scope[yshift=-0.5cm, xshift=3.3cm]
 \draw   (2,0.5) -- (1.5,1) node[left]{$a_3$};
 \draw   (2,0.5) -- (1.5,0) node[left]{$a_1$};
 \draw   (2,0.5) -- (2.5,0.5) ;
 \draw   (2.5,0.5) -- (3,1) node[right]{$a_4$};
 \draw   (2.5,0.5) -- (3,0) node[right]{$a_2$};
 \draw (3.9,0.5) node{$+$};
 \draw (2.2, -0.5) node{$C_j$};
 \endscope
 \scope[yshift=-0.5cm, xshift=6.6cm]
 \draw   (2,0.5) -- (1.5,1) node[left]{$a_4$};
 \draw   (2,0.5) -- (1.5,0) node[left]{$a_1$};
 \draw   (2,0.5) -- (2.5,0.5) ;
 \draw   (2.5,0.5) -- (3,1) node[right]{$a_2$};
 \draw   (2.5,0.5) -- (3,0) node[right]{$a_3$};
 \draw (2.2, -0.5) node{$C_k$};
 \draw (3.3,0.5) node[right]{$ \ \ = \ \ 0$};
 \endscope
 \endtikzpicture
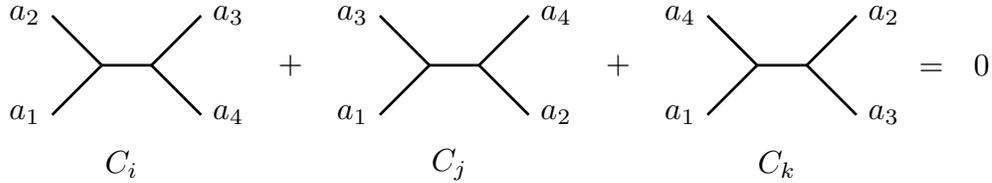
 \tikzcaption\figBCJ{The Jacobi identity implies the vanishing of the color factors associated to a triplet of cubic
 graphs, $C_i + C_j + C_k = 0$. In the above diagrams, the legs
 $a_1$, $a_2$, $a_3$ and $a_4$ may represent arbitrary subdiagrams. The BCJ duality
 states that their corresponding kinematic numerators $N_i(\ell)$ can be chosen such that $N_i(\ell) +
 N_j(\ell) + N_k(\ell) = 0$.}

As emphasized in \refs{\MafraKJ, \CachazoIEA}, the representation
\treecorr\ implies that numerators for all the other cubic diagrams
besides the $(n{-}2)!$ master graphs in \figmaster\ are determined
by the BCJ duality between color and kinematics \BernQJ: In the same
way as any triplet of graphs as shown in \figBCJ\ are related by a
group-theoretic Jacobi identity $f^{b a_1 [ a_2} f^{a_3a_4] b}=0$
among their color factors, one can arrange the kinematic dressings
of these graphs such that they satisfy the same Jacobi identities.
When computing color-ordered SYM amplitudes $A^{\rm tree}(\ldots)$
from the CHY prescription,
\eqnn\atree
$$\eqalignno{
A^{\rm tree}(\tau(1,2,\ldots,n)) &= {\cal M}^{\rm tree}_{{\rm SYM} \otimes U(N)} \, \big|_{ {\rm Tr}(t^{a_{\tau(1)}}t^{a_{\tau(2)}}   \ldots t^{a_{\tau(n)}} )  }  \cr
& = \int  \dd \mu_n^{\rm tree} \ \PT(\tau(1,2,\ldots,n)) \, {\cal K}_n^{\rm tree} &\atree
\cr
&= \! \! \sum_{\rho \in S_{n-2}}  \! \! m^{\rm tree}[  \tau(1,2,\ldots,n) \, | \,   1,\rho(2,\ldots,n{-}1),n ]  N^{\rm tree} _{1|\rho(2,3,\ldots,n{-}1)|n}  \ ,
}$$
the expansion of ${\cal K}_n^{\rm tree}$ in \treecorr\ and the form
of the doubly-partial amplitudes guarantee that each cubic-diagram
numerator is a linear combination of $N^{\rm tree}
_{1|\rho(2,\ldots,n{-}1)|n}$ with coefficients $\in \{0,1,-1\}$.
Their $(n{-}2)!$-counting agrees with the number of master
numerators under kinematic Jacobi identities, and it follows from
the arguments in \refs{\MafraKJ, \CachazoIEA} that the linear
combinations of $N^{\rm tree} _{1|\rho(2,\ldots,n{-}1)|n}$ in \atree\
satisfy kinematic Jacobi identities. In summary, the
expansion of the tree-level correlator \treecorr\ in terms of
$(n{-}2)!$ Parke--Taylor factors $\PT(\ldots)$ allows to read off a
set of BCJ master numerators.

%****************
\subsec{BCJ and KLT relations from CHY}
%****************

At tree level, a manifestly gauge invariant double-copy expression for gravity amplitudes is given by the KLT formula
\eqn\klttree{
{\cal M}^{\rm tree}_{{\rm SYM} \otimes {\rm SYM}} = \! \! \! \sum_{\rho,\tau \in S_{n-3}} \! \! \!  \tilde A^{\rm tree}(1,\rho(2,\ldots,n{-}2),n,n{-}1) \, S[\rho | \tau]_{1}  \, A^{\rm tree}(1,\tau(2,\ldots,n{-}2),n{-}1,n)
}
derived from tree-level scattering of open and closed strings \KawaiXQ.
The $(n{-}3)! \times (n{-}3)!$ matrix $S[\rho | \tau]_{1} \equiv
S[\rho(2,\ldots,n{-}2) | \tau(2,\ldots,n{-}2)]_{1}$ with entries of
order $\sim s^{n-3}$ has been firstly pinpointed to all multiplicity
in \BernSV\ and was later on studied in the momentum-kernel
formalism \BjerrumBohrTA. A recursive formula for its entries is
given by \CarrascoLDY
\eqn\momker{
S[A,j | B,j,C]_i = k_j \cdot (k_i + k_B) \, S[A | B,C]_i  \ , \ \ \ \ \ \ S[\emptyset | \emptyset]_i = 0 \ ,
}
see \mand\ for the multiparticle momenta $k_B$ associated with $B=b_1 b_2\ldots b_p$.
Permutation invariance of \klttree\ follows from BCJ relations among partial amplitudes \BernQJ
\eqn\BCJrels{
 \sum_{j=2}^{n-1} ( k_{1} \cdot k_{2 3\ldots j} )
A^{\rm tree}( 2,3,\ldots ,j,1, j{+}1, \ldots n)  = 0
}
which have been elegantly derived from monodromy properties of the
open-string worldsheet \BjerrumBohrRD. In the CHY setup, BCJ
relations emerge from the scattering equations \scatteq\ which
relate Parke--Taylor factors in complete analogy to \BCJrels\
\refs{\CachazoUQ,\CachazoGNA}
\eqn\BCJscat{
 \sum_{j=2}^{n-1} ( k_{1} \cdot k_{2 3\ldots j} )
\PT( 2,3,\ldots ,j,1, j{+}1, \ldots n)  =  0 \ {\rm mod} \ E_i \ ,
}
and they also hold for both entries of the doubly-partial amplitudes \doublypartial.
Note that the string-theory correlator \treecorr\
can be simplified to a BCJ basis of $(n{-}3)!$ worldsheet integrals
using integration by parts on the string worldsheet \nptMethod. This result
was later on identified to reproduce the structure of the KLT formula \klttree\ \BroedelTTA
\eqn\KLTcorrtree{
{\cal K}_n^{\rm tree}= \! \! \sum_{\rho,\tau \in S_{n-3}}  \! \!  \PT(1,\rho(2,\ldots,n{-}2),n,n{-}1) \, S[\rho | \tau]_{1}  \, A^{\rm tree}(1,\tau(2,\ldots,n{-}2),n{-}1,n) \ .
}
Insertion into \atree\ identifies doubly-partial amplitudes \doublypartial\ in a
suitable basis as the inverse of the momentum kernel \momker\ \refs{\BroedelTTA, \CachazoIEA},
\eqn\invMK{
m^{\rm tree}[1,\rho(2,\ldots,n{-}2),n,n{-}1 \, | \, 1,\tau(2,\ldots,n{-}2),n{-}1,n] =  S^{-1}[\rho | \tau]_{1} \ .
}
Then, the KLT formula \klttree\ follows from insertion of \KLTcorrtree\ in the CHY prescription \treeCHY,
\eqn\gravCHY{
{\cal M}^{\rm tree}_{{\rm SYM} \otimes {\rm SYM}} =  \int  \dd \mu_n^{\rm tree} \  {\cal K}_n^{\rm tree} \, \tilde {\cal K}_n^{\rm tree} \ ,
}
where it is convenient to exchange the roles of $n$ and $n{-}1$ in the
formula \KLTcorrtree\ for $\tilde {\cal K}_n^{\rm tree} $.

Notice that \gravCHY\ also makes the BCJ double-copy relations
manifest, which are equivalent to KLT relations at tree level: By
plugging \treecorr\ into \gravCHY, it follows that the
(super-)gravity amplitude is given by sum of all cubic diagrams with
numerators given by the double copy $N^{\rm tree}\,\tilde{N}^{\rm
tree}$. This is the major advantage of having a representation of
gauge-theory amplitude with numerators satisfying the BCJ
color-kinematics duality \BernQJ.

%****************
\subsec{CHY at one loop}
%****************

In the ambitwistor-string version of the CHY formalism, $g$-loop amplitudes in
various theories are written as integrals over the
moduli space of punctured genus-$g$ surfaces \AdamoTSA. At one loop,
the surface of interest is a torus with modular parameter $\tau$ in the
upper half plane such that its complex coordinate $z$ is identified with
$z{+}1$ and $z{+}\tau$. Apart from the torus punctures $z_{i=1,2,\ldots,n}$,
also the inequivalent choices of $\tau$ in the fundamental domain of
the modular group with $-{1\over 2} \leq {\rm Re}\, \tau \leq {1\over 2}$
and $|\tau|>1$ are integrated over.

However, one of the scattering equations
at genus one can be exploited \GeyerBJA\ to localize the $\tau$ integral
at the cusp $\tau \rightarrow i \infty$ where the torus degenerates to
a nodal sphere. Then, after a change of variables $\sigma = e^{2\pi i z}$,
one-loop amplitudes of (possibly supersymmetric) gravity and gauge theories
in $D$ spacetime dimensions simplify to \GeyerBJA
\eqnn\loopSUGRA
$$\eqalignno{
{\cal M}_{L \otimes R}&= \int {\dd^D \ell \over \ell^2} \int  \prod_{i=2}^n \dd \sigma_j \, \delta\Big( { (\ell \cdot k_i) \over \sigma_i} + \sum_{j=1 \atop{j \neq i}}^n {k_i \cdot k_j \over \sigma_{ij}}  \Big) \,  \widehat {\cal I}_L(\ell) \, \widehat {\cal I}_R(\ell) \ .
&\loopSUGRA
}$$
Note that translation invariance in the $z$-variable allows to insert another
integration $\dd \sigma_1$ along with a delta function e.g. $\delta(\sigma_1 - 1)$,
and the corresponding scattering equation
\eqn\loopscat{
{ (\ell \cdot k_i) \over \sigma_i} + \sum_{j=1 \atop{j \neq i}}^n {k_i \cdot k_j \over \sigma_{ij}} = 0
}
for $i=1$ does not need to be enforced separately because it follows
by adding the remaining equations for $i=2,3\ldots,n$.

For gauge theories, one of the integrands $\widehat{\cal I}_L(\ell)
\rightarrow \widehat{\cal I}_{U(N)}(\ell)$ is a sum of color
traces\foot{We suppress double traces in \fixgaugeint, and their
accompanying color-stripped amplitudes can be recovered from linear
combinations of single-trace subamplitudes \BernZX.}
\eqn\fixgaugeint{
\widehat{\cal I}_{U(N)}(\ell) = \sum_{\rho \in S_{n-1}} {\rm Tr} (t^{a_1} t^{a_{\rho(2)}} t^{a_{\rho(3)}}  \ldots t^{a_{\rho(n)}}) \, \widehat\PT^{(1)}(1,\rho(2,3,\ldots,n)) \ ,
}
accompanied by one-loop analogues $\widehat\PT^{(1)}(\ldots)$ of the Parke--Taylor factors \parketaylor,
\eqn\loopPT{
\widehat \PT^{(1)}(1,2,\ldots,n) \equiv  {1\over \sigma_{1} \sigma_{12} \sigma_{23} \ldots \sigma_{n-1,n}  } + {\rm cyc}(1,2,\ldots,n) \ .
}
The polarization-dependent integrand $\widehat{\cal I}_{\rm
SYM}(\ell)$ is the $\tau \rightarrow i \infty$ degeneration of the
genus-one correlation function
involving $n$ gauge-multiplet vertex operators $V(\sigma)$ to be
discussed in later sections \secRNS\ and \secPS,
\eqn\defloopK{
\widehat{\cal I}_{\rm SYM}(\ell) \equiv {(-1)^n {\cal K}_n(\ell) \over \sigma_1 \sigma_2\ldots \sigma_n} \ , \ \ \ \ \ \
 {\cal K}_n(\ell) \equiv \lim_{\tau \rightarrow i \infty} \langle V_1(\sigma_1) V_2(\sigma_2)\ldots V_n(\sigma_n) \rangle_\tau \ .
}
The inverse $\sigma_i$ can be traced back to the change of variables
$\sigma=e^{2\pi i z}$ with ${\rm d} z = {1\over 2\pi i } { {\rm d}
\sigma \over \sigma}$, and the prescription for evaluating the
correlation function $\langle \ldots \rangle_\tau$ is left generic
at this point to later on import results from both the RNS
and pure-spinor superstring. In terms of the two
integrands \fixgaugeint\ and \defloopK, one-loop amplitudes
\loopSUGRA\ in gauge theory and gravity are obtained as ${\cal
M}_{U(N) \otimes {\rm SYM}}$ and ${\cal M}_{{\rm SYM} \otimes {\rm
SYM}}$, respectively.

%****************
\subsec{New representations of one-loop integrals}
\par\subseclab\newellrep
%****************

\noindent It turns out that Feynman integrals arise in a
non-standard representation when integrating over the $\sigma_j$ in
\loopSUGRA: Instead of conventional propagators $(\ell+K)^2$
quadratic in $\ell$ (with some linear combination $K$ of external
momenta), the $\sigma_j$-integrals yield the results of repeated
partial fraction \GeyerBJA. The massless $n$-gon, for instance,
appears in the form of
\eqnn\ngonPF
$$\eqalignno{
&\int {2^{n-1} \ {\rm d}^D \ell \over \ell^2 (\ell{+}k_1)^2 (\ell{+}k_{12})^2 \ldots (\ell {+} k_{12\ldots n-1})^2}
= \sum_{i=0}^{n-1} \int { 2^{n-1} \ {\rm d}^D \ell \over (\ell {+} k_{12\ldots i})^2 } \prod_{j \neq i} {1\over (\ell {+} k_{12\ldots j})^2 - (\ell {+} k_{12\ldots i})^2 }
\cr
& \ \ \ \ \ = \sum_{i=0}^{n-1} \int { {\rm d}^D \ell \over \ell^2}   \prod_{j=0}^{i-1} {1\over s_{j+1,j+2,\ldots,i,-\ell}} \prod_{j=i+1}^{n-1} {1\over s_{i+1,i+2,\ldots,j,\ell} }  \ ,&\ngonPF
}$$
where the loop momentum $\ell$ in the $i^{\rm th}$ term has been
shifted by $k_{12\ldots i}$ in passing to the last line to ensure
that the only quadratic propagator is a pure $\ell^2$ in each term.
Each term in the sum over $i$ singles out one way of cutting open
the $n$-gon, and the result can be thought of as $n$ tree diagrams
involving off-shell momenta $\pm \ell$ \CachazoAOL, see \figcutting.
Each of these cubic diagrams will have a priori different kinematic
numerators, leaving a total of $n!$ inequivalent $n$-gon numerators.

\tikzpicture [scale=0.75, line width=0.30mm]
\scope[xshift=-0.8cm]
\draw (0.5,0)--(-0.5,0);
\draw (-0.5,0)--(-0.85,-0.35);
\draw [dashed](-0.85,-0.35)--(-1.2,-0.7);
\draw (0.5,0)--(1.2,-0.7);
\draw[dashed] (-1.2,-1.7)--(-1.2,-0.7);
\draw (1.2,-1.7)--(1.2,-0.7);
\draw (1.2,-1.7)--(0.85,-2.05);
\draw[dashed] (0.85,-2.05)--(0.5,-2.4);
\draw[dashed] (-0.5,-2.4)--(0.5,-2.4);
\draw[dashed] (-0.5,-2.4)--(-1.2,-1.7);
\draw (-0.5,0)--(-0.7,0.4)node[left]{$n$};
\draw (0.5,0)--(0.7,0.4)node[right]{$1$};
\draw (1.2,-0.7)--(1.6,-0.5)node[right]{$2$};
\draw (1.2,-1.7)--(1.6,-1.9)node[right]{$3$};
\draw (0,0) node{$| \! |$};
\draw (-0.25,0.2)node{$-$};
\draw (0.25,-0.2)node{$+$};
\endscope
%%%%%
%\draw[->] (2,0.5) .. controls (5.5,0.5) .. (5.5,-0.1);
%%%%%
\draw[-> ](1.7,-1.2)  -- (3.2,-1.2);
%%%%%
\scope[xshift=1.7cm, yshift=0.5cm]
\draw(11.4,-2)node{$+ \ {\rm cyclic}(1,2,\ldots,n)$};
\draw (2.9,-2)node[left]{$+\ell$} -- (5.8,-2);
\draw (7.2,-2) -- (8.1,-2)node[right]{$-\ell$};
\draw (3.5,-2) -- (3.5,-1.5)node[above]{$1$};
\draw (4.5,-2) -- (4.5,-1.5)node[above]{$2$};
\draw (5.5,-2) -- (5.5,-1.5)node[above]{$3$};
\draw[dashed] (5.8,-2) -- (7.2,-2);
\draw (7.5,-2) -- (7.5,-1.5)node[above]{$n$};
%
%\draw(3.1,-2)node{$>$}node[above]{$\ell$};
%\draw(7.9,-2)node{$<$}node[above]{$-\ell$};
\endscope
\endtikzpicture
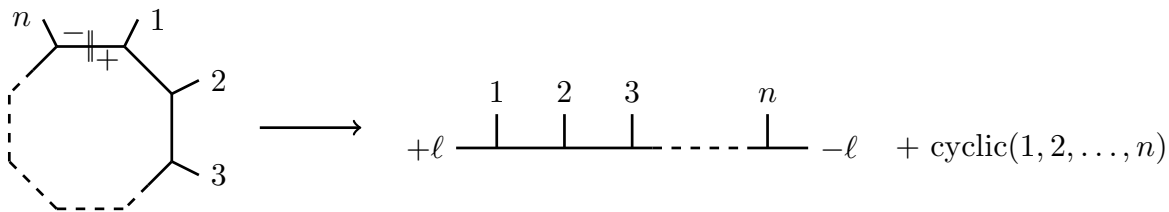
 \tikzcaption\figcutting{Interpretation of the partial-fraction representation of loop integrals as $(n{+}2)$-point tree-level diagrams.}
The manipulations in \ngonPF\ straightforwardly generalize to integrals with tree-level
subdiagrams, e.g.\ a box integral with massive momenta $k_A,k_B,k_C$ and $k_D$ allows
for the following four-term representation:
\eqn\massbox{
\int {8 \ \dd^D \ell  \over \ell^2 (\ell{+}k_A)^2 (\ell {+} k_{AB})^2 (\ell {+} k_{ABC})^2}
= \int {\dd^D \ell \over \ell^2} \Big( {1\over s_{A,\ell} s_{AB,\ell} s_{D,-\ell} } + {\rm cyc}(A,B,C,D) \Big) \ .
}
In this way, the one-loop integrand for color-ordered single-trace
amplitudes can be split into $n$ terms, similar to that of \ngonPF\
for the $n$-gon. Each of the $n$ terms can be interpreted as the
forward limit of $(n{+}2)$-point trees with off-shell momenta, e.g.
the momenta of the two legs between $n$ and $1$ being identified as
$\ell$ and $-\ell$. The off-shell momenta can be viewed as on-shell,
higher-dimensional ones, and the one-loop CHY formula \loopSUGRA\ was
obtained as the forward limit of such higher-dimensional tree
amplitudes \CachazoAOL. Although it is non-trivial to perform loop
integrations, the new representation of loop integrals has to give
the same result as the canonical Feynman integrals.

These integrals not only naturally appear in the CHY formalism, but also
play an important role in the Q-cut representation of loop amplitudes \BaadsgaardTWA.
The new representation provides a well-defined notion of ``loop integrands" for
generic, non-planar theories, which can be exploited to reveal structures of
loop amplitudes. In particular, as conjectured recently \HeMZD, in the new
representation it is natural to generalize KLT and BCJ relations, \klttree\
and \BCJrels, to one loop. In section \proofKLT, we prove these new relations, as
well as the color-kinematics duality and double copy at the one-loop level in
this new representation.

%*******************************
\newsec{BCJ and KLT at one-loop}
%*******************************
%*******************************

%****************
\subsec{One-loop correlators in generic $SL(2,\Bbb C)$ frames}
%***************

The expressions in the above review of the one-loop CHY setup are
adapted to a particular $SL(2,\Bbb C)$ frame where two additional
punctures $\sigma_+=0$ and $\sigma_- \rightarrow \infty$ are
identified on the nodal sphere and associated with momenta
$k_{\pm}=\pm \ell$. This $SL(2,\Bbb C)$-fixing is reflected in the
hat notation for the integrands $\widehat{\cal I}_{U(N)}(\ell) $ and
$\widehat{\cal I}_{\rm SYM}(\ell) $ in \fixgaugeint\ and \defloopK\
as well as the one-loop Parke--Taylor factors $\widehat
\PT^{(1)}(1,2,\ldots,n)$ in \loopPT. In this subsection, we shall
give the analogous expressions for ``unhatted'' quantities ${\cal
I}_{U(N)}(\ell), \ {\cal I}_{\rm SYM}(\ell)$ and
$\PT^{(1)}(1,2,\ldots,n)$ in a generic frame: Requiring $SL(2,\Bbb
C)$-weight two in each puncture $\sigma_{j=1,2,\ldots,n}$ and
$\sigma_{+},\sigma_{-}$ yields unique $SL(2,\Bbb C)$-covariant
uplifts, and we will introduce a method to express both ${\cal
I}_{U(N)}(\ell)$ and ${\cal I}_{\rm SYM}(\ell)$ in terms of
$(n{+}2)$-point tree-level Parke--Taylor factors \parketaylor.

For instance, $\sigma_j$-independent contributions from the correlators
${\cal K}_n(\ell)$ to the gauge-theory integrands \defloopK\ can expressed via
$SL(2,\Bbb C)$-fixed tree-level Parke--Taylor factors with
$\sigma_+=0$ and $\sigma_{-}\rightarrow \infty$ \GeyerJCH ,
\eqnn\ngon
$$\eqalignno{
\int \prod_{j=1}^n {{\rm d} \sigma_j \over \sigma_j }
&= \int \prod_{j=1}^n {{\rm d} \sigma_j \over \sigma_{j,+} }   \Big|_{\sigma_{ +} = 0}
=(-1)^n \sum_{\rho \in S_{n}} \int { {\rm d} \sigma_1 \, {\rm d} \sigma_2 \ldots {\rm d} \sigma_n \over \sigma_{+,\rho(1)} \sigma_{\rho(1),\rho(2)} \ldots \sigma_{\rho(n-1),\rho(n)}} \Big|_{\sigma_{ +} = 0}  \cr
&=(-1)^n \lim_{\sigma_- \rightarrow \infty}
\sum_{\rho \in S_{n}} \int {  {\rm d} \sigma_1 \, {\rm d} \sigma_2 \ldots {\rm d} \sigma_n \ (-\sigma_{-}^2)  \over \sigma_{+,\rho(1)} \sigma_{\rho(1),\rho(2)} \ldots \sigma_{\rho(n-1),\rho(n)} \sigma_{\rho(n),-}  \sigma_{-,+}  } \Big|_{\sigma_{ +} = 0}
\cr
&=(-1)^n \sum_{\rho \in S_{n}} \int {{\rm d} \sigma_- \, {\rm d} \sigma_+ \, \prod_{j=1}^n {\rm d} \sigma_j \over {\rm vol} \, SL(2,\Bbb C)} \PT(+,\rho(1,2,\ldots,n),-) \ , &\ngon
}$$
or in short
\eqn\covtoy{
 \prod_{j=1}^n {1 \over \sigma_j }  =  (-1)^n \lim_{\sigma_- \rightarrow \infty} (-\sigma_-^2) \lim_{\sigma_+ \rightarrow 0}
  \sum_{\rho \in S_{n}} \PT(+,\rho(1,2,\ldots,n),-) \ .
}
Likewise, one-loop Parke--Taylor factors $\widehat
\PT^{(1)}(\ldots)$ in \loopPT\ were defined in \GeyerBJA\
from their $SL(2,\Bbb C)$-covariant uplifts $\PT^{(1)}(\ldots)$,
\eqnn\loopPTcovA
\eqnn\loopPTcovB
$$\eqalignno{
\PT^{(1)}(1,2,\ldots,n) &\equiv  \PT(+,1,2,\ldots,n,-) \ +\ {\rm cyc}(1,2,\ldots,n) &\loopPTcovA \cr
\widehat \PT^{(1)}(1,2,\ldots,n) &= \lim_{\sigma_- \rightarrow \infty} (-\sigma_-^2) \lim_{\sigma_+ \rightarrow 0} \PT^{(1)}(1,2,\ldots,n) \ , &\loopPTcovB
}$$
which implies the following form for the $U(N)$ integrand in a generic $SL(2,\Bbb C)$-frame,
\eqn\gaugeint{
{\cal I}_{U(N)}(\ell) = \sum_{\rho \in S_{n-1}} {\rm Tr} (t^{a_1} t^{a_{\rho(2)}} t^{a_{\rho(3)}}  \ldots t^{a_{\rho(n)}}) \, \PT^{(1)}(1,\rho(2,3,\ldots,n)) \ .
}
As will be detailed in the next subsection, also a generic
correlator ${\cal K}_n(\ell) $ with non-trivial
$\sigma_j$-dependence admits a unique $SL(2,\Bbb C)$-covariant
uplift ${\cal I}_{\rm SYM}(\ell)$ for the SYM integrand \defloopK.
Regardless of the details of ${\cal I}_{U(N)}(\ell)$ and ${\cal
I}_{\rm SYM}(\ell)$, the one-loop CHY prescription \loopSUGRA\ in a
generic $SL(2,\Bbb C)$-frame can be boiled down to the tree-level
measure \treeCHY,
\eqnn\covSUGRA
$$\eqalignno{
{\cal M}_{L \otimes R}&= \int {\dd^D \ell \over \ell^2} \lim_{k_{\pm} \rightarrow \pm \ell} \int  \dd \mu^{\rm tree}_{n+2} \ {\cal I}_L(\ell) \,   {\cal I}_R(\ell) \ ,
&\covSUGRA
}$$
in lines with the degeneration of the torus to a nodal Riemann sphere as $\tau \rightarrow i \infty$. Note in particular that the one-loop scattering equations \loopscat\ descend from their $(n{+}2)$-point tree-level instances \scatteq\ in the limit $\sigma_- \rightarrow \infty $ and $\sigma_+ = 0$ with $k_{\pm} \rightarrow \pm \ell$,
\eqn\checkscat{
{ (k_+ \cdot k_i) \over \sigma_{i,+}} +{ (k_- \cdot k_i) \over \sigma_{i,-}} + \sum_{j=1 \atop{j \neq i}}^n {k_i \cdot k_j \over \sigma_{ij}} \Big|_{ k_{\pm} \rightarrow \pm \ell}  \rightarrow 0 \ .
}
In the same way as only $n{-}3$ scattering equations are independent
in the $n$-point tree-level prescription \treeCHY, the $n{-}1$
scattering equations in \loopSUGRA\ are sufficient for the situation
in one-loop amplitudes \covSUGRA\ with $n{+}2$ punctures.

 In theories with reduced or zero supersymmetry, the kinematic regime with
$k_{\pm} \rightarrow \pm \ell$ gives rise to singularities upon
integration over $\sigma_j$, and we will later comment on their
regularization.

%****************
\subsec{The $\sigma$-dependence of gauge-theory correlators}
\par\subseclab\polyGij
%***************

This subsection is devoted to the structure of $\sigma_j$-dependent
correlators ${\cal K}_n(\ell) $ which carry the state-dependence in
the SYM integrand \defloopK.
The expressions for ${\cal K}_n(\ell) $ can be imported from
the superstring correlator in the field-theory limit, see sections \secRNS\ and \secPS\ for explicit
examples in the RNS and pure-spinor formalism. As is well-known from
superstring theory, singularities of genus-one correlators at
generic values of $\tau$ arise from the holomorphic torus Green
function $\partial_z \log \theta_1(z,\tau)$, where $\theta_1$
denotes the odd Jacobi theta function
\eqn\jactheta{
\theta_1(z,\tau) \equiv 2iq^{1 / 8} \sin(\pi z) \prod_{j=1}^{\infty} (1-q^j) (1- e^{2\pi i z}q^j) (1- e^{-2\pi i z}q^j)  = - \theta_1(-z,\tau)
}
with a simple pole at the origin and $q \equiv e^{2\pi i\tau}$. We recall
the change of variables $\sigma=e^{2\pi i z}$ between the punctures $\sigma$
in \covSUGRA\ and the torus coordinates with identifications of $z$ with $z+1$ and $z+\tau$.
By the localization of CHY correlators at the cusp $\tau \rightarrow i\infty$, we
will only be interested in the limit \GeyerBJA
\eqn\defGij{
{1\over 2\pi i}
\lim_{ \tau \rightarrow i\infty} \partial_z \log \theta_1(z_i-z_j,\tau)  = G_{ij} \equiv  {\sigma_i+\sigma_j \over 2 \sigma_{ij}} \ .
}
In terms of the Green function $G_{ij}$, the one-loop scattering
equations \loopscat\ (and also the $\tau \rightarrow i\infty$
degeneration of integration-by-parts relations in string theory)
can be written as
\eqn\Gijscat{
(\ell \cdot k_i) +
\sum_{j=1 \atop{j \neq i}}^n (k_i \cdot k_j ) G_{ij}  = 0 \ .
}
Note that the partial-fraction identity $(\sigma_{ij} \sigma_{ik})^{-1}+ {\rm cyc}(i,j,k)=0$
among nested products of the tree-level Green function $\sigma_{ij}^{-1}$
does not carry over to $G_{ij}$,
\eqn\pfoneloop{
G_{ij} G_{ik} + {\rm cyc}(i,j,k) = { [   \sigma_{jk} (\sigma_i + \sigma_j )(\sigma_i + \sigma_k ) + {\rm cyc}(i,j,k) ] \over 4\sigma_{ij}\sigma_{ik} \sigma_{jk}} = {1\over 4}  \ .
}
This result follows from the field-theory limit of
the corresponding genus-one Fay identities
studied in \refs{\BLK,\BroedelVLA}.

 As will be proven in section \secRNS, any one-loop 
gauge-theory correlator ${\cal K}_n(\ell)$ can be written as a polynomial in
$G_{ij}$ and $\ell$, regardless of the multiplicity and the number
of supersymmetries, and even in non-supersymmetric situations. 
The degree of this polynomial will be shown to
vary with the number of supercharges, the highest power of Green
functions being $G_{ij}^{n-4}$ in presence of maximal supersymmetry,
$ G_{ij}^{n-2}$ in gauge theories with 8 or 4 supercharges and
$G_{ij}^{n}$ in non-supersymmetric cases. Of course, the $G_{ij}$ do
not appear with homogeneous degree since integration by parts
\Gijscat\ interchanges combinations of $G_{ij}$ with loop momenta, and the Fay
identity \pfoneloop\ mixes powers of $G^{k}_{ij}$, $G^{k-2}_{ij}$,
$G^{k-4}_{ij},\ldots$ along with a given $\ell$-dependence, see
the examples in section \secPS.

With less than $n$ powers of $G_{ij}$, i.e.\ in presence of at least
4 supercharges, one can furthermore use the scattering equations in
their form \Gijscat\ to eliminate closed subcycles of Green
functions such as $G_{12}^2= - G_{12} G_{21}$ and $G_{12} G_{23}
G_{31}$. In other words, when drawing an edge between vertices $i$
and $j$ for each factor of $G_{ij}$, the pattern of $G_{ij}$ in
supersymmetric ${\cal K}_n(\ell)$ can be represented as a Cayley
graph. This is always possible at any multiplicity, see appendix \appbasis\
 below (for similar algorithms at tree-level, see
\refs{\CachazoNWA}).  
%We furthermore suspect that closed subcycles 
%can still be eliminated in non-supersymmetric correlators, but the
 After the first version of this work, it was demonstrated in
\GeyerELA\ that closed subcycles can still be eliminated in non-supersymmetric 
correlators.

%****************
\subsec{Gauge-theory correlators in terms of Parke--Taylor factors}
\par\subseclab\polyGijB
%***************

The central result of this section concerns the interplay of such $G_{ij}$ with
the Parke--Taylor structure \covtoy\ seen in the case of $\sigma_j$-independent
${\cal K}_n(\ell)$, where it is convenient to define
 \eqnn\calZdefFirst
 $$\eqalignno{
\cZ_{i_1i_2i_3 \ldots i_{q-1}i_q} &\equiv {1\over \s_{i_1i_2}\s_{i_2i_3}
\ldots \s_{i_{q-1}i_q}}\,.
 &\calZdefFirst
 }$$
In the presence of $G_{ij}$ factors with no subcycles,
it will be proven in the appendix~\appGij\ that the sum in
the right-hand side of
(recall that $\s_+=0$)
\eqn\noGij{
\prod_{j=1}^{n} {1\over \sigma_j} = (-1)^n
\sum_{\rho\in S_n}\cZ_{+\rho(1,2,3, \ldots, n)}\,,
}
is modified by $\rho$-dependent signs,
 \eqn\defsgnrho{
 {\rm sgn}_{ij}^{\rho} \equiv \cases{ +1 &:
 \ \hbox{$i$ is on the right of $j$ in $\rho(1,2,\ldots,n)$} \cr
 -1 &: \ \hbox{$i$ is on the left of $j$ in $\rho(1,2,\ldots,n)$}
 }  \ .
 }
More explicitly, with $m$ factors of $G_{ij}$ without subcycles,
\eqn\anyGij{
G_{i_1j_1}G_{i_2j_2} \ldots G_{i_{m} j_{m}}\prod_{j=1}^n{1\over \sigma_j} =
{(-1)^n\over 2^m}\sum_{\rho\in S_n}\sign^\rho_{i_1j_1}\sign^\rho_{i_2j_2}
\ldots\sign^\rho_{i_{m}j_{m}}\cZ_{+\rho(1,2, \ldots,n)}\ .
}
Given that
\eqn\cZtoPT{
\cZ_{+\rho(1,2,
\ldots,n)}=\lim_{\s_-\to\infty}(-\s_-^2)P(+,\rho(1,2, \ldots,n),-) \ ,
}
the net effect of $G_{ij}$ in converting the correlator
$ {\cal K}_n (\ell) $ to a Parke--Taylor expansion of the gauge-theory
integrand \defloopK\ is captured by the prescription
$G_{ij} \rightarrow {1\over 2} {\rm sgn}^\rho_{ij}$,
\eqn\thumb{
{\cal I}_{\rm SYM}(\ell)  =   \sum_{\rho \in S_n} \PT(+,\rho(1,2,\ldots,n),-) \, \Big( {\cal K}_n (\ell) \Big|_{G_{ij} \rightarrow {1\over 2} {\rm sgn}^\rho_{ij} } \Big) \ .
}
This generalizes the expansion \treecorr\ of the tree-level
correlator in terms of $(n{-}2)!$ Parke--Taylor factors
$\PT(1,\rho(2,3,\ldots,n{-}1),n)$ to the one-loop order:
With $n{+}2$ punctures on the nodal Riemann
sphere -- $\sigma_{\pm}$ and $\sigma_{j=1,2,\ldots,n}$ --
the analogous family of Parke--Taylor factors has $n!$
elements $\PT(+,\rho(1,2,\ldots,n),-)$. By analogy with
\treecorr, it is tempting to introduce a notation
\eqn\notmaster{
N _{+|\rho(1,2,\ldots,n)|-} (\ell) \equiv {\cal K}_n (\ell) \Big|_{G_{ij} \rightarrow {1\over 2} {\rm sgn}^\rho_{ij} }
}
for the kinematic coefficients of the Parke--Taylor factors,
and it will be argued in the next subsection that the resulting expansion
\eqn\loopcorr{
{\cal I}_{\rm SYM}(\ell) = \sum_{\rho \in S_{n}} \PT(+,\rho(1,2,\ldots,n),-) \, N _{+|\rho(1,2,\ldots,n)|-} (\ell)
}
identifies the $N _{+|\rho(1,2,\ldots,n)|-} (\ell)$ in \notmaster\
as BCJ master numerators of $n$-gon graphs. The counting of Parke--Taylor
factors in \loopcorr\ matches the $n!$ inequivalent $n$-gon diagrams
in the partial-fraction representation of loop integrals, realizing
the permutations of $1,2,\ldots,n$ in \figcutting. However, \thumb\ to \loopcorr\ 
are based on representations of ${\cal K}_n(\ell)$
without any closed subcycles of $G_{ij}$ which are known to exist for
theories with at least four supercharges. 
 For cases with zero supersymmetry, representations of ${\cal K}_n(\ell)$
without any closed subcycles of $G_{ij}$ were shown to exist as well \GeyerELA\ after
the first version of this work.
%It is not clear in which way the
%results extend to non-supersymmetric cases.

%****************
\subsec{Analytic evaluation of CHY integrals and BCJ master numerators}
\par\subseclab\whymaster
%***************

Already at tree level, a central advantage of expressing the
kinematic integrand ${\cal I}^{\rm tree}_{\rm SYM}$ in terms of
Parke--Taylor factors is the availability of doubly-partial
amplitudes \doublypartial\ to evaluate the CHY integrals. Similarly,
the Parke--Taylor form of the one-loop kinematic integrand
\loopcorr\ and the $\dd \mu^{\rm tree}_{n+2} $ measure in \covSUGRA\
allow one to derive the one-loop propagators from doubly-partial
amplitudes at tree-level with $(n{+}2)$ legs
\eqnn\CHYint
$$\eqalignno{
&\int  \dd \mu^{\rm tree}_{n+2} \ \PT(\alpha(1,2,\ldots,n,+,-)) \PT(\beta(1,2,\ldots,n,+,-))  &\CHYint \cr
& \ \ \ = \lim_{k_{\pm} \rightarrow \pm \ell} m^{\rm tree}[
\alpha(1,2,\ldots,n,+,-) \, |\, \beta(1,2,\ldots,n,+,-)
] \ .
}$$
Thanks to the Berends--Giele recursion for $m^{\rm tree}[ \cdot  \, |\,  \cdot] $ \MafraLTU,
this makes analytic evaluations of gauge-theory and gravity amplitudes tractable for
a large number of external legs,
\eqnn\analyticSYM
\eqnn\analyticSUGRA
$$\eqalignno{
A(1,2,\ldots,n)&= \int {\dd^D \ell \over \ell^2}  \int \dd \mu^{\rm tree}_{n+2}
\ \PT^{(1)}(1,2,\ldots,n) \, {\cal I}_{\rm SYM}(\ell)
\cr
&= \int {\dd^D \ell \over \ell^2}  \lim_{k_{\pm} \rightarrow \pm \ell}  \sum_{\rho \in S_n}
N _{+|\rho(1,2,\ldots,n)|-} (\ell)    &\analyticSYM \cr
& \ \ \ \times  \sum_{i=0}^{n-1}  m^{\rm tree}[+,i{+}1,\ldots,n,1,2,\ldots,i,- \, |\, +,\rho(1,2,\ldots,n),-] \cr
%%%%%%%%%%
{\cal M}_{{\rm SYM} \otimes {\rm SYM}}&= \int {\dd^D \ell \over \ell^2} \lim_{k_{\pm} \rightarrow \pm \ell} \sum_{\rho,\tau \in S_n}   N _{+|\rho(1,2,\ldots,n)|-} (\ell) \, \tilde N _{+|\tau(1,2,\ldots,n)|-} (\ell)  &\analyticSUGRA   \cr
& \ \ \ \times  m^{\rm tree}[+,\rho(1,2,\ldots,n),- \, |\, +,\tau(1,2,\ldots,n),-]   \ .
}$$
It is important to perform the limit $k_{\pm} \rightarrow \pm \ell$ {\it after}
summing the permutations $\rho,\tau$ because the conspiration of different $N _{+|\rho(\ldots)|-} (\ell)$
leads to cancellations among spurious divergent propagators. In absence of maximal supersymmetry,
forward-limit divergences will arise in \analyticSYM, and a regularization scheme for
cases with at least four supercharges is given around \delicate\ as well as section \secREDUCE.

As an example for a smooth forward limit $ k_{\pm} \rightarrow \pm \ell $, let
us reproduce the scalar box integral \massbox\ in the four-point one-loop
amplitude from a sum of six-point doubly-partial amplitudes at
tree level following from \analyticSYM\ \refs{\GeyerBJA, \HeYUA, \GeyerJCH}
\eqnn\boxint
$$\eqalignno{
&\int { \dd^D \ell \over \ell^2} \lim_{k_{\pm} \rightarrow \pm \ell}  \sum_{\rho \in S_{4}} \Big( m^{\rm tree}[+,1,2,3,4,- \, |\, +,\rho(1,2,3,4),-]  + {\rm cyc}(1,2,3,4) \Big) &\boxint\cr
&= \int { \dd^D \ell \over \ell^2}  \Big( {1\over s_{1,\ell} s_{12,\ell} s_{123,\ell} } + {\rm cyc}(1,2,3,4) \Big) = \int {8 \, \dd^D \ell \over   \ell^2(\ell+k_{1})^2(\ell+k_{12})^2(\ell+k_{123})^2 }  \ .
}$$
This example illustrates that the kinematic limit must be performed
{\it after} combining the permutations $\rho$: Several
choices of $\rho$ introduce divergent tadpole propagators such as
$s_{1234}^{-1}$ in $m^{\rm tree}[+,1,2,3,4,- \, |\, +,2,1,4,3,-] =
(s_{12}s_{34} s_{12,\ell})^{-1}+(s_{12}s_{34} s_{1234})^{-1}$ which
drop out after summing over $\rho$.

A more delicate treatment is needed for half- and quarter-maximal
supersymmetry, where one factor of $G_{12}$ occurs in the three-point correlator, and \anyGij\ leads to
\eqnn\delicate
$$\eqalignno{
  \lim_{k_{\pm} \rightarrow \pm \ell}  \sum_{\rho \in S_{3}} {\rm sgn}_{12}^\rho \, m^{\rm tree}[+,1,2,3,- \, |\, +,\rho(1,2,3),-] &= { 2 \over s_{12} s_{12,\ell} } + {1\over s_{1,\ell} s_{12,\ell}}
% 2/(s[1, 2] s[1, 2, 4]) + 1/(s[1, 4] s[1, 2, 4])
 \cr
  \lim_{k_{\pm} \rightarrow \pm \ell}  \sum_{\rho \in S_{3}} {\rm sgn}_{13}^\rho \, m^{\rm tree}[+,1,2,3,- \, |\, +,\rho(1,2,3),-] &=   {1\over s_{1,\ell} s_{12,\ell} }
%   -(1/(s[2, 4] s[2, 3, 4]))
  &\delicate \cr
  \lim_{k_{\pm} \rightarrow \pm \ell}  \sum_{\rho \in S_{3}} {\rm sgn}_{23}^\rho \, m^{\rm tree}[+,1,2,3,- \, |\, +,\rho(1,2,3),-] &= {2\over s_{23} s_{1,\ell}} + {1\over s_{1,\ell} s_{12,\ell} }
  % 2/(s[1, 2] s[3, 4]) + 1/(s[3, 4] s[1, 3, 4])
}$$
via five-point doubly-partial amplitudes. In the kinematic phase space of
three massless particles, we obtain divergences from the pole in
$s_{12}= {1\over 2}[k^2_3-k_2^2-k_1^2]=0$. However, a compensating
numerator of $s_{12}$ can be extracted from the kinematic factor
along with $G_{12}$ \BergWUX, see \figminahan. Hence, in a suitable regularization
scheme due to Minahan \MinahanHA\ which is detailed in section \secREDUCE,
one can extract finite bubble contributions \BergFUI\ from the
terms $\sim ( s_{12} s_{12,\ell} )^{-1}$ and $ \sim (s_{23} s_{1,\ell})^{-1}$
in \delicate.

\tikzpicture [scale=0.8, line width=0.30mm]
\scope[xshift=3.2cm]
\draw (0,0) -- (-1,1) node[left]{$2$};
\draw (0,0) -- (-1,-1) node[left]{$1$};
\endscope
\draw (3.2,0) -- (4,0);
\draw (3.6,-0.5)node{$s_{12}$};
\draw (4,0) .. controls (4.5,1) and (6.5,1) .. (7,0);
\draw (4,0) .. controls (4.5,-1) and (6.5,-1) .. (7,0);
\draw (5.5,-0.75)node{$<$}node[above]{$\ell$};
\draw (7,0) -- (7.8,0) node[right]{$3$};
\draw (5.3,-1.7)node{numerator$ \ s_{12}(e_1 \cdot e_2)(k_1 \cdot e_3)$};
%%%%%%%%%%%%%%%%%
\draw[<->] (9.3,0.25)--(10.4,0.25);
%%%%%%%%%%%%%%%%%
\scope[xshift=8.6cm]
\scope[xshift=4cm]
\draw (0,0) -- (-1,1) node[left]{$2$};
\draw (0,0) -- (-1,-1) node[left]{$1$};
\endscope
\draw (4,0) .. controls (4.5,1) and (6.5,1) .. (7,0);
\draw (4,0) .. controls (4.5,-1) and (6.5,-1) .. (7,0);
\draw (5.5,-0.75)node{$<$}node[above]{$\ell$};
\draw (7,0) -- (7.8,0) node[right]{$3$};
\draw (5.5,-1.7)node{leftover numerator$ \ (e_1 \cdot e_2)(k_1 \cdot e_3)$};
\endscope
\endtikzpicture
 \tikzcaption\figminahan{The divergent propagator $s_{ij}^{-1}$ in external bubbles is cancelled by a
 formally vanishing factor of $s_{ij}$ in the kinematic numerator.}

 \subsubsec{The BCJ duality in the new representation of Feynman integrals}
%***************

Given that the expression \atree\ for $n$-point gauge-theory trees
is known to yield cubic-diagram numerators which satisfy kinematic
Jacobi identities \refs{\MafraKJ, \CachazoIEA}, its forward limit in
\analyticSYM\ must also realize the BCJ duality between color and
kinematics \refs{\BernQJ, \BernUE}. In particular, by restricting
the tree-level arguments of \BernYG\ to the forward limit, the
cubic-diagram numerators in the representation \analyticSUGRA\ of
the gravity amplitude are the double copies of the kinematic
gauge-theory numerators from \analyticSYM.

We emphasize that the present realization of the BCJ duality is adapted to the new
representation \ngonPF\ of Feynman integrals with all $\ell$-dependent
propagators but one linear in the loop momentum. In the original formulation of
the loop-level BCJ duality \BernUE\ with propagators quadratic in $\ell$,
each cyclically inequivalent $n$-gon graph is counted as a single
cubic diagram. As explained in section \newellrep, the results of the CHY integrals in
\CHYint\ organize one-loop amplitudes into $n$ distinct cubic diagrams per cyclically
inequivalent $n$-gon. They are interpreted as distinct tree-level diagrams with two
extra legs at the $n$ possible positions of $\ell$, and their kinematic numerators are
a priori unrelated.

Accordingly, the cubic-diagram expansion of one-loop gauge-theory
and gravity amplitudes obtained from \analyticSYM\ and \analyticSUGRA\ takes the schematic form
\eqnn\schematicBCJA
\eqnn\schematicBCJB
$$\eqalignno{
{\cal M}_{{\rm SYM} \otimes U(N)} &= \int {\dd^D \ell \over \ell^2} \sum_{ i \in \Gamma_{n+2}} { C_i \, N_i(\ell) \over \prod_{  {\rm edges} \ \alpha_i} P^2_{\alpha_i}(\ell)} &\schematicBCJA \cr
{\cal M}_{{\rm SYM} \otimes {\rm SYM}} &= \int {\dd^D \ell \over \ell^2} \sum_{ i \in \Gamma_{n+2}} {  N_i(\ell) \, \tilde N_i(\ell) \over \prod_{  {\rm edges} \ \alpha_i} P^2_{\alpha_i}(\ell)}  \ ,&\schematicBCJB
}$$
where $\Gamma_{n+2}$ denotes the set of $(n{+}2)$-point tree-level graphs $i$. The
propagators $P^{-2}_{\alpha_i}(\ell)$ are linear in $\ell$, and the color
factors $C_i$ are obtained by dressing each cubic vertex with $f^{abc}$ while contracting the two
extra legs $\pm\ell$ with a Kronecker delta. Note that all the $n$ cubic diagrams in the partial-fraction
decomposition of an $n$-gon yield identical color factors.

The numerators $N_i(\ell)$ are linear combinations of the $N_{+|\rho(1,2,\ldots,n)|-}(\ell)$ in \notmaster\ and \loopcorr\
such as to solve the kinematic Jacobi relations depicted in \figBCJ. Of course, these $N_i(\ell)$
vanish for tadpole graphs in supersymmetric theories, and also for bubble-
and triangle graphs in case of maximal supersymmetry. In summary, the expression for supersymmetric
$n$-point correlators \loopcorr\ in terms of Parke--Taylor factors identifies the
kinematic coefficients $N _{+|\rho(1,2,\ldots,n)|-}(\ell)$ as BCJ
master numerators of $n$-gon diagrams.

Since physical properties such as unitarity
cuts and UV divergences are currently more evident in the standard
representations of loop integrals in terms of propagators
$(\ell+K)^{-2}$, it would be interesting to study the systematic
recombination of the loop integrals in \analyticSYM\ and
\analyticSUGRA\ to the standard form. Moreover, it would be
desirable to preserve the color-kinematics duality in this
recombination process. We have checked that the local five-point BCJ
numerators of \MafraGJA\ for the conventional $(\ell+K)^{-2}$
propagators are reproduced in this recombination, and the situation
at six points is discussed in section \fivefive.

%****************
\subsec{Partial integrands and one-loop BCJ-relations}
%***************

The above construction of one-loop BCJ-representations was greatly
alleviated by the tight analogy with tree level. In defining
gauge invariant building blocks, however, this analogy is broken by the definition of
color-ordered one-loop amplitudes $A(\ldots)$ of SYM  through the sum
\analyticSYM\ of {\it several} $(n{+}2)$-particle Parke--Taylor factors in
$\PT^{(1)}(\ldots)$. In order to arrive at a
manifestly gauge and diffeomorphism invariant formulation of the BCJ
duality and double copy, it is convenient to study a more elementary
quantity, the {\it partial integrand}~\HeMZD
\eqn\partialint{
a(\tau(1,2,\ldots,n,+,-)) \equiv \int  \dd \mu^{\rm tree}_{n+2} \  \PT(\tau(1,2,\ldots,n,+,-)) \, {\cal I}_{\rm SYM}(\ell) \ .
}
We emphasize that this definition in the CHY framework does not require
any supersymmetry. As studied in  \refs{ \GeyerBJA,\HeYUA}  and especially in \CachazoAOL, the one-loop integrand can 
be obtained from the CHY representation of tree amplitudes in one higher dimension, no matter what the theory is. 
The contribution of each solution of the scattering equations to the one-loop integrand of a gauge theory and the partial 
integrand \partialint\ is gauge invariant before summing over all the solutions. This allows us to discard the 
singular solutions, considering that their contributions turn out to be homogeneous functions of the loop 
momentum and integrate to zero. This way, the CHY formula for one-loop amplitudes does not break 
gauge invariance like the case in \CaronHuotZT\ where the forward limit is taken in the original number of dimensions.

According to their definition \partialint, partial integrands isolate a single tree-level Parke--Taylor factor from their sum in
$\PT^{(1)}(\ldots)$ and allow to reconstruct color-stripped single-trace amplitudes \analyticSYM\ via
\eqn\backto{
A(1,2,\ldots,n)=\int  {\dd^D \ell \over \ell^2}
 \sum_{i=1}^n  a(1,2,\ldots,i,-,+, i{+}1, \ldots, n)\,.
}
Choices of $\tau \in S_{n+2}$ with non-adjacent $+$ and $-$ appear
in the CHY description of non-planar amplitudes
\eqn\nonplanar{
A(1,2,\ldots,j \, | \, j{+}1,\ldots ,n)=\int  {\dd^D \ell \over \ell^2}
 \! \! \! \! \!\sum_{\rho \in {\rm cyc}(1,2,\ldots,j) \atop
 {
 \tau \in {\rm cyc}(j{+}1,\ldots,n)
 }}  \! \! \! \! \!a(+, \rho(1,2,\ldots,j) , - , \tau(j{+}1,\ldots ,n))\,
}
associated with double traces ${\rm
Tr}(t^{1}t^{2} \ldots t^{j} ) {\rm
Tr}(  t^{j+1} \ldots t^n) $.
The partial integrands in \nonplanar\ can be reduced to the cases in \backto\
with $+,-$ adjacent via Kleiss--Kuijf relations \KleissNE
\eqn\Kleiss{
a(C,+,B,-) = (-1)^{|C|} a(+,(B \shuffle C^t),-) \ ,
}
where $C^t$ and $|C|$ denote the transpose $c_{p} \ldots c_2 c_1$
and length $p$ of the word $C\equiv c_1 c_2 \ldots c_p$, respectively. This reproduces
the amplitude relations of \BernZX\ to express double-trace contributions at one
loop in terms of single-trace amplitudes.

While the definition and Kleiss--Kuijf relations of the partial
integrand are valid in absence of supersymmetry, we shall now
explore the interplay with the Parke--Taylor organization of the
supersymmetric gauge-theory integrands. Inserting \loopcorr\ into
\partialint\ leads to the following cubic-diagram expansion
analogous to \analyticSYM,
\eqn\analyticPI{
a(\tau(1,\ldots,n,+,-))= \!  \!  \lim_{k_{\pm} \rightarrow \pm \ell}  \sum_{\rho \in S_n}  \!  N _{+|\rho(1,\ldots,n)|-} (\ell) \, m^{\rm tree}[\tau(1,\ldots,n,+,-) \, |\, +,\rho(1,\ldots,n),-] \, .
}
As an example with maximal supersymmetry, permutation invariance of the box numerator $N^{\rm box} \equiv s_{12} s_{23} A^{\rm tree}(1,2,3,4)$ \GreenSW\ gives rise to the following diagrams in the four-point partial integrand with 16 supercharges~\HeMZD
\eqnn\examplefoura
$$\eqalignno{
a^{\rm max}(1,2,3,4,-,+) &= {N^{\rm box}  \over s_{1,\ell} s_{12,\ell} s_{123,\ell}} &\examplefoura \cr
a^{\rm max}(1,2,3,-,4,+)&=  { N^{\rm box}   \over s_{1,\ell} s_{12,\ell} s_{4,\ell} } +
{ N^{\rm box}  \over s_{1,\ell} s_{12,\ell} s_{3,\ell}}
+ { N^{\rm box}  \over  s_{1,\ell} s_{14,\ell} s_{3,\ell}}
+ { N^{\rm box}  \over  s_{4,\ell} s_{14,\ell} s_{3,\ell}}  \ .
}$$
Moreover, three external gluons with polarization vectors $e_i^m$
yield the following partial integrands with half-maximal
supersymmetry \refs{\BergFUI, \HeMZD}
\eqnn\examplethree
$$\eqalignno{
a^{1/2}(1,2,3,-,+) &= { \ell_m \big[e^m_1(k_2 \! \cdot  \!e_3)(k_3\! \cdot  \!e_2) {+} (1{\leftrightarrow} 2,3) \big] \over s_{1,\ell} s_{12,\ell}}    - {(e_1\! \cdot  \!e_2)(k_1\! \cdot  \!e_3) \over s_{12,\ell}}  - {(e_2\! \cdot  \!e_3)(k_2\! \cdot  \!e_1) \over s_{1,\ell}}
 \cr
a^{1/2}(1,2,-,3,+) &=0
\ . &\examplethree
}$$
The bubble contributions $\sim (e_i\cdot e_j)$ are crucial for gauge
invariance, and they stem from the cancellation of the (formally
vanishing) invariant $s_{12}$ between the doubly-partial amplitudes in
\delicate\ and the $G_{12}$-coefficient $s_{12}(e_1\cdot
e_2)(k_1\cdot e_3)$ in ${\cal K}_3^{1/2}(\ell)$, see section \secREDUCE.

In comparison to the color-ordered amplitude $A(1,2,\ldots,n)$, the
partial integrand $a(1,2,\ldots,i,-,+,i{+}1,\ldots,n)$ only contains
the subset of cubic diagrams with the loop momentum inserted between legs
$i$ and $i{+}1$. Hence, a single partial integrand cannot
suffice to recombine to a Feynman integral with propagators of
conventional $(\ell{+}K)^{-2}$-form.

However, as a major virtue of partial integrands, they inherit BCJ
symmetry from their definition \partialint\ via Parke--Taylor
factor: In the same way as tree-level scattering equations yield the
BCJ relations \BCJscat\ for Parke--Taylor factors and thereby
$A^{\rm tree}(\ldots)$, the one-loop scattering equations as a
forward limit of their tree-level counterparts imply the one-loop
BCJ relations:
\eqn\BCJscatloop{
\sum_{i=1}^{n{-}1} (\ell \cdot k_{12\ldots i})  \, \PT(1,2,\ldots, i, +, i{+}1,\ldots, n,-)=0
}
Hence, the partial integrands \partialint\ generalize the tree-level BCJ relations \BCJrels\ to \HeMZD
\eqn\BCJrelsloop{
\sum_{i=1}^{n{-}1} (\ell \cdot k_{12\ldots i})  \, a(1,2,\ldots, i, +, i{+}1,\ldots, n,-)=0 \ ,
}
as well as another topology mixing different orders of $\ell$ \HeMZD
\eqn\moreBCJrelsloop{
\sum_{i=2}^{n{-}1} (k_1 \cdot k_{23\ldots i}) \, a(2,3,\ldots, i, 1, i{+}1, \ldots, n, -, +)   = (\ell\cdot k_1) \, a(2,3,\ldots,n,-,1,+) \ .
}
One can immediately check that these BCJ relations are obeyed by the
three- and four-point partial integrands \examplefoura\ and
\examplethree. As will be detailed in section \fivefive, the BCJ relations among partial integrands
still hold in presence of anomalies: Since permutation invariance of ${\cal I}_{\rm SYM}(\ell)$
is broken by anomalies, all partial integrands must then be defined with respect to the same
expression for ${\cal I}_{\rm SYM}(\ell)$ in \partialint.

Note that one-loop BCJ relations in the context of
conventional $(\ell+K)^{-2}$ propagators have been discussed earlier
in the field- \BoelsTP\ and string-theory literature
\TourkineBAK. As we will see in the following subsection, the
partial integrands \partialint\ along with the partial-fraction
representation of loop integrals are tailored to enter a one-loop
KLT formula. It would be interesting to reformulate the one-loop KLT
formula in terms of $(\ell+K)^{-2}$ such as to incorporate the
one-loop BCJ relations in \refs{\BoelsTP, \TourkineBAK}, possibly
along the lines of \GomezLHY.

Given that the tree-level BCJ relations leave a basis of $(n{-}3)!$ independent permutations of $A^{\rm tree}(\ldots)$ \BernQJ, one may wonder about the analogous basis dimensions for partial integrands. The forward limit of the tree-level setup implies an upper bound of $(n{-}1)!$ independent partial integrands, but already the maximally supersymmetric four-point examples in \examplefoura\ illustrate that this bound is usually not saturated: All the $a^{\rm max}(\tau(1,2,3,4,+,-))$ are proportional to $N^{\rm box} \equiv s_{12} s_{23} A^{\rm tree}(1,2,3,4)$, so they are all related by rational functions of $k_j$ and $\ell$. Similarly, we will find three linearly independent five-point partial integrands with maximal supersymmetry in section \fivefour.

%****************
\subsec{The correlator in a BCJ basis and one-loop KLT relations}
\par\subseclab\proofKLT
%***************

We recall that integration by parts or scattering equations can be
used to expand the tree-level correlator \treecorr\ in a BCJ basis
of Parke--Taylor factors, leading to the KLT form \KLTcorrtree.
These steps will now be repeated at the one-loop order, assuming a
minimum of four supercharges in one of the gauge theories.

Following the string calculations of \MafraKH, it
is convenient to perform the integration-by-parts
reduction of ${\cal I}_{\rm SYM}(\ell)$ at the level of the correlator
${\cal K}_n(\ell)$ whose $\sigma$-dependence is captured by the
Green function $G_{ij}$ in \defGij. After choosing a reference leg
1, the scattering equations \Gijscat\ allow to eliminate all
instances of $G_{1j}$ with $j=2,3,\ldots,n$, i.e.\ the correlator
${\cal K}_n(\ell)$ is rendered independent on $\sigma_1$. This
representation of ${\cal K}_n(\ell)$ without $G_{1j}$ leaves no more
freedom to apply further scattering equations without re-introducing
$\sigma_1$, so all the kinematic factors must be gauge invariant. Moreover, all
factors of ${\rm sgn}^{\rho}_{1j}=1$ disappear when
converting to ${\cal I}_{\rm SYM}(\ell)$, see \thumb.

In absence of ${\rm sgn}^{\rho}_{1j}$, in turn, the coefficients of Parke--Taylor
factors $\PT(+,\rho(1,\ldots,n),-)$ in ${\cal I}_{\rm SYM}(\ell)$ do not depend
on the position of leg 1 within $\rho(1,2,\ldots,n)$. Hence, kinematic factors
will be accompanied by
\eqnn\combsuchas
$$\eqalignno{
&\PT(+,1,2,3,\ldots,n,-) + \PT(+,2,1,3,\ldots,n,-) + \PT(+,2,3,1,\ldots,n,-) + \ldots &\combsuchas \cr
& \ \ \ \ \ \ + \PT(+,2,3,\ldots,1,n,-) + \PT(+,2,3,\ldots,n,1,-)  = -  \PT(1, +,2,3,\ldots,n,-)
}$$
and permutations in $2,3,\ldots,n$, using Kleiss--Kuijf relations in the second line.
Hence, the elimination of $G_{1j}$ in ${\cal K}_n(\ell)$ naturally leads to an $(n{-}1)!$-term
expression for the correlator,
\eqn\invcorr{
{\cal I}_{\rm SYM}(\ell)  = - \sum_{\rho \in S_{n-1}}  \PT(-,1,+,\rho(2,3,\ldots,n)) \, {\cal C}_{+|\rho(2,3,\ldots,n)|-}(\ell) \ ,
}
where ${\cal C}_{+|\rho(2,3,\ldots,n)|-}(\ell)$ can be viewed as a gauge invariant but non-local representation of an $n$-gon numerator. More precisely, ${\cal C}_{+| 2,3,\ldots,n|-}(\ell)$ accompanies all the $n$ diagrams where the external legs of the $n$-gon appear in the orders $2,3,\ldots, n$ with leg 1 inserted at an arbitrary position. The non-locality of ${\cal C}_{+|\rho(2,3,\ldots,n)|-}(\ell)$ stems from the elimination of $G_{1j}$ via scattering equations, but this only generates poles in the {\it external} Mandelstam invariants $s_{1ij\ldots p}$, i.e.\ there are no $\ell$-dependent propagators $s^{-1}_{ij\ldots p,\ell}$.
Explicit four- to six-point expressions for ${\cal C}_{+| 2,3,\ldots,n|-}(\ell)$ can be found in section \fivefour, also see section \sixthree\ for examples with reduced supersymmetry.

In order to ensure that the correct partial integrands
$a(+,\tau(2,3,\ldots,n),1,-)$ arise after performing CHY integrals
over \invcorr, the gauge invariant coefficients ${\cal
C}_{+|\rho(2,3,\ldots,n)|-}(\ell)$ must by themselves be expressible
in terms of partial integrands. The requirement is met by the expansion
\eqn\cvsa{
 {\cal C}_{+|\rho(2,3,\ldots,n)|-}(\ell) = S[\rho | \tau]_{\ell}  \, a(+,\tau(2,3,\ldots,n),1,-)
}
which reproduces the pattern of the
tree-level KLT formula upon insertion into \invcorr:
\eqn\KLTcorr{
{\cal I}_{\rm SYM}(\ell)  = \sum_{\rho,\tau \in S_{n-1}}  \PT(+,\rho(2,3,\ldots,n),-,1) \, S[\rho | \tau]_{\ell}  \, a(+,\tau(2,3,\ldots,n),1,-) \ .
}
The $(n{-}1)! \times (n{-}1)!$-matrix $S[\rho | \tau]_{\ell} \equiv
S[\rho(2,3,\ldots,n) | \tau(2,3,\ldots,n)]_{\ell}$ follows the
functional form of the tree-level momentum kernel \momker, where the
loop momentum now enters as the pivot leg. We are using that, before
performing the forward limit $k_{\pm} \rightarrow \pm \ell$ in
\CHYint, $S[\rho | \tau]_{\ell}$ is the inverse of the $(n{-}1)!
\times (n{-}1)!$ matrix of doubly-partial amplitudes $m^{\rm
tree}[+,\rho(2,3,\ldots,n),-,1 \, | \, +,\tau(2,3,\ldots,n),1,-]$,
see \invMK.

The KLT form \KLTcorr\ of the supersymmetric gauge-theory integrand
can be used to derive the analogous KLT formula for loop integrands
in supergravity. We are using the permutation symmetric and gauge invariant
definition of a supergravity integrand $m_n(\ell)$ in the
CHY framework,
\eqn\gravint{
m_n(\ell) \equiv \int \dd \mu^{\rm tree}_{n+2} \ {\cal I}_{\rm SYM}(\ell)\, \tilde {\cal I}_{\rm SYM}(\ell) \ , \ \ \ \ \ \
{\cal M}_{{\rm SYM} \otimes {\rm SYM}}
 = \int { \dd^D\ell \over \ell^2} m_n(\ell)  \ ,
}
where ${\cal I}_{\rm SYM}(\ell)$ and $\tilde {\cal I}_{\rm
SYM}(\ell)$ may refer to different gauge theories. Similar to
\backto, the definition \gravint\ amputates the overall quadratic
propagator $\ell^{-2}$ in a partial-fraction representation of
Feynman integrals \HeMZD. Next, we insert the minimal $(n{-}1)!$
form \KLTcorr\ of the left-moving and supersymmetric gauge-theory
integrand into \gravint,
\eqn\almostKLT{
m_n(\ell)  = \! \!  \sum_{\rho,\tau \in S_{n-1}}  \! \!  a(+,\rho(2,\ldots,n),1,-)  \, S[\rho | \tau]_{\ell}
\int \dd \mu^{\rm tree}_{n+2} \ \PT(+,\tau(2,\ldots,n),-,1) \, \tilde {\cal I}_{\rm SYM}(\ell) \ .
}
Then, the Parke--Taylor factor on the right-hand side suggests to
apply the definition \partialint\ of the partial integrand for
$\tilde {\cal I}_{\rm SYM}(\ell)$, whose validity does not rely on
supersymmetry. In this way, one arrives at the one-loop KLT formula \HeMZD
\eqn\oneloopKLT{
m_n(\ell)  = \! \! \sum_{\rho,\tau \in S_{n-1}}  \! \!  a(+,\rho(2,3,\ldots,n),1,-)  \, S[\rho | \tau]_{\ell}
\, \tilde a(+,\tau(2,3,\ldots,n),-,1)  \ ,
}
whose present derivation applies to any double copy of gauge
theories with at least four supercharges on one side. For the case with zero supersymmetry, we expect that \oneloopKLT\ still holds, but a careful proof including a suitable treatment of forward-limit divergences is relegated to the future.
 The results of \GeyerELA\ which appeared after the first version of this work are expected to play
a key role in adapting the above arguments to zero supersymmetry.

%************************************************************************************************
%************************************************************************************************
\newsec{One-loop RNS correlators for field-theory amplitudes}
\par\seclab\secRNS\
%************************************************************************************************
%************************************************************************************************

\noindent
In this section, we will investigate one-loop correlators \defloopK\ for
field-theory amplitudes in the RNS formulation of the underlying
ambitwistor string or superstring. We will on the one hand point out universal structures
that do not depend on the amount of supersymmetry and on the other hand describe the
simplifications in supersymmetric theories. In particular, the simple
dependence of supersymmetric correlators on the punctures which
has been central to the discussion in sections \polyGij\ and \polyGijB\
will be derived.

While external fermions will be addressed in section \secPS\ by the supersymmetric
correlators in pure-spinor superspace,
we will focus the one-loop RNS correlators for external bosons in this section.
Their multiparticle instances have been firstly discussed in \TsuchiyaVA\ for
maximally supersymmetric superstring theory (see also \refs{\StiebergerWK,
\canceltriag, \TsuchiyaNF}), and four-point string amplitudes with reduced supersymmetry
can be found in \refs{\BianchiNF, \redSUSYamps, \BergWUX}. A major challenge in the
RNS variables is to manifest the supersymmetry-induced
simplifications when combining different spin structures, the
boundary conditions for the worldsheet spinors $\psi^m(z)$ in the
RNS formulation as $z\rightarrow z+1$ and $z\rightarrow z+\tau$.

Even before performing the sum over spin structures, we find that at the
degeneration $\tau \rightarrow i\infty$ of the torus relevant for the
field-theory limit, the correlators simplify significantly: They reduce
to polynomials in the Green function $G_{ij}$ on the nodal Riemann
sphere defined in \defGij, with {\it local} functions of external polarizations
as their coefficients. After performing the
spin sum, the final form of the polynomials depends on the amount of
supersymmetry, and we present complete correlators for gauge theories
with maximal as well as half- (or quarter-)maximal supersymmetry in
various dimensions.

Scattering equations and algebraic identities of $G_{ij}$'s can be used
to reduce these monomials of $G_{ij}$ to a basis which leads to the
KLT relations described in section \proofKLT.
In this way, we obtain a basis expansion of
the correlator with {\it gauge invariant} but non-local coefficients for external bosons,
which can be nicely packaged using Berends--Giele currents. Using the supersymmetrized
version of these Berends--Giele currents (see section \secPS), we will later present explicit results
for BCJ master numerators in pure-spinor superspace whose bosonic components can be
matched with the one-loop correlators in this section.

%****************
\subsec{Structure of RNS correlators on a torus}
%***************

As a spurious difference between the correlators of the ambitwistor string and the superstring, the
bosonic worldsheet fields $x^m$ do not exhibit any two-point contractions in the former case \refs{\GomezWZA,\AdamoTSA}. At tree level this difference is known to wash out after removing double poles in $\sigma_{ij}$ via integration by parts \GomezWZA\ and
expanding the correlators in terms of Parke--Taylor factors. Since the same kind of integration by parts can be performed at arbitrary genus, there is no loss of generality in starting with the
one-loop RNS correlator of the ambitwistor string for $n$ external gluons \AdamoTSA, the same end results would have
been obtained from the superstring.

The parity-even part of the $n$-point RNS correlator ${\bf K}_n$ can be expanded in terms of
$n!$ gauge invariant terms
\eqnn\calKtau
\eqnn\calRtau
$$\eqalignno
{
&{\bf K}_n(\ell | \tau)=\sum_{\rho\in S_n}\,{\bf R}_\rho(\ell | \tau)\ , \quad {\rm with}\quad \rho=(i)\,\cdots\,(j)\,I\,\cdots\,J\ , &\calKtau \cr
&{\bf R}_{(i)\,\cdots\,(j)\,I\,\cdots\,J}(\ell | \tau)\,\equiv \,{\bf c}_i (\ell | \tau ) \,\cdots \,{\bf c}_j(\ell |\tau)\, {\rm tr}(f_I)\, \cdots \, {\rm tr}(f_J)\,{\bf G}_{I,\ldots,J} (\tau)\ ,
&\calRtau
}
$$
where the summand ${\bf R}_\rho$ is defined according to the
unique decomposition of $\rho$ into disjoint cycles\foot{For example, the sum $\rho
\in S_3$ relevant to $n=3$ is organized in terms of the cycles
$\rho=(1)(2)(3), \,(1)(23),\, (2)(31), \,(3)(12), \,(123), \,(321)$, leading to the following expansion \calKtau:
$$\eqalignno
{
{\bf K}_3(\ell | \tau)&={\bf c}_1 {\bf c}_2 {\bf c}_3 {\bf G}_{\emptyset}+ {\rm tr}(f_{(123)}) {\bf G}_{(123)}+ {\rm tr}(f_{(321)}) {\bf G}_{(321)}  \cr
& \ \ + {\bf c}_1 {\rm tr}(f_{(23)}){\bf G}_{(23)}+{\bf c}_2 {\rm tr}(f_{(31)}) {\bf G}_{(31)}+{\bf c}_3 {\rm tr}(f_{(12)}) {\bf G}_{(12)}
\ .
}
$$
}. Each length-one cycle or fixed point $(i),\cdots, (j)$ of $\rho$ contributes a factor of
\eqn\Ciitau{
{\bf c}_i (\ell | \tau) \equiv 2\pi i \,(e_i \cdot \ell )+ \sum_{j=1 \atop {j\neq i}}^n (e_i \cdot k_j) \,
\partial \log \theta_1(z_{ij},\tau) \ ,
}
which captures the contributions from the worldsheet bosons and is the only place that the loop momentum appears.
Cycles of length two and larger, on the other hand, are denoted by  $I,\ldots, J$,
e.g.\ $I=(i_1 i_2\cdots i_a)$, $J=(j_1j_2\cdots j_b)$ with $a,b\geq 2$. The associated
kinematic functions of momenta and polarizations are traces
(over Lorentz indices $m,n$) of linearized field strengths
\eqn\trf{
{\rm tr}(f_I) \equiv  - {1 \over 2}~{\rm tr}(f_{i_1}f_{i_2}\cdots f_{i_a})\,,
\qquad {\rm with} \quad f_i^{mn}=k_i^{m}e_i^{n}-e_i^{m} k_i^{n}\ .
}
Finally, the accompanying functions of the punctures
boil down to two-point contractions
\eqnn\szeg
$$\eqalignno{
S_\nu(x,\tau) &\equiv {\theta_1'(0,\tau) \theta_\nu(x,\tau) \over
\theta_1(x,\tau) \theta_\nu(0,\tau)}
&\szeg
}$$
of the worldsheet spinors
with even spin structures $\nu=2,3,4$:
\eqnn\stringG
$$\eqalignno{
{\bf G}_N(x_1,x_2,\ldots,x_N|\tau)&\equiv \sum_{\nu=2}^4  (-1)^{\nu-1} \left[  {\theta_\nu(0,\tau) }\over{\theta_1'(0,\tau)} \right]^4 S_\nu(x_1,\tau) S_\nu(x_2,\tau)\ldots S_\nu(x_N,\tau)  \ .% \ \ \ \ \ \sum_{i=1}^{N} x_i = 0 \ ,
&\stringG
}$$
More precisely, cycles $I=(i_1 i_2\cdots i_a)$, $J=(j_1j_2\cdots j_b)$ of length $a,b\geq 2$ yield
\eqn\newGcycletau{
{\bf G}_{I,\ldots,J}(\tau) \equiv  {\bf G}_{a + \ldots +b}(x_1,x_2 , \ldots, x_{a} ,\ldots,y_1,y_2,\ldots,y_{b}  |\tau)  \Big|^{x_k = z_{i_k}-z_{i_{k+1}}}
_{y_k = z_{j_k}-z_{j_{k+1}}}
}
subject to cyclic identification $i_{a+1}=i_1$ and $j_{b+1}=j_1$. By \newGcycletau, each
cycle $I,\ldots,J$ in ${\bf G}_{I,\ldots,J}(\tau)$ yields a group of arguments $x_j$ in \stringG\
which add up to zero, $\sum_{ k \in I} x_k=\sum_{ k \in J} x_k=0$. In particular, we always
have $x_1+x_2+\ldots+x_N$ in \stringG\ and can derive all cases with multiple cycles from
specializations of the single-cycle configuration.

The even Jacobi theta functions
$\theta_{\nu=2,3,4}(z,\tau)$ entering \szeg\ are regular at $z=0$ and defined in appendix \apptheta.
They conspire in the sum over the even spin structures $\nu=2,3,4$ along with the partition functions
$(  {\theta_\nu(0,\tau) \over \theta_1'(0,\tau)})^4 $. Capturing the effect of spacetime
supersymmetry, the expressions in \stringG\ simplify drastically after performing the
spin sums as in $0={\bf G}_0(\emptyset |\tau)=
{\bf G}_2(x_1,x_2 |\tau)=
{\bf G}_3(x_1,x_2,x_3 |\tau)$
as well as \TsuchiyaVA
\eqnn\stringGfive
$$\eqalignno{
{\bf G}_4(x_1,x_2,x_3,x_4 |\tau)&=1
 \ , \ \ \ \ \ \
{\bf G}_5(x_1,x_2,\ldots,x_5 |\tau)= \sum_{j=1}^5 \partial \log \theta_1(x_j,\tau) \ .
&\stringGfive
}$$
The higher-multiplicity systematics has firstly been discussed in \TsuchiyaVA\ (also see \refs{\StiebergerWK, \TsuchiyaNF}) and was later organized via combinations of Eisenstein series and elliptic functions \BroedelVLA\ explicitly known to all multiplicities. One can see from the results of the references that the spin sum in ${\bf G}_N(x_1,x_2,\ldots,x_N|\tau)$ only leaves $N{-}4$ simultaneous poles in the arguments $x_k$.

The contributions from the odd spin structure $\nu=1$ yield the parity
odd part of ${\bf K}_n$ which vanishes in spacetime dimensions
$D<9$ and will be discussed separately in section~\parodd.

%****************
\subsec{Structure of RNS correlators on a nodal Riemann sphere}
%***************

The CHY integrands \defloopK\ for maximally supersymmetric gauge theories can be obtained from the
degeneration limit
\eqn\degtau{
{\cal K}_n\equiv  {1 \over (2\pi i)^{n-4}}  \lim_{\tau \rightarrow i \infty} {\bf K}_n(\tau)
}
of \calKtau\ which preserves the expansion in terms of $n!$ separately gauge invariant terms
\eqnn\calK
\eqnn\calP
$$\eqalignno
{
&{\cal K}_n=\sum_{\rho\in S_n}\,{\cal R}_\rho\ , \quad {\rm with}\quad \rho=(i)\,\cdots\,(j)\,I\,\cdots\,J\ , &\calK \cr
&{\cal R}_{(i)\,\cdots\,(j)\,I\,\cdots\,J}\,\equiv c_i (\ell) \,\cdots \,c_j(\ell)\, {\rm tr}(f_I)\, \cdots \, {\rm tr}(f_J)\,{\cal G}_{I,\ldots,J} (\sigma)\ .
&\calP
}
$$
The unique decomposition of $\rho$ into disjoint cycles is explained below \calKtau, so it remains to investigate
the behavior of the $\tau$-dependent constituents when the toroidal worldsheet degenerates to a nodal Riemann
sphere. The loop-momentum dependent part \Ciitau\ is easily seen to degenerate into
\eqn\Cii{
c_i (\ell) \equiv {1\over 2\pi i} \lim_{\tau \rightarrow i \infty}  {\bf c}_i(\ell|\tau)
=e_i \cdot \ell + \sum_{j=1 \atop {j\neq i}}^n e_i \cdot k_j~{{\sigma_i} \over {\sigma_{ij}} }= e_i \cdot \ell + \sum_{j=1 \atop {j\neq i}}^n  e_i \cdot k_j~G_{ij}\ ,
}
which is manifestly gauge invariant on the support of the one-loop
scattering equations \Gijscat. Note that we have used momentum
conservation and transversality of $e_i$ in passing to the
representation in terms of $G_{ij}$.

The traces ${\rm tr}(f_I)$ over linearized field strengths, see \trf,
are accompanied by spin-summed correlators over worldsheet fermions
detailed around \stringG\ and \newGcycletau,
\eqn\Gcycle{
{\cal G}_{I_1,I_2,\ldots,I_k} \equiv {1 \over {(2\pi i)^{N-4} }}  \lim_{\tau \rightarrow i \infty} {\bf G}_N(x_1,x_2,\ldots,x_N|\tau) \Big|_{\sum_{i \in I_{j}} x_i=0} \ .
}
The cycles $I_1,I_2,\ldots,I_k$ track the subsets of $x_i$ that add
up to zero, and the remarkable simplifications in the $\tau \rightarrow i \infty$ limit
of the all-multiplicity spin sums \stringG\ will be presented in the next subsection. Again, any instance of \Gcycle\
with multiple cycles $I_j$ can be obtained by specializing the single-cycle configuration
\eqn\Gsingcycle{
{\cal G}_{N} \equiv {\cal G}_{(12\ldots N)}={1 \over {(2\pi i)^{N-4} } } \lim_{\tau \rightarrow i \infty} {\bf G}_N(x_1,x_2,\ldots,x_N|\tau) \Big|_{x_1+x_2+\ldots+x_N=0} \ .
}
It is useful to illustrate the expansion \calK\ of ${\cal K}_n$ with the $n=3,4$ examples,
\eqnn\exKOStwo
$$\eqalignno
{
{\cal K}_3&=c_1 c_2 c_3 {\cal G}_{\emptyset} + c_1 {\rm tr}(f_{(23)}) {\cal G}_{(23)}+c_2 {\rm tr}(f_{(31)}) {\cal G}_{(31)}+c_3 {\rm tr}(f_{(12)}) {\cal G}_{(12)} \cr
& \ \ \ \ + {\rm tr}(f_{(123)}){\cal G}_{(123)}+{\rm tr}(f_{(321)}) {\cal G}_{(321)} \ , &\exKOStwo\cr
%%%%%%%%%%%%%%%%%%%
{\cal K}_4&={ c_1 c_2 c_3 c_4 {\cal G}_{\emptyset}+ \left( c_1 c_2 {\rm tr}(f_{(34)}) {\cal G}_{(34)} + {\rm 5~more} \right) +\left( c_1 {\rm tr}(f_{(234)}) {\cal G}_{(234)}+ {\rm 7~more} \right) } \cr
& \ \ \ \ + { \left( {\rm tr}(f_{(1234)}) {\cal G}_{(1234)}+ {\rm 5~more} \right) + \left( {\rm tr}(f_{(12)}) {\rm tr}(f_{(34)}) {\cal G}_{(12)(34)} + {\rm 2~more} \right) } \ ,
}$$
and the analogous five-point expressions along with the resulting numerators are spelt out
in appendix \appexample.

Recall that the $x$'s in ${\bf G}_N$ denote differences of $z$'s on the torus; here in the $\tau\to i \infty$ limit, as a slight abuse of notation, we will continue to denote the arguments of ${\cal G}_N$ as $x_1, \ldots x_N$. Now they simply refer to $N$ pairs of labels $x_1=(i_1, j_1), \ldots, x_N=(i_N, j_N)$ determined by the cycle structure $I, \ldots, J$. For example, we have
$x_1=(1 ,2), x_2=(2,1)$ for ${\cal G}_{(12)}$, $x_1=(1,2), x_2=(2,3), x_3=(3,4), x_4=(4,1)$
for ${\cal G}_{(1234)}$ and $x_1=(1,2), x_2=(2,1), x_3=(3,4), x_4=(4,3)$
for ${\cal G}_{(12)(34)}$.

%****************
\subsec{Spin sums in maximally supersymmetric correlators}
\par\subseclab\maxspinsum
%***************

\noindent
This section is devoted to the impact of the $\tau \to i\infty$ limit on the fermionic correlators ${\bf G}_N$
in the expansion of ${\bf K}_n(\tau)$: all the elliptic functions in the expressions \stringG\ for ${\bf G}_N$
simplify drastically, and their degenerate versions ${\cal G}_N$ become {\it polynomials} in
$G_{i j}$'s. More precisely, each cyclic structure
$\{x_1,x_2,\ldots, x_N\}$ of length $N$ gives rise to symmetric
polynomials in $G_{x_1},G_{x_2},\ldots, G_{x_N}$ of degree $0\leq
k\leq N$, and it is convenient to introduce the notation
\eqn\Sgm{
\Sigma_k (x_1, x_2, \ldots, x_N)\equiv \sum_{1\leq \alpha_1< \alpha_2 <\ldots < \alpha_k\leq N} G_{x_{\alpha_1}}\,G_{x_{\alpha_2}}\,\cdots\,G_{x_{\alpha_k}}\,
}
with $\Sigma_0(x_1,\ldots,x_N)\equiv 1$. For example, the symmetric polynomials
at length $N=2$ with $x_1=(1,2), x_2=(2,1)$ are $\Sigma_1(x_1,x_2)=G_{1 2}+ G_{2 1}=0$ and
$\Sigma_2(x_1,x_2)=G_{1 2} G_{2 1}=-G_{1 2}^2$. For $N=3$ with $x_1=(1,2), x_2=(2,3), x_3=(3,1)$, we
have $\Sigma_1(x_1,x_2,x_3)=G_{12} +G_{23}+G_{31}$ and
$\Sigma_3(x_1,x_2,x_3)=G_{12}\,G_{23}\,G_{31}$, whereas
$\Sigma_2(x_1,x_2,x_3)$ is a constant by the Fay identity
\pfoneloop:
\eqn\Fay{
\Sigma_2 ( x_1,x_2, x_3 ) =G_{1 2} G_{2 3}+ G_{2 3} G_{3 1}+ G_{31} G_{12}= - {1 \over 4}\ .
}
The maximally supersymmetric spin sums \Gcycle\ turn out to yield
extremely simple linear combinations of the polynomials $\Sigma_k\equiv\Sigma_k(x_1,x_2,\ldots,x_N)$
in \Sgm. By taking the $\tau \rightarrow i\infty$ limit of the elliptic functions in ${\bf G}_N$ \BroedelVLA, we find
\eqnn\maxG
$$\eqalignno
{
{\cal G}_N  & =\Sigma_{N{-}4}+ {1\over 2} \Sigma_{N{-}6}+ {3 \over 16} \Sigma_{N{-}8} + {1 \over 16} \Sigma_{N{-}10} +\cdots \cr
&= %Sum[(m + 1)/4^m Sigma[NN - 4 - 2 m], {m, 0, Floor[(NN - 4)/2]}]
\sum_{m=0}^{  \lfloor (N-4)/2\rfloor }  {m+1 \over 4^m} \Sigma_{N-4-2m} \ ,
%\sum_{m=1}^{\lfloor N/2\rfloor-1} m~\Sigma_{N-2-2m} \ ,
&\maxG
}
$$
where the all-multiplicity conjecture has been checked up to $N=20$.
In other words, the only contributing degrees in $G_{ij}$ are $N{-}4, \, N{-}6, \, N{-}8,\ldots$, and
all cases ${\cal G}_0, {\cal G}_1,{\cal G}_2,{\cal G}_3$  below four points
vanish, ${\cal G}_N=0 \ \forall \ N<4$. Let's spell out some simple
non-vanishing examples, starting with ${\cal G}_4=\Sigma_0=1$. The
first non-trivial dependence on $x_i$ arises at $N=5$, where two
types of inequivalent cycle structures occur,
\eqn\maxGfive{
{\cal G}_5=\sum_{i=1}^5 G_{x_i}\ ,\quad {\rm e.g.}\quad {\cal G}_{(12345)}=\sum_{i=1}^5 G_{i,i{+}1}\ ,\quad {\cal G}_{(12)(345)}=G_{34}+G_{45}+G_{53} \ ,
}
in lines with the $\tau \rightarrow i\infty$ limit of \stringGfive. A complete expression for the
resulting five-point correlator is assembled in appendix \appexample.

The single-cycle expressions for $n=6, 7$ are
\eqn\spinsumsA{
{\cal G}_6=\sum_{i<j}^6 G_{x_i} G_{x_j} +{1\over 2}\ , \ \ \ \ \ \  {\cal G}_7=\sum_{i<j<k}^7 G_{x_i} G_{x_j} G_{x_k}+ {1\over 2} \sum_{i=1}^7 G_{x_i}\ ,
}
which can be specialized to multiple cycles such as
\eqnn\spinsums
$$\eqalignno{
{\cal G}_{(123)(456)}&=(G_{12}+G_{23}+G_{31})(G_{45}+G_{56}+G_{64})\ ,\cr
{\cal G}_{(12)(34)(56)}&={1\over 2}-G_{12}^2-G_{34}^2-G_{56}^2\,, &\spinsums \cr
{\cal G}_{(12)(3456)}&={1\over 2}-G_{12}^2+ \big[ G_{34} G_{45}+  {\rm cyc}(3,4,5,6) \big] + G_{34} G_{56} + G_{45}G_{63}\,,\cr
{\cal G}_{(12)(34)(567)}&=\left({1\over 2}-G^2_{12}-G^2_{34} \right) (G_{56}+G_{67}+G_{75})+G_{56}G_{67} G_{75}\,,~{\rm etc.} \ ,
}$$
where the Fay identity \Fay\ has been used in the first and last line.

As is well-known from the superstring literature, the spin sums \maxG\ expose the structure
of maximally supersymmetric field-theory correlators and manifest their power-counting of loop
momenta: The $n$-point correlator \calK\
comprises spin-summed ${\cal G}_N$ of highest power $G_{i j}^{N-4}$
along with $n{-}N$ factors of $c_i(\ell)$ defined in \Cii. Given
that each $c_i(\ell)$ is linear in $\ell$ and $G_{i j}$, the
correlator ${\cal K}_n$ is a polynomial of degree $n{-}4$ in $\ell$
and $G_{i j}$. More precisely, contributions with $k$ powers of
$\ell$ are accompanied by at most $n{-}4{-}k$ powers of $G_{i j}$.

Moreover, after evaluating the CHY integral, the bounds on $G_{i j}$ imply the
absence of triangles, bubbles and tadpoles in maximally
supersymmetric amplitudes, see e.g.\ \refs{\GreenSW, \canceltriag} for the analogous superstring discussion:
In each term of the doubly-partial amplitudes \CHYint, the number of {\it external}
propagators (i.e.\ those independent on $\ell$) is bounded by the
powers of $G_{i j}$. With at most $n{-}4$ powers of $G_{i j}$, at least three of the propagators
from the doubly-partial amplitudes depend on $\ell$, corresponding to box diagrams and higher
$n$-gons. Hence, the spin sum \maxG\ is responsible
for the famous no-triangle property \BernZX.

%****************
\subsec{CHY correlators with reduced supersymmetries}

\par\subseclab\onlynminustwo

\subsubsec{The anatomy of spin sums}
%***************

In fact, some of the simplifications seen in the previous section
even work for individual spin structures $\nu=2,3,4$ before the spin
sum, thus they also apply to cases with reduced supersymmetry. Once
we denote the cycles of fermion Green functions \szeg\ with spin structures $\nu$ by
\eqn\refine{
{\bf S}^{(\nu)}_N(x_1,x_2,\ldots,x_N|\tau) \equiv  S_\nu(x_1,\tau) S_\nu(x_2,\tau)\ldots S_\nu(x_N,\tau)  \ ,
}
it turns out that the $\tau\to i\infty$ limit of \stringG\ picks out three independent contributions
\eqnn\qtozero
$$\eqalignno{
{\cal G}^{\rm f}_N &\equiv {1\over (2\pi i)^N} {\bf S}^{(\nu=2)}_N(x_1,x_2,\ldots,x_N|\tau) \, \big|_{q^0}
&\qtozero \cr
{\cal G}^{\rm s}_N &\equiv {1\over (2\pi i)^N}  {\bf S}^{(\nu=3)}_N(x_1,x_2,\ldots,x_N|\tau) \, \big|_{q^0\phantom{/2}}
\! =+{1\over (2\pi i)^N} {\bf S}^{(\nu=4)}_N(x_1,x_2,\ldots,x_N|\tau) \, \big|_{q^0}
 \cr
{\cal G}^{\rm v}_N &\equiv {1\over (2\pi i)^N}  {\bf S}^{(\nu=3)}_N(x_1,x_2,\ldots,x_N|\tau) \, \big|_{q^{1/2}}
=-{1\over (2\pi i)^N} {\bf S}^{(\nu=4)}_N(x_1,x_2,\ldots,x_N|\tau) \, \big|_{q^{1/2}} \ ,
}$$
which remain inert for any amount of supersymmetry \refs{\GeyerBJA, \GeyerJCH}. In an expansion
w.r.t. $q\equiv e^{2\pi i \tau}$, the notation $|_{q^0}$ and
$|_{q^{1/2}}$ in \qtozero\ refers to the coefficients of $q^0$ and
$q^{1/2}$, respectively, and we have used the fact that these lowest
orders of ${\bf S}_N^{(3)}$ and ${\bf S}_N^{(4)}$ are simply related
to each other.

In terms of the symmetric polynomials $\Sigma_k$ in
\Sgm, all of ${\cal G}^{\rm f}_N,{\cal G}^{\rm s}_N$ and ${\cal G}^{\rm v}_N$
can be identified as extremely simple linear
combinations:
\eqnn\symmfsv
$$\eqalignno{
{\cal G}^{\rm f}_N&=\Sigma_N\ , \cr
{\cal G}^{\rm s}_N&=\Sigma_N+{1 \over 4}\Sigma_{N{-}2}+{1 \over 16} \Sigma_{N{-}4}+ {1 \over 64} \Sigma_{N{-}6}+\ldots
=\sum_{m=0}^{\lfloor N/2 \rfloor} { \Sigma_{N-2m} \over 4^m}
\ ,  &\symmfsv \cr
{\cal G}^{\rm v}_N&=-2 \Big( \Sigma_{N{-}2}+{1\over 2} \Sigma_{N{-}4}+ {3\over 16} \Sigma_{N{-}6}+\ldots \Big)
= -2\sum_{m=0}^{\lfloor (N-2)/2 \rfloor} {(m+1) \Sigma_{N-2-2m} \over 4^{m+1}}\ ,
}$$
as can be straightforwardly checked through the leading $q$-orders of
$S_\nu$ spelt out in \checkszeg. All contributions $\Sigma_{N{-}1},
\Sigma_{N{-}3},\Sigma_{N{-}5},\ldots$ whose parity is opposite to
$\Sigma_N$ drop out, so the sum extends down to $\Sigma_1$ for odd
$N$ and $\Sigma_0$ for even $N$.

The partition functions $(-1)^{\nu-1}\left[ {\theta_\nu(0,\tau)
}\over{\theta_1'(0,\tau)} \right]^4$ which multiply \refine\ reflect maximal
supersymmetry. Their leading orders in $q$ are spelt out in \leadpart\
and combine the building blocks
in \qtozero\ to
\eqn\Gmaxmax{
{\cal G}_N = 16( {\cal G}^{\rm s}_N -{\cal G}^{\rm f}_N) + 2 {\cal G}^{\rm v}_N
 \ .
}
By inserting the expansions \symmfsv, one recovers the organization \maxG\
of ${\cal G}_N$ in terms of symmetric polynomials $\Sigma_k$. The highest degree
$k=N{-}4$ in ${\cal G}_N$ results from cancellation of both $\Sigma_N$ and
$\Sigma_{N-2}$ due to the interplay between bosons and fermions as well as the
GSO projection in the NS sector.

%****************
%***************
\subsubsec{Super-Yang--Mills theories}

Since supersymmetry breaking only affects the one-loop correlator through
a modification of the partition function in \stringG, the structure of
the correlator \calK\ and \calP\ is universal. In scenarios with reduced
supersymmetry, we simply adjust the spin sums ${\cal G}_{I,\ldots,J} \rightarrow {\cal G}^\ast_{I,\ldots,J}$
to the particle content of the theory (indicated by the placeholder $^\ast$):
\eqnn\calKred
\eqnn\calPred
$$\eqalignno
{
&{\cal K}^{\ast}_n=\sum_{\rho\in S_n}\, {\cal R}^{\ast}_\rho\ , \quad {\rm with}\quad \rho=(i)\,\cdots\,(j)\,I\,\cdots\,J\ , &\calKred \cr
&{\cal R}^{\ast}_{(i)\,\cdots\,(j)\,I\,\cdots\,J}\,\equiv \,c_i (\ell) \,\cdots \,c_j(\ell)\, {\rm tr}(f_I)\, \cdots \, {\rm tr}(f_J)\,{\cal G}^{\ast}_{I,\ldots,J} \ .
&\calPred
}
$$
The four-point instances of the corresponding superstring amplitudes in
compactifications  with reduced supersymmetry have been discussed and
simplified in \refs{\BianchiNF, \redSUSYamps, \BergWUX}. In particular, a
systematic method to express the all-multiplicity spin sums in terms of the
Eisenstein series and elliptic functions of \refs{\TsuchiyaNF, \BroedelVLA}
has been given in \BergWUX.

The results of the previous subsection pave the way to extending the above analysis of spin sums
to cases with less or no supersymmetries.
As pointed out in \GeyerJCH, the linear combinations of ${\cal G}^{\rm f}_N,{\cal G}^{\rm s}_N$ and
${\cal G}^{\rm v}_N$ in \Gmaxmax\ can be adjusted such as to describe any number of $d$-dimensional
massless vectors, fermions and scalars running in the loop. Every scalar and fermionic degree of freedom,
for instance, contributes with $2{\cal G}^{\rm s}_N$ and $- 2{\cal G}^{\rm f}_N$ to the spin sum, respectively.
A $d$-dimensional vector, one the other hand, yields the combination $2 {\cal G}^{\rm v}_N+2(d-2){\cal G}^{\rm s}_N$ \GeyerJCH.

Accordingly, the spin sums of pure SYM theories are given by
\eqn\GSYM{
{\cal G}_N^{\a-{\rm SYM}}=2 {\cal G}^{\rm v}_N +16\a({\cal G}^{\rm s}_N- {\cal G}^{\rm f}_N)  \ , \ \ \ \ \ \ \alpha=1, {1 \over 2},{1 \over 4} \ ,
}
with $\a=1, {1 \over 2},{1 \over 4}$ for maximal, half-maximal and
quarter-maximal supersymmetry, respectively. These values of $\a$ control the number of fermions, and the
rigid combinations of ${\cal G}^{\rm s}_N- {\cal G}^{\rm f}_N$ ensure the same number of bosonic degrees of
freedom while keeping a single vector in the multiplet. Explicitly, the spin sums of half- and quarter-maximal SYM
read
\eqnn\halfSYM
\eqnn\quarterSYM
$$\eqalignno
{
{\cal G}_N^{{1\over 2}-{\rm SYM}} &= 2\Sigma_{N-2} + {3\over 2} \Sigma_{N-4} + {5 \over 8} \Sigma_{N-6} + {7 \over 32}  \Sigma_{N-8}  +\ldots  \cr
&= 2 \sum_{m=0}^{\lfloor (N-2)/2\rfloor} {2m+1 \over 4^m} \Sigma_{N-2-2m}
&\halfSYM
\cr
{\cal G}_N^{{1\over 4}-{\rm SYM}} &= 3\Sigma_{N-2} + {7\over 4} \Sigma_{N-4} + {11 \over 16} \Sigma_{N-6} + {15 \over 64}  \Sigma_{N-8}  +\ldots \cr
&= \sum_{m=0}^{\lfloor (N-2)/2\rfloor} {4m+3 \over 4^m} \Sigma_{N-2-2m}
&\quarterSYM
}
$$
and can be reconciled with the string-theory results of \BergWUX\ for certain choices of the
compactification details. From the difference of \halfSYM\ and \quarterSYM, the contributions
of a spin-${1\over 2}$ multiplet with two
scalar and fermionic degrees of freedom each is identified as
\eqn\twoplustwo{
{\cal G}_N^{2+2} = 4\big({\cal G}^{\rm s}_N- {\cal G}^{\rm f}_N\big) = \Sigma_{N-2} + {1\over 4} \Sigma_{N-4} + {1 \over 16} \Sigma_{N-6}   +\ldots
= \sum_{m=0}^{\lfloor (N-2)/2\rfloor} {1 \over 4^m} \Sigma_{N-2-2m}\; .
}
Therefore, as long as a minimum of four supercharges is preserved,
$\Sigma_N$ always drops out from \GSYM, and the degree $k$ of the
polynomials $\Sigma_k$ does not exceed $N{-}2$ (with the additional
cancellation of $\Sigma_{N{-}2}$ in case of \Gmaxmax\ with maximal
supersymmetry).

%****************
\subsubsec{Pure Yang--Mills theory}
%***************

For pure Yang--Mills theory, i.e. in absence of supersymmetry, one is
left with the bosonic truncation of \symmfsv, where the contribution
from ${\cal G}^{\rm f}_N$ is set to zero. For a single gauge boson in
$d$ spacetime dimensions, the relevant spin sum is
\eqnn\Gpure
$$\eqalignno{
{\cal G}_N^{d-{\rm YM}}&=
2 {\cal G}^{\rm v}_N+2(d-2){\cal G}^{\rm s}_N \cr
&= 2(d-2)\Sigma_N
+ \Big( {d\over2}-5 \Big) \Sigma_{N-2}
+ \Big( {d\over 8}-{9\over 4} \Big) \Sigma_{N-4}
+ \Big( {d\over 32}-{13\over 16} \Big) \Sigma_{N-6}
+ \ldots \cr
&=  2(d-2)\Sigma_N + \sum_{m=0}^{ \lfloor(N-2)/2 \rfloor} \Big( {d\over 2} - 4m - 5 \Big){ \Sigma_{N-2-2m} \over 4^m}
\ . &\Gpure
}$$
In dimensions $d>2$, one can see that $\Sigma_N$ no longer drops out, and the spin sum
of pure YM is a degree-$N$ polynomial.

%****************
\subsubsec{Implications for the power counting of loop momenta}
%***************

The theory-dependent spin sums ${\cal G}^{\ast}_{I,\ldots,J}$ comprising $N$ legs
are accompanied by $n{-}N$ factors of $c_i(\ell)$ which are linear in $\ell$
and $G_{ij}$ by \Cii. For half- or quarter-maximal SYM,
${\cal G}^{{1\over 2},{1\over 4}-{\rm SYM}}_{N=0}$ and ${\cal G}^{{1\over 2},{1\over 4}-{\rm SYM}}_{N=1}$
vanish by the cancellation of $\Sigma_{N}$, and ${\cal K}^{{1\over 2},{1\over 4}-{\rm SYM}}_n$
can be identified as polynomials of degree $n{-}2$ in $\ell$ and $G_{ij}$.
Similarly,  ${\cal K}^{d-{\rm YM}}_n$ of pure YM with ${\cal G}^{d-{\rm YM}}_{N}$
of degree $N$ in $G_{ij}$ are polynomials of degree $n$ in $\ell$ and
$G_{ij}$. For example, the four-point correlator is of degree four
for the pure Yang--Mills case, of degree two for half- or quarter-maximal
SYM and constant for maximal SYM.

Accordingly, by evaluation of the CHY integrals, an $n$-gon
numerator in half- and quarter-maximal SYM can have a maximum power
of $n{-}2$ loop momenta. Tadpole diagrams are suppressed by this
power counting, and external bubbles cancel when combining the
partial integrands to a color-ordered single-trace amplitude as in
\backto. In absence of supersymmetry, however, any $n$-gon diagram
may appear with $n$ loop momenta in the numerator.

%****************
\subsec{The correlator in a basis of worldsheet functions}
\par\subseclab\sectfourfive
%***************

As we have shown, the one-loop correlators \calKred\ with any amount of supersymmetry are written as a polynomial of $G_{ij}$ with $\{i,j\}\in\{1,2, \ldots,n\}$. However, different monomials in these functions are not independent due to scattering equations and the Fay identity \pfoneloop. In addition,
higher-point correlators \calKred\
also contain subcycles of propagators such as $G_{ij}^2$ or $G_{ij} G_{jk} G_{ki}$. Such subcycles
do not allow an immediate application of the methods in section \polyGijB, but we will see that they can always
be eliminated from supersymmetric correlators via scattering equations, e.g.
\eqn\twosubcycle{
G_{ij}^2 = { G_{ij} \over s_{ij} } \sum_{k \neq i,j}^n s_{jk}G_{jk}+ { G_{ij} \over s_{ij} }\,\ell\cdot k_j
}
for a length-two cycle\foot{In the corresponding superstring computation, such a double pole in the worldsheet
variables appears in combination $\ap \partial^2 \log \theta_1(z_{ij},\tau) + s_{ij} (\partial \log \theta_1(z_{ij},\tau))^2$,
see appendix B.1 of \MafraNWR. The result of integration by parts
$$\eqalignno
{
&\Big[ \partial^2 \log \theta_1(z_{ij},\tau) + \ap s_{ij} (\partial \log \theta_1(z_{ij},\tau))^2 \Big] {\cal I}_6= - \partial_j \Big[ \partial \log \theta_1(z_{ij},\tau) {\cal I}_6\Big]
\cr
& \ \ \ \ \ \ \ \ \ \ \ \ \ \ \ \ \ \ +  \partial \log \theta_1(z_{ij},\tau) \sum_{k \neq i,j}^n \ap s_{jk}  \partial \log \theta_1(z_{jk},\tau) {\cal I}_6
}
$$
with the six-point Koba--Nielsen factor ${\cal I}_6=\prod_{p<q} | \theta_1(z_{pq},\tau) |^{2\ap s_{pq}}$
then degenerates to the right hand side of \twosubcycle\ and exemplifies that the correlators of the ambitwistor
string and the superstring are identical after elimination of subcycles.}.

However, the elimination of length-$m$ subcycles $G_{i_1 i_2}G_{i_2 i_3}\ldots G_{i_m i_1}$
generally introduces poles in the $m$-particle Mandelstam invariants $s_{i_1 i_2\ldots i_m}$ which generalize the factor
of $s_{ij}^{-1}$ on the right hand side of \twosubcycle.  Hence, in absence of numerator factors 
$\sim s_{i_1 i_2\ldots i_m}$, the treatment of length-$m$ subcycles requires
a momentum phase space of at least $m{+}2$ massless on-shell particles to keep $s_{i_1 i_2\ldots i_m}\neq0$ 
and avoid singularities. Given that supersymmetric $n$-point correlators
involve a maximum of $n{-}2$ powers of $G_{ij}$ by the discussion of section \onlynminustwo, their subcycles of
maximum length $n{-}2$ are all compatible with this phase-space constraint.  It remains to find a suitable treatment of length-$n$ subcycles in the $n$-point correlator of pure Yang--Mills, possibly along
the lines of \CachazoAOL.  The possibility to eliminate closed subcycles in non-supersymmetric
correlators without introducing any singularities follows from the results of \GeyerELA\ which appeared after the first
version of this work.

In the appendix~\appbasis\ we describe an algorithm to expand arbitrary
polynomials in $G_{ij}$ in a basis of
worldsheet functions which do not depend on $\sigma_1$.
A central role in our choice of basis is played by the following combinations of
Green functions \MafraKH,
\eqn\xxx{
X_{a_1a_2} \equiv s_{a_1a_2}G_{a_1a_2},\qquad
X_{a_1a_2\cdots a_m}\equiv\prod_{p=2}^m\Big(\sum_{q=1}^{p-1}X_{a_qa_p}
\Big) \ ,
}
whose simplest instances at $m=3,4$ read
\eqn\2{
 X_{234}\equiv X_{23}(X_{24}+X_{34}),
 \quad X_{2345}\equiv X_{23}(X_{24}+X_{34})(X_{25}+X_{35}+X_{45})\,.
}
The choice of \xxx\ is motivated by the simple action of scattering equations
which can be used iteratively to eliminate any appearance of $a_i=1$, see appendix \appbasis\
(in particular \eqthree) for further details.
Moreover, their symmetry properties such as $X_{23}=-X_{32}$ as well as $X_{234}=-X_{324}$ and
$X_{234} + {\rm cyc}(2,3,4)=0$ shared by nested commutators $[t^2,t^3]$ and $[[t^2,t^3],t^4]$
leave $(m{-}1)!$ independent permutations of $X_{a_1a_2\cdots a_m}$ \MafraKH.

When written in terms of a basis of functions $X_{a_1a_2\ldots a_m}$ with $a_i \in \{2,3,4,\ldots,n\}$,
the one-loop correlators ${\cal K}^{\ast}_n$ for supersymmetric theories take the schematic form,
\eqnn\exp
$$\eqalignno{
{\cal K}^{\ast}_n\,=&\,{\cal C} (\ell) + \sum_{2\leq i<j} {\cal C}_{i, j} (\ell) X_{ij}
+ \sum_{2\leq i<j, k} {\cal C}_{i, j, k } (\ell) X_{i j k}
&\exp\cr
&+ \sum_{2\leq i<j} \sum_{i<k<l} {\cal C}_{i,j; k, l} (\ell) X_{i j} X_{k l}
+ \sum_{2\leq i<j,k,l} {\cal C}_{i, j, k, l} (\ell) X_{i j k l}
+ \cdots\, ,
\cr
}$$
where the terms in the ellipsis involve at least three powers of $G_{ij}$.
Since the worldsheet functions form a basis, the
coefficients ${\cal C}(\ell)$ are gauge invariant kinematic factors.
They build up the gauge invariant $n$-gon numerators
${\cal C}_{+|\rho(23\ldots n)|-}(\ell)$ in \invcorr\ and \cvsa\ through the
dictionary $G_{ij} \rightarrow {1\over 2} {\rm sgn}^{\rho}_{ij}$ of section \polyGijB.

Recall from \calKred\ that ${\cal K}^{\ast}_n$ are polynomials in $\ell$ and $G_{ij}$ of total degree $n{-}4$
for maximal supersymmetry, $n{-}2$ for reduced supersymmetry and $n$ for zero supersymmetry.
Accordingly, each accompanying factor of $X_{ij}$ reduces the maximum power of $\ell$ in the kinematic
factors ${\cal C}_{\ldots}$ of \exp\ by one. Given that the subleading symmetric polynomials $\Sigma_k$ in the
spin sums reduce the
homogeneity degree in $\ell$ and $G_{ij}$ by $2,4,6,\ldots$, the $n$-point kinematic factors along with
$p$ powers of $X_{ij}$ have an expansion of the schematic form
\eqnn\expC
$$\eqalignno{
(n{-}p) \ {\rm even}:\quad &{\cal C}_{I}(\ell)=C_{I} + \ell_{m} \ell_{n}~C_{I} ^{mn} +\ell_{m} \ell_{n}\ell_p\ell_q~C_{I} ^{mnpq} + \ldots   \,,\cr
(n{-}p) \ {\rm odd}: \quad &{\cal C}_{I}(\ell)=\ell_{m}~C_{I}^{m}+ \ell_{m} \ell_{n} \ell_{p}~C_{I}^{mnp}
+ \ell_{m} \ell_{n} \ell_{p}\ell_q\ell_r~C_{I}^{mnpqr}+\ldots  \,.
}$$
The subscript $I$ collectively refers to the labels of the accompanying $p$ factors of $X_{ij}$, and the highest
powers of $\ell$ is $n{-}4{-}p$, $n{-}2{-}p$ or $n{-}p$ for maximal, reduced or zero supersymmetry.

\bigskip

\tikzpicture [line width=0.30mm]
\draw(-1,0)--(10.6,0);
\draw(0,0.5)--(0,-3.7);
\draw(0,0)--(-1,0.5);
\draw(-1,0.2)node{$n$};
\draw(-0.3,0.5)node{$k$};
\draw(1.2,0.5)--(1.2,-3.7);
\draw(2.4,0.5)--(2.4,-3.7);
\draw(3.6,0.5)--(3.6,-3.7);
\draw(5.0,0.5)--(5.0,-3.7);
\draw(6.4,0.5)--(6.4,-3.7);
\draw(7.8,0.5)--(7.8,-3.7);
\draw(9.2,0.5)--(9.2,-3.7);
\draw(-0.5,-0.4)node{$2$};
\draw(-0.5,-1.0)node{$3$};
\draw(-0.5,-1.6)node{$4$};
\draw(-0.5,-2.2)node{$5$};
\draw(-0.5,-2.8)node{$6$};
\draw(-0.5,-3.4)node{$7$};
\draw(0.6,0.3)node{$0$};
\draw(0.6,-0.4)node{$1$};
\draw(0.6,-1.0)node{$1$};
\draw(0.6,-1.6)node{$1$};
\draw(0.6,-2.2)node{$1$};
\draw(0.6,-2.8)node{$1$};
\draw(0.6,-3.4)node{$1$};
\draw(1.8,0.3)node{$1$};
\draw(1.8,-0.4)node{$$};
\draw(1.8,-1.0)node{$1$};
\draw(1.8,-1.6)node{$3$};
\draw(1.8,-2.2)node{$6$};
\draw(1.8,-2.8)node{$10$};
\draw(1.8,-3.4)node{$15$};
\draw(3.0,0.3)node{$2$};
\draw(3.0,-0.4)node{$$};
\draw(3.0,-1.0)node{$$};
\draw(3.0,-1.6)node{$2$};
\draw(3.0,-2.2)node{$11$};
\draw(3.0,-2.8)node{$35$};
\draw(3.0,-3.4)node{$85$};
\draw(4.3,0.3)node{$3$};
\draw(4.3,-0.4)node{$$};
\draw(4.3,-1.0)node{$$};
\draw(4.3,-1.6)node{$$};
\draw(4.3,-2.2)node{$6$};
\draw(4.3,-2.8)node{$50$};
\draw(4.3,-3.4)node{$225$};
\draw(5.7,0.3)node{$4$};
\draw(5.7,-0.4)node{$$};
\draw(5.7,-1.0)node{$$};
\draw(5.7,-1.6)node{$$};
\draw(5.7,-2.2)node{$$};
\draw(5.7,-2.8)node{$24$};
\draw(5.7,-3.4)node{$274$};
\draw(7.1,0.3)node{$5$};
\draw(7.1,-0.4)node{$$};
\draw(7.1,-1.0)node{$$};
\draw(7.1,-1.6)node{$$};
\draw(7.1,-2.2)node{$$};
\draw(7.1,-2.8)node{$$};
\draw(7.1,-3.4)node{$120$};
\draw(8.5,0.3)node{$\#{\cal C}_{\rm max}$};
\draw(8.5,-0.4)node{$0$};
\draw(8.5,-1.0)node{$0$};
\draw(8.5,-1.6)node{$1$};
\draw(8.5,-2.2)node{$7$};
\draw(8.5,-2.8)node{$46$};
\draw(8.5,-3.4)node{$326$};
\draw(9.9,0.3)node{$\#{\cal C}_{\rm red}$};
\draw(9.9,-0.4)node{$1$};
\draw(9.9,-1.0)node{$2$};
\draw(9.9,-1.6)node{$6$};
\draw(9.9,-2.2)node{$24$};
\draw(9.9,-2.8)node{$120$};
\draw(9.9,-3.4)node{$720$};
\endtikzpicture

\medskip
{\leftskip=20pt\rightskip=20pt\noindent\ninepoint
{\bf Table 1.} The numbers $S_{n{-}1, n{-}k{-}1}$ of independent degree-$k$ polynomials in $G_{ij}$. Summing over the ranges $0\leq k\leq n{-}4$ and $0\leq k\leq n{-}2$ admitted by maximal and reduced supersymmetry yields the tabulated numbers $\#{\cal C}_{\rm max}$ and $\#{\cal C}_{\rm red}=(n{-}1)!$ of worldsheet functions in \exp, respectively.
\par }
\bigskip
Before ending, let us record the number of independent worldsheet functions in \exp\ for supersymmetric theories.
The counting is governed by unsigned Stirling numbers $S_{N, r}$ of the first kind (see table 1)
which count the number of ways to distribute $N$ elements into $r$ cycles.
Scattering equations together with the symmetry properties of $X_{a_1a_2\ldots a_m}$ leave $S_{n{-}1, n{-}k{-}1}$
independent polynomials in $G_{ij}$ of degree $k$. Then, the range $0\leq k\leq n{-}2$ for reduced supersymmetry
yields a total of $\#{\cal C}_{\rm red}=\sum_{k=0}^{n{-}2} S_{n{-}1, n{-}k{-}1}=(n{-}1)!$ terms in \exp. Maximal supersymmetry,
however, only allows for $0\leq k\leq n{-}4$, and the resulting numbers
$\#{\cal C}_{\rm max}=\sum_{k=0}^{n{-}4} S_{n{-}1, n{-}k{-}1}$ of basis functions are gathered in table 1.

Even though the methods of this section only give access to their bosonic components, we will provide
the maximally supersymmetric completions for the $(n{\leq}6)$-point kinematic factors ${\cal C}(\ell)$ in
the next section. These kinematic factors in pure-spinor superspace are conveniently organized
in terms of Berends--Giele currents, and we will spell out the corresponding Berends--Giele description of their gluon components
which follows from the basis reduction described in this section.

%****************
\subsec{Parity-odd contributions to RNS correlators}
\par\subseclab\parodd
%***************

The above discussion has been tailored to the parity-even contributions to one-loop gauge-theory amplitudes.
However, the running of chiral fermions in the loop yields additional parity-odd terms proportional to the $d$-dimensional
Levi--Civita tensor $\epsilon_{d}$. They arise from the single odd spin structure of the worldsheet fermions whose
correlation functions in an ambitwistor setup have been described in \AdamoTSA. By the integral over fermionic zero modes,
these correlators are bound to vanish for multiplicities smaller than ${d\over 2}$.

As manifested by the expressions of \AdamoTSA\ reviewed in appendix \appparityodd,
the parity-odd correlators are polynomials in
$\ell$ and $\partial \log \theta_1(z_{ij},\tau)$ of degree $n{+}1{-}{d\over 2}$ after integration over
fermionic zero modes. Hence, their degeneration \defGij\
at $\tau \rightarrow i\infty$ is manifestly a polynomial in $G_{ij}$ and $\ell$, in complete analogy to the
above parity-even results. However, from the additional zero modes in the ghost sector of this spin structure,
a picture changing operator introduces a spurious dependence on its insertion point $\sigma_0$ via $G_{0j}$.
Since BRST invariance of the RNS ambitwistor string guarantees that the final result is independent on $\sigma_0$,
one can always eliminate any appearance of $G_{0j}$ through a sequence of Fay identities and scattering equations.

The conclusion from the parity-even sector is therefore unchanged: The parity-odd contributions to the $n$-point
correlator due to chiral fermions can be expressed as degree-$(n{+}1{-}{d\over 2})$ polynomials in $\ell$ and $G_{ij}$
with $i,j\neq0$. In particular, chiral theories in $d=10$ and $d=6$ dimensions lead to the degrees $n{-}4$
and $n{-}2$ familiar from the parity-even sectors with maximal and half-maximal supersymmetry, respectively.

Since analogous statements hold for the RNS superstring, we will translate results for string
correlators with all dependence on $\sigma_0$ eliminated to the ambitwistor setup. In case of ten-dimensional
SYM, the contributions from chiral fermions vanishes below five points\foot{The factor of $i$ reflects our conventions $\epsilon_d^{m_1m_2\ldots m_d}\epsilon_d^{m_1m_2\ldots m_d} = +d!$ for the normalization of the Levi--Civita tensor.}
\eqn\paroddA{
{\cal K}_{n\leq 4}^{\epsilon_{10}} = 0 \ , \ \ \ \ \ \
{\cal K}_{5}^{\epsilon_{10}}  = i  \ell_m\epsilon^m_{10}(e_1,k_2,e_2,k_3,e_3,k_4,e_4,k_5,e_5) \ ,
}
where the shorthand $\epsilon_{10}^m(e_1,k_2,e_2,k_3,e_3,k_4,e_4,k_5,e_5)
= \epsilon_{10}^{mnpqrsabcd}e^n_1k^p_2e^q_2k^r_3e^s_3k^a_4e^b_4k^c_5e^d_5$ avoids proliferation of indices.
While the five-point correlator does not allow any contribution with $G_{0j}$ on kinematic grounds, a long calculation
is needed to demonstrate the disappearance of $G_{0j}$ from the six-point correlator. The
manifestly $\sigma_0$-independent
superstring correlators of \refs{\ClavelliFJ, \MafraNWR} then degenerate into
\eqnn\paroddB
$$\eqalignno{
&{\cal K}_{6}^{\epsilon_{10}}  =  i \big[ (\ell\cdot e_2) \epsilon_{10}(\ell,e_1,k_3,e_3,k_4,e_4,k_5,e_5,k_6,e_6) + (2\leftrightarrow 3,4,5,6) \big] &\paroddB \cr
& \ \ \ +i \big[ G_{12} \ell_m E^m_{12|3,4,5,6}
+(2\leftrightarrow 3,4,5,6)
\big]
+i\big[ G_{23} \ell_m E^m_{1|23,4,5,6}
+(2,3|2,3,4,5,6)
\big] \ ,
}$$
with vectorial kinematic factors
\eqnn\paroddC
\eqnn\paroddD
$$\eqalignno{
E^m_{12|3,4,5,6} &= (e_1 \cdot k_2)\epsilon_{10}^m(e_2,k_3,e_3,\ldots,k_6,e_6)
-(e_2 \cdot k_1)\epsilon_{10}^m(e_1,k_3,e_3,\ldots,k_6,e_6) \cr
&- (e_1 \cdot e_2)\epsilon_{10}^m(k_2,k_3,e_3,\ldots,k_6,e_6)
&\paroddC
\cr
E^m_{1|23,4,5,6}  &= (e_2 \cdot k_3)\epsilon_{10}^m(e_1,k_{23},e_3,\ldots,k_6,e_6)
-(e_3 \cdot k_2)\epsilon_{10}^m(e_1,k_{23},e_2,\ldots,k_6,e_6) \cr
&-(e_2 \cdot e_3)\epsilon_{10}^m(e_1,k_{2},k_3,\ldots,k_6,e_6)
-(k_2 \cdot k_3)\epsilon_{10}^m(e_1,e_{2},e_3,\ldots,k_6,e_6)
&\paroddD
}$$
and $k_4,e_4,k_5,e_5$ in the ellipsis. The notation $(i_1,{\ldots } , i_p | i_1,{\ldots} ,i_q)$ on
the right hand side of \paroddB\ with $q>p$ instructs to sum over all possibilities to choose $p$ elements $i_1,\ldots
,i_p$ out of the larger set $\{i_1,{\ldots} ,i_q\}$, for a total of ${q\choose p}$ terms.

For chiral SYM in six dimensions, the minimal multiplicity is shifted by two such that
\eqn\paroddA{
{\cal K}_{n\leq 2}^{\epsilon_{6}} = 0 \ , \ \ \ \ \ \
{\cal K}_{3}^{\epsilon_{6}}  = i \ell_m\epsilon^m_{6}(e_1,k_2,e_2,k_3,e_3) \ .
}
The above expressions for ${\cal K}_{6}^{\epsilon_{10}} $ including the mechanisms for the decoupling
of $G_{0j}$ have been generalized to arbitrary even dimensions in \BergWUX. Accordingly, the
six-dimensional four-point correlator
\eqnn\paroddF
$$\eqalignno{
&{\cal K}_{4}^{\epsilon_{6}}  =  i \big[ (\ell\cdot e_2) \epsilon_{6}(\ell,e_1,k_3,e_3,k_4,e_4) + (2\leftrightarrow 3,4) \big] &\paroddF \cr
& \ \ \ +i\big[ G_{12} \ell_m E^m_{12|3,4}
+(2\leftrightarrow 3,4)
\big]
+i \big[ G_{23} \ell_m E^m_{1|23,4}
+(2,3|2,3,4)
\big] \ ,
}$$
follows the structure of ${\cal K}_{6}^{\epsilon_{10}} $ with kinematic factors resembling \paroddC\ and \paroddD,
\eqnn\paroddG
\eqnn\paroddH
$$\eqalignno{
E^m_{12|3,4} &= (e_1 \cdot k_2)\epsilon_{6}^m(e_2,k_3,e_3,k_4,e_4)
-(e_2 \cdot k_1)\epsilon_{6}^m(e_1,k_3,e_3,k_4,e_4) \cr
&- (e_1 \cdot e_2)\epsilon_{6}^m(k_2,k_3,e_3,k_4,e_4)
&\paroddG
\cr
E^m_{1|23,4}  &= (e_2 \cdot k_3)\epsilon_{6}^m(e_1,k_{23},e_3,k_4,e_4)
-(e_3 \cdot k_2)\epsilon_{6}^m(e_1,k_{23},e_2,k_4,e_4) \cr
&-(e_2 \cdot e_3)\epsilon_{6}^m(e_1,k_{2},k_3,k_4,e_4)
-(k_2 \cdot k_3)\epsilon_{6}^m(e_1,e_{2},e_3,k_4,e_4)  \ .
&\paroddH
}$$
The pure-spinor superspace expressions for the ten-dimensional correlators to be discussed
in the following section automatically combine both the parity-even and the parity-odd components.
The BCJ master numerators in $n\leq 4$-point amplitudes of chiral SYM in six dimensions
will be given in section \secREDUCE.

%***********************************************************
\newsec{Pure-spinor representations for the gauge multiplet}
\par\seclab\secPS\
%***********************************************************

\noindent
In this section we present the field-theory limit of the
superstring one-loop correlators that have been
computed using the pure-spinor formalism \refs{\psf,\MPS}.
By the arguments of \GomezWZA, identical results are obtained
when performing the computation with the loop-level prescription
\AdamoHOA\ of the pure-spinor ambitwistor string \BerkovitsXBA.

%****************
\subsec{Review of pure-spinor superspace}
\par\subseclab\fiveone
%***************

Supersymmetric scattering amplitudes in ten dimensions admit compact representations
in the language of pure-spinor superspace. This new type of
superspace arises naturally within the pure-spinor formalism of the
superstring and its properties played an important role in recent
advances in the computation of string scattering amplitudes.

A super-Poincar\'e invariant description of ten-dimensional SYM theory
uses four types of superfields
\eqn\symsup{
A_\a(x,\t),\; A_m(x,\t),\; W^\a(x,\t),\; F_{mn}(x,\t)
}
that depend on the superspace coordinates $x^m,\t^\a$ with vector indices $m=0,
\ldots,9$ and spinor indices $\a=1, \ldots,16$ of the ten-dimensional Lorentz-group.
They satisfy the following (linearized) equations of motion
\wittentwistor
\eqn\SYM{
\eqalign{
D_{(\a} A_{\b)}  &= \phantom{{1\over 4}}\g^m_{\a\b} A_m,\cr
D_\a A_m &= (\g_m W)_\a + \partial_m A_\a ,\cr
}
\qquad\eqalign{
D_\a W^{\b}  &= {1\over 4}(\g^{mn})_\a{}^\b F_{mn}\cr
D_\a F_{mn} &= \partial_{[m} (\g_{n]} W)_\a\,,
}}
where $D_\a \equiv \p_\a + \half \p_m (\g^m \t)_\a$ is the supersymmetric
covariant derivative and $\gamma^m_{\a \b}=\gamma^m_{\b \a}$ denote $16\times 16$ Pauli matrices\foot{They often appear in antisymmetrized combinations subject to $\gamma^{mnp}_{\alpha \beta}=-\gamma^{mnp}_{\beta \alpha }$
and $\gamma^{mnpqr}_{\alpha \beta}=\gamma^{mnpqr}_{\beta \alpha }$ with normalization conventions such as $\gamma^{mn}{}_{\a}{}^{\b} \equiv {1\over 2} (\gamma^m\gamma^n - \gamma^n \gamma^m)_{\a}^{\b}$.}.
The $\t$-expansions of the superfields \symsup\ are written\foot{For
historical reasons, we omit the factor of $i$ in the plane wave
expansion.} in terms of gluon polarizations $e_m$, gluino wavefunctions $\chi^\alpha$
as well as the field-strength $f_{mn} = 2k_{[m} e_{n]}$ \thetaExp:
\eqnn\thetaexpan
$$\eqalignno{
A_{\a}(x,\t)&=\Big( {1\over 2}e_m(\g^m\t)_\a -{1\over 3}(\chi\g_m\t)(\g^m\t)_\a
-{1\over 32}f_{mn}(\g_p\t)_\a (\t\g^{mnp}\t) + \cdots \Big) {\rm e}^{k\cdot x}   \cr
A_{m}(x,\t) &= \Big( e_m - (\chi \g_m\t) - {1\over 8}(\t\g_m\g^{pq}\t)f_{pq}
         + {1\over 12}(\t\g_m\g^{pq}\t)k_p(\chi\g_q\t) + \cdots \Big) {\rm e}^{k\cdot x} \cr
W^{\a}(x,\t) &= \Big( \chi^{\a} - {1\over 4}(\g^{mn}\t)^{\a} f_{mn}
           + {1\over 4}(\g^{mn}\t)^{\a} k_m(\chi\g_n\t)
%	   + {1\over 48}(\g^{mn}\t)^{\a}(\t\g_n\g^{pq}\t)\p_m F_{pq}
	   + \cdots \Big) {\rm e}^{k\cdot x} &\thetaexpan \cr
F_{mn}(x,\t) &= \Big( f_{mn} - 2 k_{[m}(\chi\g_{n]}\t) + {1\over
4}(\t\g_{[m}\g^{pq}\t)k_{n]}f_{pq} + \cdots \Big) {\rm e}^{k\cdot x} \ .
}$$
Pure-spinor superspace expressions are defined as
expansions of the form \psf
\eqn\pssdef{
\langle \l^\a\l^\b \l^\g f_{\a\b\g}(\t)\rangle,
}
where $f_{\a\b\g}(\t)$ denotes an arbitrary function of the
superfields \symsup\ and encodes the information about the
polarizations of the particles participating in the scattering.
For example, the three-particle scattering of SYM states
is described by $f_{\a\b\g}(\t) = A^1_\a(\t) A^2_\b(\t) A^3_\g(\t)$.
In the above definition \pssdef, the variables $\l^\a$ are the
zero modes of a pure spinor subject to $(\l \gamma^m \l)=0$,
and the angular bracket is defined by
\refs{\psf,\pssnorm}
\eqn\psbracket{
\langle (\lambda \gamma^m \theta) (\lambda \gamma^n \theta)  (\lambda \gamma^p \theta)
(\theta \gamma_{mnp} \theta) \rangle = 2880\,,
}
while expressions of different degrees $\lambda^{\neq 3}$ or $\theta^{\neq 5}$ yield
a vanishing bracket. The prescription \psbracket\ is motivated by supersymmetry and the cohomology of the
BRST operator $Q =\lambda^\alpha D_\alpha$: BRST invariant superfields
$Q(\lambda^\a \lambda^\b \lambda^\g f_{\a\b\g}(\t))=0$ are mapped to supersymmetric and gauge invariant
components $ \langle \lambda^\a \lambda^\b \lambda^\g f_{\a\b\g}(\t)\rangle$. BRST exact superfields, on the
other hand, are annihilated, i.e.\ $ \langle Q( \lambda^\a \lambda^\b g_{\a\b}(\t))\rangle=0$ for any choice of $g_{\a \b}(\t)$.

%****************
\subsec{Review of one-loop building blocks}
\par\subseclab\fivetwo
%***************

The superspace description of SYM theory can be generalized to
a multiparticle setup which is convenient to describe the scattering
of a high number of external particles. The so-called
{\it multiparticle superfields} have been defined using recursion
relations both in local and non-local forms \refs{\EOMBBs,\LeeUPY}.
For example, in a notation
where uppercase latin letters $P=123 \ldots p$ encompass the labels
of $p$ external legs, the non-local recursion relations are given by
\eqnn\SFa
\eqnn\SFb
\eqnn\SFc
\eqnn\SFd
$$\eqalignno{
\cA^{P}_\a &= {1 \over 2 s_P} \sum_{XY=P} \bigl[  \cA^{Y}_\a (k^{Y}\cdot  \cA^X)
+  \cA^{Y}_m (\g^m \cW^X)_\a - (X\leftrightarrow Y)\bigr]  &\SFa\cr
\cA^{P}_m &=  {1 \over 2 s_P}  \sum_{XY=P}   \bigl[ \cA^{Y}_m (k^Y\cdot  \cA^{X}) + \cA^{Y}_n  \cF^X_{mn}
+ ( \cW^{X}\g_m \cW^Y)
- (X \leftrightarrow Y)\bigr] &\SFb\cr
\cW_{P}^\a &= {1 \over 2 s_P}  k_P^m  \gamma_m^{\alpha \beta} \sum_{XY=P}
 \big[ \cA_X^n (\gamma_n \cW_Y)_\beta  - (X \leftrightarrow Y) \big]
&\SFc\cr
\cF^{mn}_P &= k_P^m \cA_P^n - k_P^n \cA_P^m
- \sum_{XY=P}\!\!\big( \cA_X^m \cA_Y^n - \cA_X^n \cA_Y^m \big) \,,
&\SFd\cr
}$$
and they give rise to a supersymmetric generalization of the
Berends--Giele currents \BerendsME.
In the above formulae, the summation over $XY=P$ denotes
a sum over the deconcatenations of $P=123 \ldots p$ into
into non-empty words $X=12\ldots j$ and $Y=j{+}1\ldots p$ with $j=1,2,\ldots,p{-}1$.
The propagators $1/s_P$ in the above recursions
identify the tree-level subdiagrams described by the currents
and characterize their non-local nature.

The above Berends--Giele supercurrents constitute the fundamental
building blocks for kinematic factors in ten-dimensional one-loop superstring amplitudes.
They have been systematically assembled in \MafraGSA\
by closely following the zero-mode saturation rules in the pure-spinor
formalism \MPS. For example, from the definitions
\eqnn\Mscal
$$\eqalignno{
M_{A,B,C} &\equiv {1\over 3}(\l\g^m \cW_A)(\l\g^n \cW_B)  \cF_C^{mn}
+ (A\leftrightarrow B,C)\,, &\Mscal\cr
\cW^m_{A,B,C,D} &\equiv {1\over 12}(\l\g_n \cW_A)(\l\g_p
\cW_B)(\cW_C\gamma^{mnp}\cW_D) + (A,B|A,B,C,D)\,,
}$$
it follows that
\eqnn\Mmdef
$$\eqalignno{
M^m_{A,B,C,D} &\equiv  \bigl[  \cA_A^m M_{B,C,D}
+ (A\leftrightarrow B,C,D)\bigr] + \cW^m_{A,B,C,D} &\Mmdef\cr
M^{mn}_{A,B,C,D,E}&\equiv
\cA^n_A M^m_{B,C,D,E} + \cA^m_A \cW^n_{B,C,D,E} + (A \leftrightarrow B,C,D,E)\,
}$$
exhibit covariant BRST transformations, and they naturally appear in string scattering computations at one loop.

%****************
\subsubsec{BRST invariant combinations}
%***************

Generalizing the above structures paves the way for the definition of
kinematic BRST invariants and so-called pseudo-invariants
of arbitrary tensor rank \MafraGSA. For example, using $M_P\equiv \l^\a
\cA_\a^P$ one can recursively\foot{See \MafraGSA\ for the
explicit form of the recursion and associated definitions. To keep
the presentation short, here we chose to write down a few examples
of their outcome.} define scalar BRST invariants such as:
\eqnn\scalarBRST
$$\eqalignno{
C_{1|2,3,4} &\equiv M_1 M_{2,3,4}\cr
C_{1|23,4,5} &\equiv M_1 M_{23,4,5} + M_{12}M_{3,4,5} - M_{13}M_{2,4,5}\,, \cr
C_{1|234,5,6} &\equiv M_1 M_{234,5,6} + M_{12}M_{34,5,6} + M_{123}M_{4,5,6} - M_{124}M_{3,5,6}\cr
&{}- M_{14}M_{23,5,6} - M_{142}M_{3,5,6} + M_{143}M_{2,5,6}\,, &\scalarBRST \cr
C_{1|23,45,6}
&\equiv M_1 M_{23,45,6} + M_{12}M_{45,3,6} - M_{13}M_{45,2,6} + M_{14}M_{23,5,6} - M_{15}M_{23,4,6}\cr
&{}+\big[ M_{124}M_{3,5,6} - M_{134}M_{2,5,6}+ M_{142}M_{3,5,6} - M_{143}M_{2,5,6}
- (4\leftrightarrow 5)\big]\,.\cr
}$$
In addition, one can define pseudo-invariants\foot{{\it Pseudo}-invariants are
defined to be expressions whose BRST variation, instead
of vanishing, gives rise to {\it anomalous} superfields \MafraGSA\ that carry the fingerprints of the
hexagon anomaly of ten-dimensional SYM.
For a prominent example of their use, see \MafraNWR.}  of arbitrary tensor
ranks and multiplicity.
In this paper we will be concerned with explicit amplitudes up to multiplicity six,
for which the following definitions suffice
\eqnn\tensC
$$\eqalignno{
C^m_{1|2,3,4,5} &\equiv M_1 M^m_{2,3,4,5} + \big[ k_2^m M_{12} M_{3,4,5}
+ (2\leftrightarrow 3,4,5) \big]\,, &\tensC\cr
C^m_{1|23,4,5,6} &\equiv
M_1 M^m_{23,4,5,6} + M_{12} M^m_{3,4,5,6} - M_{13} M^m_{2,4,5,6}\cr
&\quad{} +\big[ k^m_3 M_{123}M_{4,5,6} + (3\leftrightarrow 4,5,6)\bigr]
- \big[ k^m_2 M_{132}M_{4,5,6} + (2\leftrightarrow 4,5,6)\bigr]\cr
&\quad{}
+\big[ k^m_4 M_{14}M_{23,5,6} + k^m_4 M_{142}M_{3,5,6} - k^m_4
M_{143}M_{2,5,6} + (4\leftrightarrow 5,6)\bigr]\,,\cr
C^{mn}_{1|2,3,4,5,6} &\equiv M_1 M^{mn}_{2,3,4,5,6} + 2 \big[ k_2^{(m} M_{12} M^{n)}_{3,4,5,6}
+ (2\leftrightarrow 3,4,5,6) \big]\cr
&\quad{} + 2 \big[ k_2^{(m} k_3^{n)} (M_{123}+M_{132}) M_{4,5,6} + (2,3|2,3,4,5,6) \big].
}$$

%****************
\subsubsec{Pure-spinor superspace versus gluon components\foot{This
subsection was written by Carlos Mafra with the aid of \PSS.}}
%***************

The above kinematic expressions are written in pure-spinor
superspace. While compact superspace expressions suffice for most
purposes, one might still want to obtain the explicit component form of the
amplitudes written in terms of the physical gluon and gluino
polarizations. Fortunately, the properties of the pure-spinor superspace
measure \psbracket\ can be exploited to easily
automate this task. In addition, with the techniques advanced in
\LeeUPY\ the results take an elegant and compact form even at
the level of components. To see this one defines
Berends--Giele currents for the gluon polarization and field-strength
in component form, starting with the single-particle cases
$\ce^m_i = e^m_i$ and $\cf^{mn}_i = k^m_i e^n_i - k^n_i e^m_i$:
\eqnn\recone
\eqnn\rectwo
$$\eqalignno{
\ce^m_P &\equiv {1 \over 2 s_P } \sum_{XY=P}  \bigl[ \ce_{Y}^m (k^Y\cdot  \ce^{X})
+ \ce^{Y}_n  \cf_X^{mn} - (X \leftrightarrow Y)\bigr] &\recone\cr
\cf^{mn}_P &\equiv k_P^m \ce_P^n - k_P^n \ce_P^m
- \sum_{XY=P}\!\!\big( \ce_X^m \ce_Y^n - \ce_X^n \ce_Y^m \big)  \ .
&\rectwo
}$$
This can be viewed as a truncation of \SFb\ and \SFd\ where the fermionic variables are suppressed,
and it is straightforward to generalize \recone\ and \rectwo\ to include gluino polarizations. Using the
multiparticle Harnad--Shnider gauge introduced in \LeeUPY, one can
show that the gluon components of the above BRST pseudo-invariants
can be compactly written in terms of the $t_8$-tensor \teightMG:
\eqnn\teights
$$\eqalignno{
t_{A,B,C,D} &\equiv \cf^{mn}_A \cf^{np}_B \cf^{pq}_C \cf^{qm}_D - {1\over 4}
(\cf^{mn}_A \cf^{nm}_B) ( \cf^{pq}_C \cf^{qp}_D) + {\rm cyc}(B,C,D)
= t_8(\cf_A,\cf_B,\cf_C,\cf_D)\cr
t^m_{A,B,C,D,E} &\equiv \big[ \ce^m_A t_{B,C,D,E}  + (A \leftrightarrow B,C,D,E) \big]
+{i\over 2} \epsilon_{10}^{m}(\ce_A,\cf_B,\cf_C,\cf_D,\cf_E)
&\teights \cr
t^{mn}_{A,B,C,D,E,F} &\equiv 2 \big[ \ce^{(m}_A \ce^{n)}_B t_{C,D,E,F}  + (A,B| A,B,C,D,E,F)  \big] \cr
& \ \ \  +
i \big[ \ce_B^{(m}\epsilon_{10}^{n)}(\ce_A,\cf_C,\cf_D,\cf_E,\cf_F) + (B\leftrightarrow C,D,E,F)\big]\; .
}$$
Moreover, we have used the shorthand $\epsilon_{10}^{m}(\ce_A,\cf_B,\cf_C,\cf_D,\cf_E)\equiv
\epsilon_{10}^{mnpqrsabcd}\ce^n_A\cf_B^{pq}\cf_C^{rs}\cf_D^{ab}\cf_E^{cd}$ to avoid proliferation of
indices in the parity-odd contributions from section \parodd. Motivated by the simple examples
\eqnn\Cfourfive
$$\eqalignno{
-16\, \langle C_{1|2,3,4} \rangle &= t_{1,2,3,4} &\Cfourfive\cr
-16\, \langle C_{1|23,4,5} \rangle &= t_{12,3,4,5} + t_{1,23,4,5} - t_{13,2,4,5}\,,\cr
-16\,\langle C^m_{1|2,3,4,5}\rangle &= t^m_{1,2,3,4,5}
+ \bigl( k_2^m t_{12,3,4,5} + 2\leftrightarrow 3,4,5\bigr)\cr
-16\, \langle C^{mn}_{1|2,3,4,5,6}\rangle &= t^{mn}_{1,2,3,4,5,6}
+ \bigl(2k_2^{(m}t^{n)}_{12,3,4,5,6} + (2\leftrightarrow 3,4,5,6)\bigr)\cr
&\quad{}- \bigl(2k_2^{(m}k_3^{n)}t_{213,4,5,6} + (2,3|2,3,4,5,6)\bigr) \ ,\cr
}$$
one can verify that the
translation from pseudo-BRST invariants \scalarBRST\ and \tensC\
to their gluonic components can be obtained as follows\foot{It is important to stress that
the validity of the map \trivialMap\ is checked {\it within} BRST
(pseudo-)invariants as its contact-term mismatch cancels in such cases.}
\eqnn\trivialMap
$$\eqalignno{
-16\,\langle M_A M_{B,C,D}\rangle &\rightarrow t_{A,B,C,D}
\cr
-16\,\langle M_A M^m_{B,C,D,E}\rangle &\rightarrow t^m_{A,B,C,D,E} &\trivialMap \cr
-16\,\langle M_A M^{mn}_{B,C,D,E,F}\rangle &\rightarrow
t^{mn}_{A,B,C,D,E,F} \ .
}$$

%****************
\subsec{Maximally supersymmetric one-loop correlators from string theory}
\par\subseclab\fivethree
%***************

In string theory, the supersymmetric one-loop integrands at four,
five and six points have been
computed using the pure-spinor formalism in
\refs{\MPS,\MafraKH,\MafraNWR}. In the field-theory
limit they can be written as\foot{This particular representation using
the explicit loop momentum $\ell^m$ is based on unpublished work \wipL.},
\eqnn\PSfour
\eqnn\PSfive
\eqnn\PSsix
$$\eqalignno{
{\cal K}_4 &= \langle C_{1|2,3,4} \rangle  &\PSfour\cr
{\cal K}_5 &= \langle \ell_m C^m_{1|2,3,4,5} + [X_{23}  C_{1|23,4,5} + (2,3|2,3,4,5) ] \rangle  &\PSfive \cr
{\cal K}_6&= \langle {1 \over 2} \ell_m \ell_n C^{mn}_{1|2,3,4,5,6} + \ell_m \big[
X_{23} C^{m}_{1|23,4,5,6} + (2,3|2,3,4,5,6) \big]
\cr
&+ \big[ X_{23}\, X_{34}  C_{1|234,5,6}
-X_{23} \, X_{24} C_{1|324,5,6}  -X_{24} \, X_{34} C_{1|243,5,6} + (2,3,4|2,3,4,5,6) \big]
 \cr
&+  \big[ X_{23}\, X_{45}  C_{1|23,45,6} + (2,3|4,5|2,3,4,5,6) \big]
- {1\over 4} k_1^m k_1^n C^{mn}_{1|2,3,4,5,6}  \rangle  \ ,
&\PSsix
}$$
see the previous subsection for their gluon component expansions. We note that the last term in the six-point
correlator without any accompanying factors of $X_{ij}$ or $\ell$ is permutation symmetric and in fact
proportional to the six-point tree-level amplitude of Born--Infeld
theory\foot{We thank Carlos Mafra for several discussions on finding
a compact representation for the LHS of \Born.}:
\eqnn\Born
$$\eqalignno{
&- {1\over 4} k_1^m k_1^n \langle C^{mn}_{1|2,3,4,5,6}  \rangle =M^{\rm tree}_{\rm BI}(1,2,3,4,5,6) &\Born \cr
& \ \ \ = \sum_{\rho \in S_4} s_{1\rho(2)} (s_{1\rho(3)}+s_{\rho(23)})(s_{\rho(45)}+s_{\rho(4)6}) s_{\rho(5)6} A^{\rm tree}(1,\rho(2,3,4,5),6)
}$$
The representation of the Born--Infeld amplitude is based on its double-copy structure \CachazoXEA\ involving
gauge-theory trees and the BCJ master numerators for the NLSM of \refs{\CarrascoLDY, \CarrascoYGV}.

%****************
\subsec{Pure-spinor representations of BCJ numerators and partial integrands}
\par\subseclab\fivefour
%***************

As discussed in section \whymaster, one can identify BCJ master numerators at
one loop by rewriting the correlator in terms of Parke--Taylor factors.
Applying the dictionary \thumb\ to the pure-spinor correlators \PSfour\ to \PSsix\
and exploiting the absence of $G_{1j}$ in our
basis of functions leads to a manifestly supersymmetric CHY integrand of the form \invcorr.
We obtain the following supersymmetric BCJ master numerators ${\cal C}_{+|\rho(2, \ldots,n)|-}(\ell)$ for the $n$-gon
diagrams:
\eqnn\masterfour
\eqnn\masterfive
\eqnn\mastersix
$$\eqalignno{
{\cal C}_{+|\rho(2,3,4)|-}(\ell) &= \langle C_{1|2,3,4}\rangle = s_{12}s_{23}A^{\rm tree}(1,2,3,4)&\masterfour\cr
{\cal C}_{+|\rho(2, \ldots,5)|-}(\ell) &= \ell_m \langle C^m_{1|2,3,4,5}\rangle+
\half\Big[ s_{23}\sign_{23}^\rho \langle C_{1|23,4,5}\rangle +
(2,3|2,3,4,5)\Big] &\masterfive\cr
{\cal C}_{+|\rho(2, \ldots,6)|-}(\ell) &=\half\ell_m\ell_n\langle
C^{mn}_{1|2,3,4,5,6}\rangle +
\half\ell_m\Big[s_{23}\sign_{23}^\rho\langle C_{1|23,4,5,6}^m\rangle
+ (2,3|2,3,4,5,6)\Big] \cr
&\quad{}+{1\over4}\Big[s_{23}s_{45}\sign_{23}^\rho\sign_{45}^\rho\langle
C_{1|23,45,6}\rangle + (2,3|4,5|2,3,4,5,6)\Big]&\mastersix\cr
&\quad{}+{1\over4}\Big[
s_{23}s_{34}\sign_{23}^\rho\sign_{34}^\rho\langle C_{1|234,5,6}\rangle
-s_{23}s_{24}\sign_{23}^\rho\sign_{24}^\rho\langle C_{1|324,5,6}\rangle\cr
&\qquad{} \ \ \ - s_{24}s_{34}\sign_{24}^\rho\sign_{34}^\rho\langle C_{1|243,5,6}\rangle
+ (2,3,4|2,3,4,5,6)\Big]\cr
&\quad{}-{1\over 4}k^1_mk^1_n\langle C^{mn}_{1|2,3,4,5,6}\rangle \ .
}$$
An explicit form of the bosonic components in terms of recursive Berends--Giele currents
is readily obtained via \trivialMap. The notation $+
(2,3|4,5|2,3,4,5,6)$ in the second line of \mastersix\ means a sum over all pairs $\{i,j\}$
and $\{p,q\}$ such that $i,j,p,q\in\{2,3,4,5,6\}$ and
$i<j$, $p<q$ and $i<p$. In absence of factors of $\sign_{1j}^\rho$, the above $n$-gon numerators
do not depend on the position of leg 1, and by the kinematic Jacobi identities, numerators with
leg 1 involved in a tree-level subdiagram vanish.

Note that the above BCJ numerators yield the following partial integrands
\eqnn\pintfour
\eqnn\pintfive
\eqnn\pintsix
$$\eqalignno{
a(1,2,3,4,-,+) &=  {  \langle C_{1|2,3,4} \rangle \over s_{1,\ell}s_{12,\ell}s_{123,\ell} }
&\pintfour \cr
a(1,2,\ldots,5,-,+) &= {  \langle \ell_m C^m_{1|2,3,4,5}  - {1\over 2} [s_{23}C_{1|23,4,5} +(2,3|2,3,4,5) ]\rangle \over s_{1,\ell} s_{12,\ell} s_{123,\ell} s_{1234,\ell}} &\pintfive \cr
&\ \ \
- { \langle C_{1|23,4,5} \rangle \over s_{1,\ell} s_{123,\ell}  s_{1234,\ell}}
- { \langle C_{1|34,2,5} \rangle \over s_{1,\ell}  s_{12,\ell} s_{1234,\ell}}
- { \langle C_{1|45,2,3} \rangle \over s_{1,\ell} s_{12,\ell} s_{123,\ell}}
 \cr
a(1,2,\ldots,6,-,+) &= {{1\over 2}  \langle \ell_m \ell_n C^{mn}_{1|2,3,4,5,6}  -   \ell_m [s_{23}C^m_{1|23,4,5,6} +(2,3|2,3,4,5,6) ]\rangle \over s_{1,\ell} s_{12,\ell} s_{123,\ell} s_{1234,\ell} s_{12345,\ell}}
\cr
&\! \! \! \! \! \! \! \! \! \! \! \! \! \! \! \! \! \! \! \! \! \! \! \! \! \! \! \! \! \! \! \! \! \! \! \! \! \! \! \! \! \! + {  \langle C_{1|2;3;4;5;6} \rangle   \over s_{1,\ell} s_{12,\ell} s_{123,\ell} s_{1234,\ell} s_{12345,\ell}}   + {  \langle C_{1|23;4;5;6} - \ell_m C^m_{1|23,4,5,6} \rangle   \over s_{1,\ell}   s_{123,\ell} s_{1234,\ell} s_{12345,\ell}}
+ {  \langle C_{1|2;34;5;6} - \ell_m C^m_{1|2,34,5,6} \rangle   \over s_{1,\ell} s_{12,\ell}  s_{1234,\ell} s_{12345,\ell}}
\cr
&\! \! \! \! \! \! \! \! \! \! \! \! \! \! \! \! \! \! \! \! \! \! \! \! \! \! \! \! \! \! \! \! \! \! \! \! \! \! \! \! \! \!  + {  \langle C_{1|2;3;45;6} - \ell_m C^m_{1|2,3,45,6} \rangle   \over s_{1,\ell} s_{12,\ell} s_{123,\ell}   s_{12345,\ell}}
+ {  \langle C_{1|2;3;4;56} - \ell_m C^m_{1|2,3,4,56} \rangle   \over s_{1,\ell} s_{12,\ell} s_{123,\ell} s_{1234,\ell}  }
+ {\langle C_{1|234,5,6} \rangle \over s_{1,\ell}  s_{1234,\ell} s_{12345,\ell}}&\pintsix
\cr
%%%%%%%%%
&\! \! \! \! \! \! \! \! \! \! \! \! \! \! \! \! \! \! \! \! \! \! \! \! \! \! \! \! \! \! \! \! \! \! \! \! \! \! \! \! \! \!
+ {\langle C_{1|2,345,6} \rangle \over s_{1,\ell} s_{12,\ell} s_{12345,\ell}}
+ {\langle C_{1|2,3,456} \rangle \over s_{1,\ell} s_{12,\ell} s_{123,\ell} }
+ {\langle C_{1|23,45,6} \rangle \over s_{1,\ell}   s_{123,\ell}  s_{12345,\ell}}
+ {\langle C_{1|23,4,56} \rangle \over s_{1,\ell}   s_{123,\ell} s_{1234,\ell}  }
+ {\langle C_{1|2,34,56} \rangle \over s_{1,\ell} s_{12,\ell}   s_{1234,\ell}  }
}$$
with the scalar hexagon numerator
\eqnn\scalarhex
$$\eqalignno{
4 C_{1|2;3;4;5;6} &= -k^1_mk^1_n C^{mn}_{1|2,3,4,5,6}  + \big[ s_{23} s_{45} C_{1|23,45,6} + (2,3|4,5|2,3,4,5,6) \big]
&\scalarhex \cr
&+\big[ s_{23}s_{34} C_{1|234,5,6}- s_{23}s_{24} C_{1|324,5,6} - s_{24}s_{34} C_{1|243,5,6} + (2,3,4|2,3,4,5,6) \big]
}$$
and scalar pentagons such as
\eqnn\scalarpent
$$\eqalignno{
2C_{1|23;4;5;6} &=  s_{45}C_{1|23,45,6} + s_{46} C_{1|23,46,5} + s_{56} C_{1|23,4,56} &\scalarpent \cr
&+ \big[
s_{34} C_{1|234,5,6} - s_{24} C_{1|324,5,6}
+(4\leftrightarrow 5,6)
\big]\ .
}$$
The partial integrands \pintfour\ to \pintsix\ have been checked to follow from the partial-fraction decomposition of the
Feynman integrals in the color-ordered amplitudes of \MafraGJA. Given that the scalar invariants $\langle C_{1|A,B,C} \rangle$ can be expanded in a BCJ basis of SYM tree amplitudes $A^{\rm tree}(\ldots)$ \refs{\MafraKH,\EOMBBs}, the five-point kinematic factors $\langle \ell_m C^m_{1|2,3,4,5} \rangle$ and $\langle C_{1|ij,k,l} \rangle$ allow for three linearly independent permutations of the partial integrand \pintfive. This is another example of how maximal supersymmetry introduces extra degeneracies beyond the upper bound of $(n{-}1)!$ linearly independent $n$-point partial integrands.

%****************
\subsec{Reconciling the hexagon anomaly with the BCJ duality}
\par\subseclab\fivefive
%***************

\subsubsec{Deviation from BCJ relations in the literature}

Among the one-loop integrands constructed in \MafraGJA\ from BRST invariance and locality, only
the five-point numerators were found to obey the BCJ duality. It deserves clarification why the six-point
amplitude of \MafraGJA\ incorporated deviations from the BCJ duality even though it gives rise to the
same partial integrand \pintsix\ as the BCJ master numerators \mastersix. Generally speaking, the
puzzle is resolved by the different bookkeeping of cubic diagrams resulting from the new representation
of Feynman integrals reviewed in section \newellrep. As explained in subsection \whymaster, this leaves
more flexibility to tune the numerators such as to satisfy the kinematic Jacobi relations.

An example for a kinematic Jacobi relation which has been violated in the six-point amplitude representation
of \MafraGJA\ is depicted in \figcounter: Since each cyclically inequivalent pentagon in the reference is associated
with a single numerator, diagrams with different positions of $\ell$ among the internal edges are interlocked through shifts
such as $\ell \rightarrow \ell- k_{23}$. The resulting numerator for the rightmost diagram in \figcounter\ was found to
violate the depicted Jacobi relation \MafraGJA.

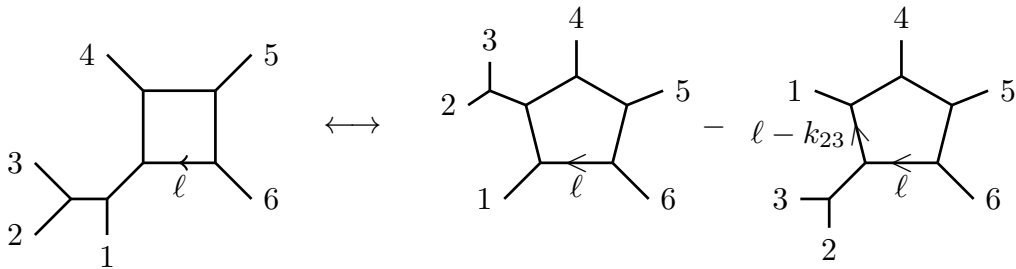
 \tikzpicture [scale=0.8,line width=0.30mm]
 % BCJ box
 \scope[yshift=-4.3cm, xshift=-0.6cm]
 \draw (0,0) -- (1,0) ;
 \draw (0,0) -- (0,1) ;
 \draw (0,1) -- (1,1) ;
 \draw (1,0) -- (1,1) ;
 \draw[->] (0.51,0)--(0.5,0)node[below]{$\ell$};
 \draw (0,0) -- (-0.5,-0.5) ;
 \draw (-0.5,-0.5) -- (-1,-0.5) ;
 \draw (-1,-0.5) -- (-1.5,-1)node[left]{$2$};
 \draw (-1,-0.5) -- (-1.5,0)node[left]{$3$};
 \draw (-0.5,-0.5) -- (-0.5,-1) node[below]{$1$} ;
 \draw (0,1) -- (-0.5,1.5) node[left]{$4$};
 \draw (1,1) -- (1.5,1.5) node[right]{$5$};
 \draw (1,0) -- (1.5,-0.5) node[right]{$6$} ;
 \endscope
 %%%%%%%%%
 \scope[xshift= 4.9cm,yshift=-4.3cm]
 \draw (0.5,0) node{$<$} node[below]{$\ell$};
 \draw (0,0) -- (1,0) ;
 \draw (0,0) -- (-0.2,0.8) ;
 \draw (-0.2,0.8) -- (0.5,1.2) ;
 \draw (1.2,0.8) -- (0.5,1.2) ;
 \draw (1,0) -- (1.2,0.8) ;
 \draw (-2.6,0.5) node{$\longleftrightarrow$};
 \draw (0,0) -- (-0.5,-0.5) node[left]{$1$};
 \draw (1,0) -- (1.5,-0.5) node[right]{$6$};
 \draw (-0.2,0.8) -- (-0.7,1);
 \draw (-0.7,1)-- (-1,0.8) node[left]{$2$};
 \draw (-0.7,1)-- (-0.7,1.4) node[above]{$3$};
 \draw (1.2,0.8) -- (1.7,1)node[right]{$5$} ;
 \draw (0.5,1.2) -- (0.5,1.7)node[above]{$4$} ;
 \endscope
 %%%%%%%%%
 \scope[xshift= 9.4cm,yshift=-4.3cm]
 \draw (0.5,0) node{$<$} node[below]{$\ell$};
 \draw (-0.1,0.4) node[rotate=105]{$>$} node[left]{$\ell-k_{23}$};
 \draw (0,0) -- (1,0) ;
 \draw (0,0) -- (-0.2,0.8) ;
 \draw (-0.2,0.8) -- (0.5,1.2) ;
 \draw (1.2,0.8) -- (0.5,1.2) ;
 \draw (1,0) -- (1.2,0.8) ;
 \draw (-2.1,0.5) node{$- $};
 \draw (0,0) -- (-0.5,-0.5);
 \draw(-0.5,-0.5) -- (-0.9,-0.5) node[left]{$3$};
 \draw(-0.5,-0.5) -- (-0.5,-0.9) node[below]{$2$};
 \draw (1,0) -- (1.5,-0.5) node[right]{$6$};
 \draw (-0.2,0.8) -- (-0.7,1) node[left]{$1$};
 \draw (1.2,0.8) -- (1.7,1)node[right]{$5$} ;
 \draw (0.5,1.2) -- (0.5,1.7)node[above]{$4$} ;
 \endscope
 \endtikzpicture
 \tikzcaption\figcounter{Counterexample for kinematic Jacobi relations at six points.}

In the present context with most propagators linear in $\ell$, however, the numerators for the pentagon diagrams on the right
hand side are both given by differences ${\cal C}_{+|23456|-}(\ell)-{\cal C}_{+|32456|-}(\ell)$ of hexagon
numerators \mastersix. In other words, the numerator of the rightmost diagram is {\it not} given 
by ${\cal C}_{+|45623|-}(\ell{-}k_{23})-{\cal C}_{+|45632|-}(\ell{-}k_{23})$ as one might naively think by 
tracking the momentum in the edge adjacent to leg 1. We exploit that the hexagon numerators
\mastersix\ do not depend on the position of leg 1 in the diagram, and the box numerator on the left hand side of \figcounter\
with leg 1 in a massive corner vanishes accordingly.

\subsubsec{The hexagon anomaly from a partial integrand}

It has been speculated in \MafraGJA\ that the deviations from six-point kinematic Jacobi relations
in the representation of the reference are related to the hexagon anomaly of ten-dimensional SYM.
We will propose a treatment of the anomaly which preserves both the BCJ duality and the KLT relations
for supergravity amplitudes.

At the level of the partial integrand \pintsix, the hexagon anomaly can be seen from the tensor hexagon numerator
${1\over 2} \ell_m \ell_n C^{mn}_{1|2,3,4,5,6}$ whose non-zero BRST variation $\sim \ell_m \ell_n \eta^{mn}$
\MafraGSA\ signals a breakdown of linearized gauge invariance. The gauge variations \MafraNWR
\eqn\anomalyZ{
a(1,2,\ldots,6,-,+)  \, \big|_{e_1\rightarrow k_1} = {   \ell_m \ell_n \eta^{mn} i \epsilon_{10}(k_2,e_2,k_3,e_3,\ldots,k_6,e_6) \over s_{1,\ell} s_{12,\ell} s_{123,\ell} s_{1234,\ell} s_{12345,\ell}}
}
of the partial integrands combine to a rational term in the ten-dimensional color-stripped single-trace
amplitude after undoing the partial-fraction rearrangement of the hexagon:
\eqnn\anomalyY
$$\eqalignno{
A(1,2,\ldots,6)  \, \big|_{e_1\rightarrow k_1}&=     i \epsilon_{10}(k_2,e_2,k_3,e_3,\ldots,k_6,e_6) \int {\dd^{10} \ell \over \ell^2}  \cr
& \ \ \ \ \times \bigg\{
{  \ell_m \ell_n \eta^{mn} \over s_{1,\ell} s_{12,\ell} s_{123,\ell} s_{1234,\ell} s_{12345,\ell}} + {\rm cyc}(1,2,3,4,5,6)
\bigg\}
 \cr
&=   32 i \epsilon_{10}(k_2,e_2,k_3,e_3,\ldots,k_6,e_6) \int \dd^{10} \ell &\anomalyY
\cr
& \ \ \ \ \bigg\{
{  \ell_m \ell_n \eta^{mn} \over \ell^2(\ell{+}k_1)^2 (\ell{+}k_{12})^2  \ldots (\ell {+}k_{12345})^2}
-{  1 \over (\ell{+}k_1)^2 (\ell{+}k_{12})^2  \ldots (\ell {+}k_{12345})^2}
\bigg\} \cr
&=  i \epsilon_{10}(k_2,e_2,k_3,e_3,\ldots,k_6,e_6)  {(2\pi)^5 \over 5!}
}$$
In dimensional regularization with $\dd^{10} \ell  \rightarrow \dd^{10-2\varepsilon} \ell $, the rational result can be understood from the difference between the ten-dimensional
components $\ell_m \ell_n \eta^{mn}$ in the numerator and the $(10{-}2\varepsilon)$-dimensional loop momenta in the
propagators \ChenEVA.

\subsubsec{BCJ and KLT relations in presence of anomalies}

Given that the partial integrand \pintsix\ only exhibits a gauge variation \anomalyZ\ in the first leg but stays invariant
under $e_j\rightarrow k_j$ for the remaining ones $j=2,3,\ldots,6$, it cannot stem from a permutation invariant CHY integrand. Indeed, the
breakdown of permutation symmetry in the string-theory correlator underlying \PSsix\ has been identified as a boundary
term in moduli space \MafraNWR\ which translates into
\eqn\anomalyX{
{\cal K}_6 - ({\cal K}_6 \big|_{1\leftrightarrow 2} ) =  \ell_m \ell_n \eta^{mn}
i\epsilon_{10}(e_1,e_2,k_3,e_3,k_4,e_4,k_5,e_5,k_6,e_6) \ .
}
Strictly speaking, one-loop $(n\geq 6)$-point correlators single out one external leg\foot{In the opening line for the
computation of the superstring amplitudes, one leg enters through the unintegrated vertex operator in the pure-spinor
formalism or in the $-1$ superghost picture in the parity-odd sector of the RNS setup.} which carries the violation of
linearized gauge invariance by a rational term \anomalyY\ \refs{\BerkovitsBK, \MafraNWR}. Keeping track of the singled-out leg $j$ in the correlator through an additional
superscript ${\cal K}_{n}^{(j)}$ (and ${\cal I}_{\rm SYM}^{(j)}$ according to \thumb), partial integrands also need to be
defined with a reference leg,
\eqn\anomalyW{
a^{(j)}(\tau(1,2,\ldots,n,+,-)) =  \int  \dd \mu^{\rm tree}_{n+2} \  \PT(\tau(1,2,\ldots,n,+,-)) \, {\cal I}^{(j)}_{\rm SYM}(\ell) \ ,
}
which is taken to be $j=1$ in the above expressions.
However, this dependence on the reference leg $j$ does not alter the BCJ relations \BCJrelsloop\ and \moreBCJrelsloop\
among $a^{(j)}(\tau(1,2,\ldots,n,+,-))$ with different permutations $\tau$, provided that $j$ is the same for each term in the
BCJ relations: They are a sole consequence of the scattering equations relating the Parke--Taylor factors in \anomalyW,
regardless of the permutation properties of the accompanying ${\cal I}^{(j)}_{\rm SYM}(\ell)$. By a similar argument,
kinematic Jacobi relations are not affected by the dependence \anomalyX\ of the underlying correlators on the reference leg.

Accordingly, there is no obstruction in constructing six-point integrands for ten-dimensional supergravity from the double-copy
of the BCJ numerators \mastersix\ or from the partial integrands \pintsix\ along with the one-loop KLT relations. Regardless
of the relative chirality of the fermions in the two gauge-theory copies, the resulting supergravity is known to have
no hexagon anomaly \AlvarezGaumeIG.

In the context of the double-copy approach, anomaly cancellation suggests that the integrated supergravity amplitude
\eqn\anomalyV{
M_6=
\int  {\dd^{10} \ell \over \ell^2}
\sum_{\rho,\tau \in S_{5}}  \! \!  a^{(j)}(+,\rho(2,3,4,5,6),1,-)  \, S[\rho | \tau]_{\ell}
\, \tilde a^{(j)}(+,\tau(2,3,4,5,6),-,1)
}
does not depend on the choice of the reference leg $j$ in the SYM constituents. It would be interesting to verify this by
explicitly integrating the hexagon contributions along the lines of \anomalyY\ which carry the spurious sensitivity to $j$.

%************************************************************************************************
%************************************************************************************************
\newsec{Bosonic correlators with reduced supersymmetry}
\par\seclab\secREDUCE\
%************************************************************************************************
%************************************************************************************************

\noindent
This section is devoted to explicit and simplified representations of the CHY correlators for half-maximal
and parity-even parts of quarter-maximal SYM. The three- and four-point results of this section coincide with the field-theory
limits of superstring one-loop correlators with reduced supersymmetry, with the results of \refs{\BergWUX,
\BergFUI} as a starting point.

Most of the subsequent expressions for the correlators\foot{The additional normalization factor of $-{1\over 2}$ as compared to \calKred\ is chosen for convenience to arrive at more natural expressions for kinematic factors in \tsix\ and later equations.}
\eqn\reno{
{\cal K}_n^{1/2} \equiv -{1\over 2}\, {\cal K}_n^{{1\over 2}-{\rm SYM}}.
}
are tailored to chiral SYM in six dimensions with eight supercharges. Their dimensional reductions and quarter-maximally 
supersymmetric counterparts in four dimensions are straightforwardly obtained by dropping the parity-odd 
contributions $\sim \epsilon_6$ and rescaling the scalar box numerator in the four-point correlator of section \sixtwo.

%****************
\subsec{Review of Minahaning}
\par\subseclab\sixone
%***************

As a consequence of the spin sums \halfSYM\ and \quarterSYM, parity even parts of $n$-point CHY correlators of 
half-maximal and quarter-maximal SYM are polynomials in $\ell$ and $G_{ij}$ of degree $n{-}2$. The symmetry properties 
of the resulting BCJ master numerators \notmaster\ give rise to triangle and bubble diagrams in the partial integrands 
\analyticPI. This includes bubbles in the external legs as depicted in figure \figmoreminahan, where one of the 
propagator $\sim s^{-1}_{12\ldots n-1} = k_n^{-2}$ formally diverges in the phase space of $n$ massless particles.

\tikzpicture [scale=0.8, line width=0.30mm]
\scope[xshift=3.2cm]
\draw (0,0) -- (-1,1) node[left]{$2$};
\draw (0,0) -- (-1,-1) node[left]{$1$};
\endscope
\draw (3.2,0) -- (4,0);
\draw (3.6,-0.5)node{$s_{12}$};
\draw (4,0) .. controls (4.5,1) and (6.5,1) .. (7,0);
\draw (4,0) .. controls (4.5,-1) and (6.5,-1) .. (7,0);
\draw (5.5,-0.75)node{$<$}node[above]{$\ell$};
\draw (7,0) -- (7.8,0) node[right]{$3$};
\draw (5.5,1.5)node{$s_{12}(e_1 \cdot e_2)(k_1 \cdot e_3)$};
%%%%%%%%%%%%%%%%%
\draw[<->] (9.5,0.25)--(10.7,0.25);
%%%%%%%%%%%%%%%%%
\scope[xshift=9.4cm]
\scope[xshift=4cm]
\draw (0,0) -- (-1,1) node[left]{$2$};
\draw (0,0) -- (-1,-1) node[left]{$1$};
\endscope
\draw (4,0) .. controls (4.5,1) and (6.5,1) .. (7,0);
\draw (4,0) .. controls (4.5,-1) and (6.5,-1) .. (7,0);
\draw (5.5,-0.75)node{$<$}node[above]{$\ell$};
\draw (7,0) -- (7.8,0) node[right]{$3$};
\draw (5.9,1.5)node{$(e_1 \cdot e_2)(k_1 \cdot e_3)$};
\endscope
\endtikzpicture
 \tikzcaption\figmoreminahan{The divergent propagator $s_{ij}^{-1}$ in external bubbles is cancelled by a
 formally vanishing factor of $s_{ij}$ in the kinematic numerator.}

The external-bubble numerators derived from the CHY- or superstring correlators turn out to vanish with $s_{12\ldots n-1}$.
The resulting ``0/0'' indeterminate can be regularized by relaxing momentum conservation in intermediate steps,
following the proposal of Minahan in 1987 \MinahanHA\ and the recent four-point implementation in
\refs{\BergWUX, \BergFUI}. The idea is to use no relation among Mandelstam invariants other than
$\sum_{1\leq i<j}^n s_{ij}=0$ which amounts to a lightlike deformation of momentum conservation
\eqn\minahanA{
k_1 +k_2+\ldots+ k_n = p \ , \ \ \ \ \ \ p^2=0 \ , \ \ \ \ \ \ (e_i \cdot p)=0 \ .
}
In this regularization scheme for infrared divergences, the three-point correlator \exKOStwo\ with the spin
sums \halfSYM, \quarterSYM\ and parity-odd part
\paroddA\ is evaluated as
\eqnn\minahanB
$$\eqalignno{
{\cal K}_3^{1/2} &= \ell_m \big[ e_1^m (e_2\cdot k_3)(e_3\cdot k_2) + (1\leftrightarrow 2,3)  \big] + i\epsilon_6(\ell,e_1,k_2,e_2,k_3,e_3)
\cr
& \ \ + \big[ G_{12} s_{12} (e_1 \cdot e_2) (k_1 \cdot e_3) + {\rm cyc}(1,2,3) \big] \ , &\minahanB
}$$
see \refs{\MinahanHA, \BergWUX, \BergFUI} for the superstring ancestors. The deformation \minahanA\ temporarily assigns
nonzero values such as $s_{12} = {1\over 2} (k_1+k_2)^2 =  {1\over 2} (k_3+p)^2 =(k_3\cdot p)$ to the three-particle
Mandelstam invariants, and the resulting triangle numerators \notmaster\ are given by
\eqnn\minahanC
$$\eqalignno{
&N_{+|123|-}(\ell) =  \ell_m \big[ e_1^m (e_2\cdot k_3)(e_3\cdot k_2) + (1\leftrightarrow 2,3)  \big] + i\epsilon_6(\ell,e_1,k_2,e_2,k_3,e_3)  \cr
& \ \ - {1\over2} \big[
s_{12} (e_1\cdot e_2) (k_1 \cdot e_3)  +s_{13} (e_1\cdot e_3) (k_1 \cdot e_2)  +s_{23} (e_2\cdot e_3) (k_2 \cdot e_1)
\big] \ . &\minahanC
}$$
After dressing with the doubly-partial amplitudes \delicate, all potential divergences from propagators $s_{ij}^{-1}$
are compensated by the numerator factors of $\sim s_{ij}$ in second line. In other words, the limit $p\rightarrow 0$ and thereby $s_{ij}\rightarrow 0$ is taken in the last step of
\eqnn\minahanD
$$\eqalignno{
&a^{1/2}(1,2,3,-,+) = \lim_{s_{ij} \rightarrow0} \lim_{k_{\pm} \rightarrow \pm \ell} \sum_{\rho \in S_3} m^{\rm tree}[+,1,2,3,- | +,\rho(1,2,3),- ]  \, N_{+|\rho(123)|-}(\ell)  \cr
& \ \ =  \lim_{s_{ij} \rightarrow0} \bigg\{ -{1\over 2}  \Big[ {2\over s_{12} s_{12,\ell} }+ {1\over s_{1,\ell} s_{12,\ell} }  \Big] s_{12} (e_1\cdot e_2) (k_1 \cdot e_3) - {1\over 2}  {1\over s_{1,\ell} s_{12,\ell} }  s_{13} (e_1\cdot e_3) (k_1 \cdot e_2)
\Big.
\cr
&\ \  \ \ \ \ \ \ \ \ \ \ \ \  \ \,  -{1\over 2}   \Big[ {2\over s_{23} s_{1,\ell} }+ {1\over s_{1,\ell} s_{12,\ell} } \Big] s_{23} (e_2\cdot e_3) (k_2 \cdot e_1)   +  { \ell_m N^m_{1,2,3} \over s_{1,\ell} s_{12,\ell}} \bigg\} &\minahanD \cr
& \ \ = { \ell_m N^m_{1,2,3} \over s_{1,\ell} s_{12,\ell}}    - {(e_1\! \cdot  \!e_2)(k_1\! \cdot  \!e_3) \over s_{12,\ell}}  - {(e_2\! \cdot  \!e_3)(k_2\! \cdot  \!e_1) \over s_{1,\ell}}  \ ,
}$$
see \delicate\ for the doubly-partial amplitudes.
For the external bubble adjacent to leg 3, the cubic-diagram numerator
$N_{+|123|-}(\ell)-N_{+|213|-}(\ell)=-s_{12} (e_1\cdot e_2) (k_1 \cdot e_3)$ cancels
the divergent propagator\foot{A similar interplay between
divergent propagators and vanishing numerators has been
observed in the four-point four-loop amplitude of maximal SYM \BernUF. Their finite net contribution
plays an important role to obtain the expected UV divergence.} $s_{12}^{-1}$, and the situation is depicted
in  \figmoreminahan. The vector triangle contribution $\ell_m N^m_{1,2,3}$ refers to the first line of \minahanC\
which is unaffected by the limit $s_{ij} \rightarrow 0$.

The analogous discussion with propagators quadratic in $\ell$ can be found in \BergFUI, and
in both the reference and in \minahanD, gauge invariance of the integrands relies on the bubble contributions.
Although the partial-fraction representation of
the external bubbles manifest that they integrate to zero in color-stripped single-trace amplitudes \backto, it would obscure
gauge invariance to drop them at the level of partial integrands.

Before discussing the cancellation of divergent propagators in four-point partial integrands analogous to \minahanD,
we describe the correlator \minahanB\ in a Berends--Giele framework and set the stage for kinematic factors
at higher multiplicity.

%****************
\subsec{Berends--Giele representation of reduced-supersymmetry correlators}
\par\subseclab\sixtwo
%***************

Kinematic factors in maximally supersymmetric correlators are conveniently expressed in terms of the
Berends--Giele currents in \teights\ and their supersymmetrizations. In the same way, the following building blocks
are tailored to describe the polarization dependence in gluonic one-loop amplitudes with reduced supersymmetry \refs{\BergWUX, \BergFUI},
\eqnn\tsix
$$\eqalignno{
t_{A,B} &\equiv - {1\over 2} \cf^{mn}_A \cf_B^{mn}
\cr
t^m_{A,B,C}&\equiv \big[ \ce^m_A t_{B,C}  + (A \leftrightarrow B,C) \big]  + {i \over 4} \epsilon_6^m(\ce_A , \cf_B, \cf_C) &\tsix
\cr
t^{mn}_{A,B,C,D} &\equiv 2 \big[ \ce^{(m}_A \ce^{n)}_B t_{C,D}  + (A,B| A,B,C,D)   \big]
+{i \over 2} \big[  \ce^{(m}_B \epsilon_6^{n)}(\ce_A , \cf_C, \cf_D)  + (B \leftrightarrow C,D) \big]\,.
}$$
By inserting the recursive definitions \recone\ and \rectwo\ of the Berends--Giele currents $\ce^m_A$ and $\cf^{mn}_B$, the kinematic factor of the external bubble in \figmoreminahan\ is reproduced by
\eqnn\tsixexample
$$\eqalignno{
t_{12,3} &= (e_1\cdot e_2) (k_1 \cdot e_3) \ . &\tsixexample
}$$
The cancellation of the pole $\cf^{mn}_{12} \sim s_{12}^{-1}$ in $t_{12,3}$ follows from the infrared regularization scheme
in \minahanA. One can analogously show that the four-point scalars $t_{12,34}$ and $t_{123,4}$
only have simple poles in $s_{ij}$ \BergFUI\ in spite of the spurious pole structure $\sim (s_{ij}s_{123})^{-1}$
of $\cf^{mn}_{123}$.

In complete analogy to their maximally supersymmetric counterparts \scalarBRST\ and \tensC, the bosonic building blocks
\tsix\ enter one-loop amplitudes in their gauge invariant combinations \refs{\BergWUX, \BergFUI}
\eqnn\redinv
$$\eqalignno{
C^{1/2}_{1|23} &\equiv t_{1,23} + t_{12,3} - t_{13,2}\,, \cr
C^{1/2}_{1|234} &\equiv t_{1,234} + t_{12,34} + t_{123,4} - t_{124,3} - t_{14,23} - t_{142,3} + t_{143,2}\cr
C^{m,1/2}_{1|2,3} &\equiv t^m_{1,2,3} +  k_2^m t_{12,3}+  k_3^m t_{13,2}
\,, &\redinv\cr
C^{m,1/2}_{1|23,4} &\equiv
t^m_{1,23,4} + t^m_{12,3,4} - t^m_{13,2,4}
+k^m_3 t_{123,4} - k^m_2 t_{132,4} + k^m_4 \big[  t_{14,23} - t_{214,3} + t_{314,2} \bigr]
 \ .
}$$
The parity-odd part of the tensorial generalization
\eqn\redpseudo{
C^{mn,1/2}_{1|2,3,4} \equiv t^{mn}_{1,2,3,4} + 2 \big[ k_2^{(m} t^{n)}_{12,3,4}
+ (2\leftrightarrow 3,4) \big]- 2 \big[ k_2^{(m} k_3^{n)} t_{213,4} + (2,3|2,3,4) \big]
}
which will appear in a box numerator gives rise to an anomalous gauge variation
\eqn\boxanom{
C^{mn,1/2}_{1|2,3,4}  \, \big|_{e_1\rightarrow k_1} = 2i \eta^{mn} \epsilon_6(k_2,e_2,k_3,e_3,k_4,e_4) \ , \ \ \ \ \
C^{mn,1/2}_{1|2,3,4}  \, \big|_{e_j\rightarrow k_j} = 0 \ , \ \ \  j=2,3,4 
}
analogous to \anomalyZ\ due to the ten-dimensional tensor hexagon in pure-spinor superspace.
The kinematic factors \redinv\ and \redpseudo\ have been noticed in the simplification of the superstring
correlators \BergWUX\ as well as the resulting field-theory limits \BergFUI, and the scalar instances coincide with the tree-level amplitudes,
$C^{1/2}_{1|23\ldots n} = A^{\rm tree}(1,2,3,\ldots,n)$.

In terms of the kinematic variables in \tsix, the three-point correlator \minahanB\ in half-maximally supersymmetric SYM
and its four-point counterpart are given by
\eqnn\halfthree
\eqnn\halffour
$$\eqalignno{
{\cal K}_3^{1/2} &= \ell_m t^m_{1,2,3} + [ X_{12} t_{12,3} + {\rm cyc}(1,2,3)  ]   &\halfthree \cr
{\cal K}_4^{1/2} &= {1 \over 2} \ell_m \ell_n t^{mn}_{1,2,3,4} + \ell_m [X_{12}  t^m_{12,3,4} + (1,2|1,2,3,4) ]  \cr
& \ \ +  [X_{12}(X_{13}+X_{23}) t_{123,4} + X_{13}(X_{12}+X_{32}) t_{132,4} + (4\leftrightarrow 3,2,1) ]  \cr
& \ \ + [ X_{12} X_{34} t_{12,34} + {\rm cyc}(2,3,4)] + {1\over 4} t_8(\cf_1,\cf_2,\cf_3,\cf_4)  \ ,&\halffour
}$$
see \BergWUX\ for the superstring antecedent of the latter with the loop momentum integrated out. Following the
spin sums in \halfSYM\ and \twoplustwo, the relative factor between the last term $t_8(\cf_1,\cf_2,\cf_3,\cf_4)$ and the remaining correlator depends on the particle content, also see section 5.2 of \BergFUI\ for a discussion in a string-theory context.
The use of scattering equations explained in section \sectfourfive\ leads to the manifestly gauge invariant rewritings
\eqnn\halfgthree
\eqnn\halfgfour
%\eqnn\halfgfive
$$\eqalignno{
{\cal K}_3^{1/2} &=  \ell_m  C^{m, 1/2}_{1|2,3} + X_{23} C^{1/2}_{1|23}
 &\halfgthree \cr
{\cal K}_4^{1/2} &=
 {1 \over 2} \ell_m \ell_n C^{mn, 1/2}_{1|2,3,4} + \ell_m [X_{23}  C^{m,1/2}_{1|23,4} + {\rm cyc}(2,3,4) ]  &\halfgfour \cr
& \ \ + [  X_{23} X_{34} C^{1/2}_{1|234}-X_{23} X_{24} C^{1/2}_{1|324}-X_{24} X_{34} C^{1/2}_{1|243}]  - {1\over 4} s_{23} s_{34} C^{1/2}_{1|234}  \ ,
 \cr
}$$
and we have rewritten the last term using $t_8(\cf_1,\cf_2,\cf_3,\cf_4)= - s_{23} s_{34} C^{1/2}_{1|234}$. 
The manipulations in section \parodd\ and appendix \appbasis\ allow to express higher-multiplicity correlators
in a similar basis of functions. Note the close structural similarity to the maximally supersymmetric five- and 
six-point correlators in \PSfive\ and \PSsix.

%****************
\subsec{BCJ numerators and partial integrands with reduced supersymmetry}
\par\subseclab\sixthree
%***************

The correlators \halfgthree\ and \halfgfour\ translate into the following gauge invariant BCJ master numerators
\eqnn\masterhalfthree
\eqnn\masterhalffour
$$\eqalignno{
{\cal C}^{1/2}_{+|\rho(2,3)|-}(\ell) &= \ell_m C^{m,1/2}_{1|2,3} + {1\over 2} s_{23}\sign_{23}^\rho C^{1/2}_{1|23} &\masterhalfthree \cr
{\cal C}^{1/2}_{+|\rho(2,3,4)|-}(\ell) &= {1\over 2} \ell_m\ell_n C^{mn,1/2}_{1|2,3,4} + {1\over 2} \ell_m \big[ s_{23}\sign_{23}^\rho C^{m,1/2}_{1|23,4} + {\rm cyc}(2,3,4)\big]  - {1\over 4} s_{23} s_{34} C^{1/2}_{1|234} &\masterhalffour \cr
&\! \! \! \! \! \! \! \! \! + {1\over 4} \big[ s_{23}\sign_{23}^\rho s_{34}\sign_{34}^\rho C^{1/2}_{1|234}
-s_{23}\sign_{23}^\rho s_{24}\sign_{24}^\rho C^{1/2}_{1|324}
-s_{24}\sign_{24}^\rho s_{34}\sign_{34}^\rho C^{1/2}_{1|243}\big]
}$$
for triangle- and box diagrams, respectively. These numerators result in the following expressions for
the three- and four-point partial integrands \HeMZD\
\eqnn\partthreepoint
\eqnn\partfour
$$\eqalignno{
&a^{1/2}(1,2,3,-,+)= { \ell_m C^{m,1/2}_{1|2,3} \over s_{1,\ell} s_{12,\ell} } - { C^{1/2}_{1|23} \over s_{1,\ell} } &\partthreepoint
\cr
 & \ \ \ \ = {\ell_m \big[ e^m_1 (k_2\cdot e_3)(k_3\cdot e_2) + (1\leftrightarrow 2,3) \big] +  \big[ (\ell \cdot k_2) (e_1\cdot e_2)(k_1\cdot e_3) + (2\leftrightarrow 3) \big]  \over s_{1,\ell} s_{12,\ell} } \cr
& \ \ \ \ \ \ \ \ +
{ i\epsilon_6(\ell,e_1,k_2,e_2,k_3,e_3) \over s_{1,\ell} s_{12,\ell} }
+ {(e_1\cdot e_2)(k_1 \cdot e_3)+(e_2\cdot e_3)(k_2 \cdot e_1)+(e_1\cdot e_3)(k_3 \cdot e_2) \over s_{1,\ell} }
\cr
%%%%%
%%%%%
&a^{1/2}(1,2,3,4,-,+)=  {  C_{1|234} ^{1/2}\over s_{1,\ell}}
-{ \ell_m C^{m,1/2}_{1|23,4} \over s_{1,\ell} s_{123,\ell}}
-{ \ell_m C^{m,1/2}_{1|34,2} \over s_{1,\ell} s_{12,\ell}}
-{ s_{23} s_{34} C^{1/2}_{1|234} \over  2 s_{1,\ell} s_{12,\ell} s_{123,\ell} }
 \cr
&  \ \ \ \ \ \ \ \   + {\ell_m \ell_n C^{mn,1/2}_{1|2,3,4} - \ell_m\big[ s_{23} C^{m,1/2}_{1|23,4} + s_{24} C^{m,1/2}_{1|24,3} + s_{34} C^{m,1/2}_{1|34,2}\big]  \over 2 s_{1,\ell} s_{12,\ell} s_{123,\ell}  }  &\partfour
%+{1\over 2} t_8(1,2,3,4)
}$$
cf.\ \minahanD\ for the three-point case. We have exploited permutation invariance of $s_{23}s_{34}C_{1|234} ^{1/2}$ to identify the scalar box numerator in the first line of \partfour, and its prefactor $-{1\over 2}$ is specific to the spin sum \halfSYM\ of a single vector multiplet in the loop. In presence of $n_{\rm vec}$ vector multiplets and $n_{\rm hyp}$ hypermultiplets, the partial integrand generalizes to
\eqnn\partfourgen
$$\eqalignno{
&a^{1/2}_{n_{\rm vec},n_{\rm hyp}}(1,2,3,4,-,+)=  \Big(  n_{\rm vec}   + { n_{\rm hyp} \over 2} \Big) \bigg\{
  {  C_{1|234} ^{1/2}\over s_{1,\ell}}
-{ \ell_m C^{m,1/2}_{1|23,4} \over s_{1,\ell} s_{123,\ell}}
-{ \ell_m C^{m,1/2}_{1|34,2} \over s_{1,\ell} s_{12,\ell}}
&\partfourgen \cr
&  \ \ \ \    + {\ell_m \ell_n C^{mn,1/2}_{1|2,3,4} - \ell_m\big[ s_{23} C^{m,1/2}_{1|23,4} + s_{24} C^{m,1/2}_{1|24,3} + s_{34} C^{m,1/2}_{1|34,2}\big]  \over 2 s_{1,\ell} s_{12,\ell} s_{123,\ell}  }  \bigg\}   -{n_{\rm vec}  s_{23} s_{34} C^{1/2}_{1|234} \over  2 s_{1,\ell} s_{12,\ell} s_{123,\ell} } 
%+{1\over 2} t_8(1,2,3,4)
}$$
by virtue of the additional spin sum \twoplustwo, and the vanishing of the scalar box numerator with $n_{\rm vec}$ has been noticed in \BergFUI.

%In identifying the scalar box numerator in the first line of \partfour, we have used that $s_{23}s_{34}C_{1|234} ^{1/2}$ is permutation symmetric in $2,3,4$.

Note that Kleiss--Kuijf relations \Kleiss\ imply the vanishing of non-planar partial integrands at three points,
\eqn\nothreept{
a^{\beta-{\rm SYM}}(1,2,-,3,+) = 0 \ , \ \ \ \ \ \ \beta=1,{1\over 2},{1\over 4} \ .
}
Accordingly, the three-point supergravity integrand from the KLT formula \oneloopKLT\ involving at least one supersymmetric
gauge-theory copy $\beta=1,{1\over 2},{1\over 4} $ is identically zero,
\eqn\threeptKLT{
m^{{\rm (S)YM} \otimes \beta-{\rm SYM}}_3(\ell)  = \! \! \sum_{\rho,\tau \in S_{2}}  \! \!  a^{{\rm (S)YM}}(+,\rho(2,3),1,-)  \, S[\rho | \tau]_{\ell}
\, \tilde a^{\beta-{\rm SYM}}(+,\tau(2,3),-,1) = 0
\ .
}
At four points, the anomalous gauge variation \boxanom\ of the tensor building block yields
\eqn\anomalyZZ{
a^{1/2}(1,2,3,4,-,+)  \, \big|_{e_1\rightarrow k_1} = { \ell_m \ell_n \eta^{mn} i \epsilon_{6}(k_2,e_2,k_3,e_3,k_4,e_4) \over s_{1,\ell} s_{12,\ell} s_{123,\ell} }
}
in analogy to \anomalyZ. For a six-dimensional color-stripped single-trace
amplitude, one can follow the manipulations of \anomalyY\ to undo the
partial-fraction rearrangement of the box and to identify the anomaly as a purely rational term:
\eqnn\anomalyYY
$$\eqalignno{
A^{1/2}(1,2,3,4)  \, \big|_{e_1\rightarrow k_1}&=    i \epsilon_{6}(k_2,e_2,k_3,e_3,k_4,e_4)  {(2\pi)^3 \over 3!} \ .
&\anomalyYY
}$$
The representation of $A^{1/2}(1,2,3,4)$ constructed in \BergFUI\ from gauge invariance and locality is equivalent to
the partial integrand \partfour, but it was observed in the reference to deviate from the BCJ duality. By the arguments
of section \fivefive, organizing the loop integrand in terms of cubic diagrams with propagators linear in $\ell$ (cf.\ section \whymaster) alleviates the task of finding BCJ numerators. Hence, there is no contradiction in presenting BCJ master
numerators \masterhalffour\ in terms of the same building blocks seen in the BCJ violating setup of \BergFUI\ since
the cubic diagrams in the reference were tailored to propagators quadratic in $\ell$.

Similar to the maximally supersymmetric six-point correlator, the anomalous four-point correlator \halfgfour\ also violates
permutation invariance, cf.\ \anomalyX. Following the reasoning around \anomalyW, a fully accurate labelling of the partial
integrand \partfour\ would involve an additional superscript $a^{1/2}(1,2,3,4,-,+) \rightarrow a^{1/2,(j=1)}(1,2,3,4,-,+)$
to indicate that linearized gauge invariance is violated in the $j^{\rm th}$ leg, see \boxanom. Finally, the dependence on $j$ is expected to disappear after integrating the supergravity amplitude from the one-loop KLT formula \oneloopKLT\ over $\ell$.

%************************************************************************************************
%************************************************************************************************
\newsec{Conclusions}
%************************************************************************************************
%************************************************************************************************

In this paper we studied new BCJ representations of one-loop scattering amplitudes in supersymmetric gauge-theory and gravity amplitudes, which are largely inspired by both the CHY/ambitwistor-string formulation and superstring theory. Based on the CHY-inspired representation for supersymmetric amplitudes, we give a general proof of one-loop BCJ and KLT relations for the {\it partial integrands} proposed in \HeMZD. In the RNS incarnation of this new representation, we bring one-loop correlators on a nodal Riemann sphere into a form which makes BCJ numerators accessible for all multiplicities.  The method works for external bosons in presence of any nonzero number of supercharges and for both parity-even and parity-odd sectors. Moreover, from the field-theory limit of pure-spinor superstrings, we supersymmetrized the $(n\leq 6)$-point BCJ numerators to include external fermions as well.

We would like to highlight three intriguing features of our results. First, the manifestly gauge- and diffeomorphism-invariant BCJ and KLT relations can be proved solely based on structural results on one-loop CHY formulae, without referring to the explicit form of the BCJ numerators. Second, correlators with maximal and reduced supersymmetry are shown to be degree-$(n{-}4)$ and degree-$(n{-}2)$ polynomials in loop momentum $\ell$ and the Green function on the nodal sphere, manifesting the powercounting of $\ell$ including the no-triangle property for maximal supersymmetry. Last but not least, since we naturally obtain one-loop amplitudes with linear propagators, our BCJ numerators satisfy the color-kinematics duality in a slightly different organization scheme of cubic diagrams as compared to its original loop-level formulation \BernUE, see section \whymaster. However, to our best knowledge, this is the first $D$-dimensional, all-multiplicity control of one-loop BCJ numerators which can be directly double copied to give supergravity integrands.

Although we have only considered supersymmetric gauge theories and gravity, we expect our results to hold for non-supersymmetric theories as well. Besides, our main results naturally apply to other theories as well: the one-loop KLT formula with the NLSM and (super-)Yang--Mills theory yields integrands of Born--Infeld theory along with supersymmetric extensions to Dirac--Born--Infeld--Volkov--Akulov theories. As will be elaborated elsewhere, the one-loop amplitude relations for EYM partial integrands~\HeMZD\ can be proved using CHY representations. Using explicit results for the correlators, one can obtain BCJ numerators for one-loop amplitudes of the NLSM and for EYM in a similar way.

There are several directions to investigate in the future. Already at one loop, it would be highly desirable to determine higher-point supersymmetric correlators from the field-theory limit of the pure-spinor formalism. We expect the results to be expanded in a basis of worldsheet functions as explained in section 4, with coefficients given by BRST pseudo-invariants, which have been studied in \MafraGSA. Moreover, it would be interesting to incorporate $\alpha'$-corrections of the superstring using the same approach and to study one-loop BCJ numerators and KLT relations for amplitudes from higher-dimensional operators as well as those in Z-theory \refs{\CarrascoLDY, \MafraMCC, \CarrascoYGV}.

A particularly exciting direction is to generalize the new BCJ representations and their applications to higher loops. For example, a natural follow-up question is how to construct BCJ numerators and derive KLT formulae at higher loops. We expect that a strategic path forward is to again organize $g$-loop correlators on the nodal Riemann spheres in terms of Parke--Taylor factors with $g$ pairs of double points $\sigma_{\pm}$. Although a systematic study such higher-loop correlators, KLT relations and BCJ numerators will be given in the future, we would like to display the two-loop four-point correlator as an encouraging example.

%*******************************************
\subsec Preview example: The two-loop four-point correlator on the nodal sphere

A central ingredient of genus-$g$ correlators are the global holomorphic one-forms $\omega_J$ with $J=1,2,\ldots,g$ which degenerate as follows on nodal Riemann spheres:
\eqn\deltwozero {
\omega_{J}(\sigma_i) = { ( \sigma_{J+}- \sigma_{J-}) \, \dd \sigma_i \over ( \sigma_i - \sigma_{J+}) ( \sigma_i - \sigma_{J-}) } \ .
}
They enter the genus-two superstring correlators of \DHokerVCH\ through the antisymmetric combinations
\eqn\deltwo{
\Delta_{i,j} \equiv \omega_1(\sigma_i)\omega_2(\sigma_j)
-\omega_2(\sigma_i)\omega_1(\sigma_j) = \varepsilon^{IJ} \omega_I(\sigma_i)\omega_J(\sigma_j) \ ,
}
in lines with modular invariance. Moreover, the moduli-space measure introduces differences of the double points $\sigma_{1\pm}$ and $\sigma_{2\pm}$ into the correlator on the nodal sphere \GeyerWJX,
\eqn\Yfull{
\prod_{j=1}^{4} \dd\sigma_j \, {\cal I}_4^{\rm 2-loop} =     { s_{12} \Delta_{4, 1} \Delta_{2, 3} +
   s_{23} \Delta_{1, 2} \Delta_{3, 4}  \over (\sigma_{1+}{-}\sigma_{2+})(\sigma_{1+}{-}\sigma_{2-})(\sigma_{1-}{-}\sigma_{2+})(\sigma_{1-}{-}\sigma_{2-})
   %(1^+ 2^-)(2^- 1^-)(1^- 2^+)(2^+ 1^+)
   } \ ,
}
where the overall kinematic factor $ t_8(f_1,f_2,f_3,f_4)$ is suppressed.
This result can be expanded in terms of eight-point Parke Taylor factors involving $\sigma_{5,6}\equiv \sigma_{1\pm}$ and $\sigma_{7,8}\equiv \sigma_{2\pm}$:
\eqnn\twofour
$$\eqalignno{
 {\cal I}_4^{\rm 2-loop}  &=  s_{12}\big[\PT(7,1,2,5,3,4,6,8) + \PT(7,1,2,6,3,4,5,8)\big] &\twofour \cr
&\ \ +s_{23} \big[ \PT(7,1,5,2,3,6,4,8) +\PT(7,1,6,2,3,5,4,8) \big] \cr
& \ \ + s_{12} \big[\PT(7,5,1,2,6,3,4,8) + \PT(7,6,1,2,5,3,4,8)\big]
+ {\rm perm}(1,2,3,4)  \ .
}$$
Based on this 144-term sum, it would be very interesting to study two-loop KLT formulae as well as BCJ numerators, for maximally supersymmetric Yang-Mills and gravity amplitudes. Of course, more work is needed to obtain the parental string correlators at higher multiplicity and loop order as well as reduced supersymmetry for generic points in the moduli space of the relevant Riemann surface.

%************************************************************
\bigskip \noindent{\bf Acknowledgements:} We are indebted to Carlos Mafra for a variety of
enlightening discussions, his participation in parts of
the project -- in particular his proof of equation \anyGij\ -- and
valuable comments on a draft. We would like to thank Nima Arkani-Hamed, Marcus Berg, Freddy Cachazo, Hao Fu, Xiangrui Gao, Yvonne Geyer, Ricardo Monteiro, Ellis Yuan and Minshan Zheng for combinations of useful discussions and helpful comments on a draft. This research was supported in part by the National Science Foundation under Grant No.\ NSF PHY11-25915, and we are grateful to the KITP Santa Barbara as well as the organizers of the workshop ``Scattering Amplitudes and Beyond'' for providing stimulating atmosphere, support and hospitality. S.H.'s research is supported in part by the Thousand Young Talents program and the Key Research Program of Frontier Sciences of CAS. Y.Z.'s research is partly supported by NSFC Grants No. 11375026 and 11235003.

%*************************************************
\appendix{A}{One-loop basis of worldsheet functions}
\applab\appbasis
%*************************************************

\noindent The goal of this appendix is to arrive at a basis of worldsheet
functions for field-theory amplitudes at one loop. Following the discussion of section \sectfourfive,
we will describe how to achieve this in two steps:
\smallskip
\item{1.} Eliminating all subcycles of propagators $G_{a_1a_2}G_{a_2a_3} \ldots G_{a_n a_1}$
\item{2.} Eliminating the dependence on the position of leg $1$ from
any $G_{ij}$

%*******************************************
\subsec Eliminating subcycles of propagators

Since multiple subcycles can be recursively reduced to cases with fewer subcycles, it is sufficient to
consider the case with one subcycle, say
$G_{a_1a_2}G_{a_2a_3}\cdots G_{a_m a_1}$. The algorithm to break
it open selects a subset of its propagators
(therefore this is not a cycle by itself) and rewrites it in a
basis of ``IBP functions'' $X_{a_1a_2\ldots a_m}$ defined in \xxx. For example,
\eqnn\ggxx
\eqnn\ggxxII
$$\eqalignno{
G_{12}G_{13}&={1\over s_{123}}\Bigg\{{s_{23}\over 4}
+{X_{123}\over s_{12}}+{X_{132}\over s_{13}}\Bigg\}\,, &\ggxx\cr
G_{12}G_{13}G_{14}&={1\over s_{1234}}\Bigg\{
\Big[{X_{1234}\over s_{12}s_{123}}+{\rm symm}(2,3,4)\Big] &\ggxxII\cr
&\;\;+\Big[{s_{34}\over 4}\Big({1\over s_{12}}
+{1\over s_{134}}\Big)X_{12} + {1\over 4}\Big(
{s_{24}\over s_{124}} -
{s_{34}\over s_{134}}\Big)
X_{23} + {\rm cyc}(2,3,4)\Big]\Bigg\}\,,
}$$
and such inverse relations exist for any monomial of propagators
without subcycles. One can check
these relations by plugging in \xxx\ and by using the Fay identity
\pfoneloop. Alternatively, we will sketch how to derive such
relations below.

For example, to break the subcyle $G_{12}G_{23}G_{13}$ we
rewrite $G_{12}G_{13}$ in terms of IBP functions as shown
in \ggxx. Since both labels $2$ and $3$
appear in  $X_{123}$ and $X_{132}$, one uses an IBP relation to
rewrite $X_{123}=X_{12}(X_{34}+X_{35}+\cdots X_{3n}+\ell\cdot k_3)$
and similarly,
$X_{132}=X_{13}(X_{24}+X_{25}+\cdots X_{2n}+\ell\cdot k_2)$.
Then we have no subcycles left:
\eqn\gggx{
G_{12}G_{23}G_{13}={G_{23}\over s_{123}}\bigg\{{s_{23}\over 4}
+{G_{12}\Big(\sum_{p=4}^ns_{3p}G_{3p}+\ell\cdot k_3\Big)
+{G_{13}}\Big(\sum_{p=4}^ns_{2p}G_{2p}+\ell\cdot
k_2\Big)}\bigg\}\,,
}
and the same idea can be applied to any other subcycle. Apart from loop momenta and terms
with fewer powers of $G_{ij}$ (such as the contribution of ${1\over 4}s_{23}G_{23}$ in \gggx)
which are intrinsic to genus one,
the elimination of $G_{a_1a_2}G_{a_2a_3}\cdots G_{a_m a_1}$ largely follows the tree-level
techniques to address products of Parke--Taylor factors (see e.g.\ \CachazoNWA).

After eliminating all subcycles, we are left with products of $X$
functions with overlapping labels, such as $X_{\cdots a_1
\cdots}G_{a_ma_1}$ and $G_{12}G_{23}G_{3p}$ with $p=4,\cdots,n$
above. By using \ggxx, \ggxxII\ and generalizations, we can again rewrite them in terms of
products of functions without overlapping labels, which are suitable
for integration by parts. One
therefore obtains a polynomial of IBP functions $X_{a_1a_2\ldots a_m}$ where in
every monomial each particle label appears at most once as a subscript.

\subsec{Eliminating the dependence on $\s_1$}

After eliminating subcycles, the resulting $X$
functions are not yet linearly independent; it is
straightforward to see that one can still eliminate
their dependence on particle $1$ using scattering equations,
\eqn\eqthree{
X_{123}=X_{12} (X_{13}+X_{23})=(X_{2 3}+\ldots +X_{2 n}+ k_2 \cdot \ell)
(X_{34}+\ldots + X_{3 n} + k_3 \cdot \ell)\,,
}
where, as we mentioned, no subcycle will appear and again we recast
e.g. $X_{23} X_{34}$ into $X_{234}$ and $X_{243}$ using e.g.\ \ggxx.
By repeating this process we obtain a basis of $X_{a_1a_2\ldots a_m}$ functions where
particle $1$ is eliminated. Moreover,
one can always fix the first subscript of $X$ to be the
smallest\foot{This follows from the fact that \xxx\ satisfies Lie
symmetries \MafraKH.% In general, $X_{PiQ} = - X_{i\ell(P)Q}$ where
%$\ell(P)$ denotes the Dynkin bracket of $P$ \reutenauer.}
}, for example $X_{342} = - X_{234} + X_{243}$.

There is a straightforward way to count the degree directly in terms
of $X$ functions. It is convenient to introduce $X_i\equiv 1$ ($i
\neq 1$) for labels that did not appear in a monomial such that
after inserting them, each label $2,\ldots, n$ appears exactly once.
For example, we write the identity $1=\prod_{i=2}^n X_i$, $X_{23} X_{45}= X_{23}
X_{45} X_6$ for $n=6$, and $X_{234} X_{56} = X_{234} X_{56} X_7
X_8$ for $n=8$ etc.. After inserting these
identities, we see that the degree of $G_{ij}$'s is given by
$n{-}1$ minus the total number of $X$ functions. In the examples
above, the degree is $0$, $5-3=2$ and $7-4=3$, respectively.

Now we are ready to count the number of basis elements of IBP
friendly functions, for $n$ points with a given degree in $G_{i
j}$'s. Since label $1$ is eliminated, the number of independent
monomials in $X$ functions with degree $0\leq k\leq n{-}2$ is given by
the number of ways to distribute $n{-}1$ labels into $n{-}k{-}1$
disjoint, non-empty sets, where labels in each set form a cycle
(including length-$1$ cycles). The solution to this counting problem is known as the Stirling
number of the first kind, $S_{n{-}1, n{-}k{-}1}$, see table 1. For
example, for $k=0$, $S_{n{-}1,n{-}1}=1$ corresponds to the identity
$1$. For $k=1$, choosing $n=4$ and $n=5$ allows for $3$ and $6$ elements
$X_{2 3}, X_{2 4}, X_{3 4}$ and $X_{2 3}, X_{2 4}, X_{2 5},\ldots, X_{4
5}$, respectively. Finally, $k=2$, $n=4$ gives rise to the 2 basis elements $X_{2 3 4}$ and
$X_{2 4 3}$.

For correlators with reduced supersymmetry, the degree of the polynomial in $\ell$ and $G_{ij}$ is $n{-}2$,
thus the total number of basis elements for $n$ points is given by
$\sum_{k=0}^{n{-}2} S_{n{-}1, n{-}k{-}1}=(n{-}1)!$ (see table 1 in section \sectfourfive). For example,
for $n=5$, in addition to the elements with $k=0,1$ above, we have
11 elements for $k=2$: $X_{2 3}X_{4 5}, \, X_{2 4}X_{3 5}, \, X_{2 5}X_{3
4}$ and $X_{2 3 4}, \, X_{2 4 3}$ along with their images under ${\rm cyc}(2,3,4,5)$.
Finally, $n=5$ and $k=3$ introduces the six permutations of $X_{2 3 4 5}$ in
$3,4,5$ which altogether yields $1+6+11+6=24$ basis elements at $n=5$.

Maximal supersymmetry allows for maximum degree $n{-}4$ in $X$ functions, thus the total number of
basis elements is $\sum_{k=0}^{n{-}4} S_{n{-}1, n{-}k{-}1}\equiv a_n$ (see table 1).
Here $a_n$ counts the number of $(n{-}1)$-permutations with at least 3 cycles
(sequence {\bf A067318} of \oeis), e.g. $a_4=1$, $a_5=7$ and $a_6=46$. For example, the $7$-element
basis for $n=5$ consists of $1$ (along with $\ell$) as well as $X_{23}, X_{24}, X_{25},X_{34}, X_{35}, X_{45}$.
For $n=6$, we have $1$ (along with $\ell^2$), $X_{2 3}, \, X_{2 4}, \ldots, X_{5 6}$ (along with $\ell$) as well as $X_{2 3}X_{4 5}, \, X_{2 4}X_{3 5}, \, X_{2 5}X_{3 4}$
plus $(2345\leftrightarrow 2346,2356,2456,3456)$ and  $X_{234}, \, X_{243}$ plus
$(234 \leftrightarrow 235,236, \ldots, %245,246,256,345,346,356,
456)$, altogether 46 elements.

%******************************************************
\appendix{B}{Combinatoric proof of the formula \anyGij.}
\applab\appGij
%******************************************************

\noindent In this appendix\foot{This
appendix was written by Carlos Mafra.} we prove the formula \anyGij, namely
\eqn\equivGij{
G_{i_1i_2} \ldots G_{i_{2p-1} i_{2p}}\prod_{j=1}^n{1\over \sigma_j} =
{(-1)^n\over 2^p}\sum_{\rho\in S_n}\sign^\rho_{i_1i_2}
\ldots\sign^\rho_{i_{2p-1}i_{2p}}\cZ_{0\rho(1,2, \ldots,n)} \ ,
}
where $\s_0\equiv \s_+\equiv 0$, and $\cZ_P$ was defined in
\calZdefFirst. For convenience, define the shorthands
\eqn\SigmaP{
\Sigma_{123 \ldots n} \equiv {1\over \s_1\s_2 \ldots \s_n},
\qquad
\Sigma^{i}_{123 \ldots n} \equiv \s_i \Sigma_{123 \ldots n}
={1\over \s_1\s_2 \ldots
\widehat\s_i \ldots \s_n} \ ,
}
where $\hat\s_i$ denotes the absence of $\s_i$, and the generalization to
multiparticle $\Sigma^Q_P$ is obvious. Note
that $\Sigma^Q_P$ is totally symmetric in $P$ and $Q$.
Recalling the auxiliary variable $\s_0=0$
and denoting a sum over permutations of the indices
in $P$ by $(P)$
one can show that\foot{The proof \elemII\ is as follows:
$\cZ_{0(Pj)}(2\cZ_{jk}+\cZ_{k0}) =  2 \cZ_{kj0(P)} + \cZ_{k0(Pj)}
= 2\cZ_{0\{jk\shuffle (P)\}}- \cZ_{0\{k\shuffle (jP)\}} =
\cZ_{0\{jk\shuffle (P)\}} - \cZ_{0\{kj\shuffle (P)\}}
=\sign_{jk} \cZ_{0(jkP)}$,
since one factor of $\cZ_{0\{jk\shuffle (P)\}}$ is cancelled by the
permutations in $- \cZ_{0\{k\shuffle (jP)\}}$
in which the labels $j$ and $k$ are in the same order as $jk$.
Also note that $\cZ_{0(Pj)} = - \cZ_{j0(P)}$ was used in the first
equality above.
%(recall that $\cZ_{j0(P)} = -
%\cZ_{0(Pj)}$ and that $(P)$ denotes permutations of $P$).
},
\eqnn\SigmaZ
\eqnn\lemma
\eqnn\elemI
\eqnn\elemII
$$\eqalignno{
\Sigma^Q_P &= (-1)^{\len{P}-\len{Q}}\cZ_{0(P\setminus Q)}\,,&\SigmaZ\cr
\cZ_{0(Q)}\Sigma^Q_P
%= (-1)^{\len{P}-\len{Q}} \cZ_{0(Q)}\cZ_{0(P\setminus Q)} = (-1)^{\len{P}-\len{Q}}
%\cZ_{(Q)0(P\setminus Q)}
&= (-1)^{\len{P}-\len{Q}}\cZ_{0(P)}\,&\lemma\cr
\cZ_{0(P)}\cZ_{0(Q)} &= \cZ_{0(PQ)} &\elemI\cr
\cZ_{0(Pj)}(2\cZ_{jk}+\cZ_{k0}) &= \cZ_{0(jkP)}{\rm sign}_{jk},
\quad\hbox{if $k\cap P =
\emptyset$}\,  , &\elemII\cr
}$$
where we used that
$\cZ_{PiQ} = (-1)^\len{P}\cZ_{i\tilde P\shuffle Q}$. Note that
\SigmaZ\ is also valid when $Q=\emptyset$, i.e.,
$\Sigma_{123 \ldots n} = (-1)^n
\cZ_{0\{1\shuffle 2\shuffle 3 \shuffle \ldots \shuffle n\}}$.

Now let us consider products of $G_{ij}\Sigma_P^Q$ with a single or
no overlap between $ij$ and $Q$:
\eqnn\GSigI
\eqnn\GSigII
$$\eqalignno{
G_{12}\Sigma_P &= G_{12}\cZ_{10}\cZ_{20}\Sigma^{12}_P =
-\half(\cZ_{120}-\cZ_{210})\Sigma^{12}_P
= \half {\rm sign}_{12}\cZ_{0(12)}\Sigma^{12}_P &\GSigI\cr
G_{23}\Sigma^2_P&= G_{23}\cZ_{30}\Sigma^{23}_P = -\half(2\cZ_{23} +
\cZ_{30})\Sigma^{23}_P \ , &\GSigII
}$$
where we used $\Sigma_{123 \ldots n} = \cZ_{i0}\Sigma^i_{123 \ldots n} =
\cZ_{i0}\cZ_{j0}\Sigma^{ij}_{123 \ldots n} = \ldots$
etc as well as
\eqnn\twocases
$$\eqalignno{
G_{12}\cZ_{10}\cZ_{20} &= -\half\({\s_{10}\over \s_{12}}
+ {\s_{20}\over \s_{12}}\){1\over \s_{10}\s_{20}}=
-\half\({1\over \s_{12}\s_{20}}+{1\over \s_{12}\s_{10}}\)=
-\half(\cZ_{120}-\cZ_{210})\cr
G_{23}\cZ_{30} &=
-\half\({\s_{20}\over \s_{23}}
+ {\s_{30}\over \s_{23}}\){1\over \s_{30}}=-\half
\({2\over \s_{23}}+{1\over \s_{30}}\) = -\half(2\cZ_{23}+\cZ_{30})\,.
&\twocases
}$$
The general case $G_{i_1i_2} \ldots G_{i_{2p-1}i_{2p}}\Sigma_P$ in
\equivGij\ can be proven by induction
using \GSigI, \GSigII, \elemI, \elemII\ and starting
with \GSigI
\eqn\start{
G_{12}\Sigma_P = \half \sign_{12}\cZ_{0(12)}\Sigma^{12}_P =
\half \sign_{12}\cZ_{0(P)}\,.
}
The induction step leads to two cases for
an additional propagator $G_{ij}$ multiplying the left-hand side of \start.
When there is no overlap between $G_{ij}$ and the previous propagators,
\eqnn\secc
$$\eqalignno{
G_{12}G_{34}\Sigma_P &= \half \sign_{12}\cZ_{0(12)}\(G_{34}\Sigma^{12}_P\)
= {1\over 4}\sign_{12}\sign_{34}\cZ_{0(12)}\cZ_{0(34)}\Sigma^{1234}_P\cr
&={1\over 4} \sign_{12}\sign_{34}\cZ_{0(1234)}\Sigma^{1234}_P
= {1\over 4} \sign_{12}\sign_{34}\cZ_{0(P)}\,,&\secc
}$$
where we used \elemI\ and \lemma\ on the last line.
If there is an overlap with the previous propagators one gets instead,
\eqnn\tercc
$$\eqalignno{
G_{12}G_{23}\Sigma_P &= \half \sign_{12}\cZ_{0(12)}\(G_{23}\Sigma^{12}_P\) =
-{1\over 4} \sign_{12}
\cZ_{0(12)}(2\cZ_{23}+\cZ_{30})\Sigma^{123}_P\cr
&=-{1\over 4} \sign_{12}\sign_{23}\cZ_{0(123)}\Sigma^{123}_P
={1\over 4} \sign_{12}\sign_{23}\cZ_{0(P)}\,, &\tercc
}$$
where we used \elemII\ to arrive at the second line.
Since these steps can be freely iterated, it is now easy to see that each
additional propagator $G_{ij}$ leads to a factor of
$\half \sign_{ij}$ on the right-hand side of \equivGij,
finishing its proof.~\qed

%%%%%%
\appendix{C}{Theta functions and $q$ series}
\applab\apptheta
%******************************************************

\noindent
Even Jacobi theta functions are defined by
\eqnn\thetafct
$$\eqalignno{
\theta_2(z,\tau) &=
2q^{1/8}\cos( \pi z) \prod_{j=1}^{\infty}  ( 1 -  q^{j} ) ( 1+e^{2\pi i z} q^{j}) ( 1+ e^{-2\pi i z} q^{j} )
\cr
\theta_3(z,\tau) &=  \prod_{j=1}^{\infty}  ( 1 -  q^{j} ) ( 1+e^{2\pi i z} q^{j-1/2}) ( 1+ e^{-2\pi i z} q^{j-1/2} ) &\thetafct
\cr
\theta_4(z,\tau) &=  \prod_{j=1}^{\infty}  ( 1 -  q^{j} ) ( 1-e^{2\pi i z} q^{j-1/2}) ( 1- e^{-2\pi i z} q^{j-1/2} )
\ .
}$$
These definitions yield the following leading $q$-orders for the Szeg\"o kernel \szeg,
\eqnn\checkszeg
$$\eqalignno{
S_2(z_{ij},\tau) |_{q^0} &= i \pi {\sigma_i+\sigma_j\over \sigma_i-\sigma_j} \cr
S_3(z_{ij},\tau) |_{q^0} &= 2 \pi i {\sqrt{\sigma_i\sigma_j} \over \sigma_i-\sigma_j}   &\checkszeg \cr
S_3(z_{ij},\tau) |_{q^{1/2}} &=  2\pi i {\sigma_i-\sigma_j\over\sqrt{\sigma_i\sigma_j}} \ ,
}$$
where $\sigma_j = e^{2\pi i z_j}$.

The contributions to the $\tau \rightarrow i \infty$ limit of the ambitwistor-string and superstring correlators
are selected by the partition functions
\eqn\leadpart{
\left[ {\theta_2(0,\tau)
}\over{\theta_1'(0,\tau)} \right]^4 = {1 \over (2\pi i)^4} \big[ 16+ {\cal O}(q) \big]
\ , \ \ \ \ \
\left[ {\theta_{3,4}(0,\tau)
}\over{\theta_1'(0,\tau)} \right]^4 = {1 \over (2\pi i)^4} \left[ {1\over \sqrt{q}} \pm 8 + {\cal O}(q^{1/2}) \right] \ .
}

\appendix{D}{Five-point example with maximal supersymmetry}
\applab\appexample
%******************************************************

\noindent This appendix is devoted to a maximally supersymmetric five-point example to illustrate the procedure of section \secRNS\
to express one-loop CHY integrands as a polynomial in $\ell$ and $G_{ij}={\sigma_i+\sigma_j \over 2\sigma_{ij}}$.
The starting point is the $5!$-term expansion \calK\ of the five-point correlator,
\eqnn\appCA
$$\eqalignno{
{\cal K}_5 &=
 c_1 c_2 c_3 c_4 c_5{\cal G}_{\emptyset} + \left( c_1 c_2 c_3 {\rm tr}(f_{(45)}) {\cal G}_{(45)} + {\rm 9~more} \right) + \left( c_1 c_2 {\rm tr}(f_{(345)}) {\cal G}_{(345)} + {\rm 19~more} \right)\quad & \cr
& \ \ \ \  +\left( c_1 {\rm tr}(f_{(2345)}) {\cal G}_{(2345)}+ {\rm 29~more} \right)+\left( c_1 {\rm tr}(f_{(23)}){\rm tr}(f_{(45)}) {\cal G}_{(23)(45)}+ {\rm 14~more} \right) &\appCA \cr
& \ \ \ \  + \left( {\rm tr}(f_{(12345)}) {\cal G}_{(12345)}+ {\rm 23~more} \right)+ \left( {\rm tr}(f_{(12)}) {\rm tr}(f_{(345)}) {\cal G}_{(12)(345)} + {\rm 19~more} \right) \,,
}$$
see \trf\ and \Cii\ for the polarization-dependent ingredients ${\rm tr}(f_{I})$ and $c_i(\ell)$. In case of
maximal supersymmetry, spin sums ${\cal G}_{I,J}$ with three or fewer particles in the union of the cycles $I,J$ vanish, see
section \maxspinsum. Their four-point instances in turn are given by ${\cal G}_{(ij)(kl)}= {\cal G}_{(ijkl)}=1$, and
five-point cases give rise to linear functions \maxGfive\ in $G_{ij}$. Hence, one can collect the coefficients
of $\ell$ and $G_{ij}$ in \appCA:
\eqnn\appCB
\eqnn\appCC
\eqnn\appCD
$$\eqalignno{
{\cal K}_5 &= \ell_m T^m_{1,2,3,4,5} + \big[  G_{12} T_{12,3,4,5} + (1,2|1,2,3,4,5) \big] &\appCB
\cr
T^m_{1,2,3,4,5}  &= e^m_1 \Big({1\over 4}{\rm tr}(f_{2}f_{3}) {\rm tr}(f_{4}f_{5}) - {\rm tr}(f_{2}f_{3}f_{4}f_{5}) + {\rm cyc}(3,4,5) \Big) + (1\leftrightarrow 2,3,4,5)  &\appCC
\cr
T_{12,3,4,5} &=
(e_1 \cdot k_2)\Big({1\over 4} {\rm tr}(f_{2}f_{3}) {\rm tr}(f_{4}f_{5}) - {\rm tr}(f_{2}f_{3}f_{4}f_{5})+ {\rm cyc}(3,4,5) \Big)  &\appCD
\cr
&- (e_2 \cdot k_1)\Big( {1\over 4}{\rm tr}(f_{1}f_{3}) {\rm tr}(f_{4}f_{5}) - {\rm tr}(f_{1}f_{3}f_{4}f_{5})+ {\rm cyc}(3,4,5) \Big) \cr
%
%&+ \Big( {1\over 2}{\rm tr}(f_{1}f_{2}f_{3}) {\rm tr}(f_{4}f_{5})-  {\rm tr}(f_{1}f_{2}f_{3}f_{4}f_{5})   + {\rm perm}(3,4,5) \Big)
&+ \Big( {1\over 2}{\rm tr}(f_{1}f_{2}f_{3}) {\rm tr}(f_{4}f_{5}) + {\rm cyc}(3,4,5)\Big) - \Big(  {\rm tr}(f_{1}f_{2}f_{3}f_{4}f_{5})   + {\rm perm}(3,4,5) \Big)
 \ .&
}$$
These expressions can be streamlined using the two-particle field-strength
\eqnn\appCE
$$\eqalignno{
s_{12}\cf^{mn}_{12}&=
e_1 \cdot e_2(k_2^m k_1^n -k_1^m k_2^n)
+s_{12}(e_2^m e_1^n- e_1^m e_2^n)&\cr
&
+\Big(k_2 \cdot e_1 (k_2^m e_2^n -e_2^m k_2^n+ k_1^m e_2^n -e_2^m k_1^n)- (1\leftrightarrow2)\Big)
&\appCE
}
$$
obtained as a special case of \rectwo\ as well as the definition $t_8$-tensor in \teights:
\eqnn\appCF
\eqnn\appCG
$$\eqalignno{
T^m_{1,2,3,4,5}  &= e_1^mt_8( \cf_{2} ,  \cf_{3} , \cf_{4} , \cf_{5} )+ (1\leftrightarrow 2,3,4,5)
&\appCF
\cr
T_{12,3,4,5} &= s_{12} t_8( \cf_{12} ,  \cf_{3} , \cf_{4} , \cf_{5} ) = s_{12} t_{12,3,4,5}  \ .
&\appCG
}$$
The dictionary \notmaster\ then implies the pentagon numerator
\eqnn\appCH
$$\eqalignno{
N_{+|12345|-} &= \ell_m T^m_{1,2,3,4,5} - {1\over 2} (T_{12,3,4,5} + T_{13,2,4,5}+T_{14,2,3,5}+T_{15,2,3,4} &\appCH
\cr
& \ \ \ +T_{23,1,4,5}+T_{24,1,3,5} + T_{25,1,3,4}+T_{34,1,2,5}+T_{35,1,2,4}+T_{45,1,2,3}) \ ,
}$$
and the corresponding box numerators determined by the BCJ duality collapse to
\eqn\boxfivept{
N^{\rm box}_{12} =  N_{+|12345|-}  - N_{+|21345|-}  = - T_{12,3,4,5}
}
for the box diagram with legs 1 and 2 in a massive corner. The resulting partial integrand can be assembled via
\analyticPI\ and comprises four box diagrams, see example C of \HeMZD:
\eqnn\appCI
$$\eqalignno{
a(1,2,3,4,5,-,+) &= { N_{+|12345|-}  \over s_{1,\ell} s_{12,\ell} s_{123,\ell} s_{1234,\ell}}  - {T_{12,3,4,5} \over s_{12} s_{12,\ell} s_{123,\ell} s_{1234,\ell} }
- { T_{1,23,4,5} \over s_{23} s_{1,\ell} s_{123,\ell}  s_{1234,\ell}}   \cr
& \ \ \  - { T_{1,2,34,5} \over s_{34} s_{1,\ell}  s_{12,\ell} s_{1234,\ell}}
- { T_{1,2,3,45} \over s_{45} s_{1,\ell} s_{12,\ell} s_{123,\ell}}  &\appCI
}$$
After eliminating any $G_{1j}$ via scattering equations, the functions in \appCB\ are converted to a basis. Their coefficients
are then gauge invariant and match the bosonic components \Cfourfive\ of the five-point correlator \PSfive\ in
pure-spinor superspace. Hence, the same conclusions can be obtained by taking the $\tau \rightarrow i\infty$ limit
of superstring correlators in the RNS formalism \refs{\TsuchiyaVA, \StiebergerWK, \canceltriag} or the pure-spinor
formalism \refs{\MafraKH}.

%%%%%%
\appendix{E}{Parity-odd correlators from the ambitwistor string}
\applab\appparityodd
%******************************************************

\noindent
As explained in \AdamoTSA, the parity-odd contributions to the $d$-dimensional RNS correlators of
section \parodd\ can be represented as a Pfaffian
\eqn\oddPFAFF{
{\cal K}_n^{\epsilon_d} = i\int \dd^d \Psi_0 \ {\rm Pf}\pmatrix{
A &-C^T\cr
C&B
} \ .
}
The Grassmann integral requires the saturation of all the $d$ zero-mode components $\Psi^m_0$ of the worldsheet fermions
in their odd spin structure,
\eqn\zmint{
\int \dd^d \Psi_0 \ \Psi_0^{m_1} \,\Psi_0^{m_2}\, \ldots \, \Psi_0^{m_{d}} = \epsilon_d^{m_1m_2\ldots m_d} \ .
}
In the $\tau \rightarrow i\infty$ limit, the entries of the $n\times n$ blocks $A, B$ and $C$ in \oddPFAFF\ are given as follows: In the off-diagonal cases with $i\neq j$, we have
\eqnn\pfaO
$$\eqalignno{
&A_{i\,j}=k_i\cdot k_j \, G_{ij}
%S_{1}(\sigma_i, \sigma_j ,q)
+k_i\cdot \Psi_0 \, k_j\cdot \Psi_0\,,\quad{\rm for} ~i,j\neq 1& \cr
&B_{i\,j}=e_i \cdot e_j \, G_{ij}+e_i\cdot \Psi_0 \, e_j\cdot \Psi_0\,,& \cr
&C_{i\,j}=e_i \cdot k_j \, G_{ij}+e_i\cdot \Psi_0 \, k_j\cdot \Psi_0\,,\quad{\rm for}~ i\neq 1 \ , &\pfaO
}$$
while the diagonal entries are given by
\eqnn\pfaOd
$$\eqalignno{
&A_{i\,i}=B_{i\,i}=0\,, &\cr
& C_{i\,i}=-e_i \cdot \ell-\sum^n_{j\neq i} e_i \cdot k_j \, G_{ij}
-e_i\cdot \Psi_0 \, k_i\cdot \Psi_0\,,\quad{\rm for}~ i\neq 1 \ .&\pfaOd
}$$
In the first row or column with $i=1$, the entries of $A$ and $C$ are modified to
\eqnn\pfaOdd
$$\eqalignno{
&A_{1\,j}=P(\sigma_0)\cdot k_j  \, G_{0j} +P(\sigma_0)\cdot \Psi_0 \, k_i\cdot \Psi_0\,,\quad{\rm for}~ j\neq 1&\cr
&C_{j\,1}=e_j \cdot P(\sigma_0) \, G_{j0} + e_j\cdot \Psi_0\,P(\sigma_0)\cdot \Psi_0  \ , &\pfaOdd \cr
}$$
where the picture changing operator of the RNS string contributes a factor of
\eqn\picchange{
P^m(\sigma_0) = \ell^m + \sum_{j=1}^n  k_j^m G_{0j} \ .
}
Since the Pfaffian in \oddPFAFF\ is a polynomial of degree $n{+}1$ in $G_{ij},\ell$ and $(\Psi_0 \Psi_0)$, the zero-mode integral \zmint\ leaves a polynomial of degree $n{+}1-{d\over 2}$ in $G_{ij}$ and $\ell$. Note that the correlator ${\bf K}_n^{\epsilon_d}(\tau)$ at finite values of $\tau$ can be easily obtained from \pfaOd\ and \pfaOdd\ by replacing $G_{ij} \rightarrow \partial \log \theta_1(z_{ij})$.

\listrefs

\bye